\documentclass{jfm}
\usepackage{graphicx}
\usepackage{epstopdf, epsfig}
\usepackage{amsmath}
\usepackage{adjustbox}
\usepackage{upgreek}
\usepackage{comment}
\usepackage{booktabs}
\usepackage{amssymb}

\shorttitle{LESs of wind farms operating in CNBLs}
\shortauthor{L. Lanzilao and J. Meyers}

\title{A parametric large-eddy simulation study of wind-farm blockage and gravity waves in conventionally neutral boundary layers}

\author{L. Lanzilao\aff{1} \corresp{\email{luca.lanzilao@kuleuven.be}}
	\and J. Meyers\aff{1}}

\affiliation{\aff{1}KU Leuven, Department of Mechanical Engineering, Celestijnenlaan 300 – box 2421, B-3001 Leuven, Belgium} 

\begin{document}

\maketitle

\begin{abstract}
We present a suite of large-eddy simulations of a wind farm operating in conventionally neutral atmospheric boundary layers (CNBLs). A fixed 1.6 GW wind farm is considered for 40 different atmospheric stratification conditions to investigate effects on wind-farm efficiency and blockage, as well as related gravity-wave excitation. A tuned Rayleigh damping layer and a wave-free fringe region method (Lanzilao \& Meyers, Bound. Layer Meteor. 186, 2023) are used to avoid spurious excitation of gravity waves, and a domain-size study is included to evaluate and minimize effects of artificial domain blockage. A fully neutral reference case is also considered, to distinguish between a case with hydrodynamic blockage only, and cases that include hydrostatic blockage excited by gravity waves. We discuss in detail the dependence of gravity-wave excitation, flow fields, and wind-farm blockage on capping-inversion height, strength and free-atmosphere lapse rate. In all cases, an unfavourable pressure gradient is present in front of the farm, and a favourable pressure gradient in the farm, with hydrostatic contributions arising from gravity waves at least an order of magnitude larger than hydrodynamic effects. Using respectively non-local and wake efficiencies $\eta_\textit{nl}$ and $\eta_\textit{w}$ (Allaerts \& Meyers, Bound. Layer Meteor. 166, 2018), we observe a strong negative correlation between unfavourable upstream pressure rise and $\eta_\textit{nl}$, and a strong positive correlation between the favourable pressure drop in the farm and $\eta_\textit{w}$. Using a simplified linear gravity-wave model, we formulate a simple scaling for $\eta_\textit{nl}/\eta_\textit{w}$, which matches reasonably well with the LES results.
\end{abstract}


\section{Introduction}\label{sec:intoduction}
Conventionally neutral boundary layers (CNBLs) often occur in offshore conditions, with air temperatures adapting to the sea-water temperature given a sufficiently large offshore fetch \citep{Csanady1974,Smedman1997,Lange2004}. Such boundary layers are characterized by a neutral stratification, but with a boundary layer height that is often capped by a strong stably stratified inversion layer (the capping inversion) and a stably stratified free atmosphere aloft; conditions that are driven by larger weather-scale circulation patterns. When large wind farms are operated in a CNBL, they may excite gravity waves consisting of two-dimensional interface waves on the capping inversion and three-dimensional internal waves in the atmosphere above \citep{Smith2010,Allaerts2017}. This can lead to so-called wind-farm blockage, leading to a significant slow-down of wind speeds in front of the farm, and reducing the overall wind-farm efficiency \citep{Allaerts2018,Bleeg2018}. With the current and future plans for large offshore wind-farm developments across the world, a better understanding of the interaction of wind farms with CNBLs is necessary.  

To date, the number of large-eddy simulation (LES) studies of wind farms operating in conventionally neutral boundary layers (CNBLs) is limited to a handful of cases. This is mainly due to two facts. First, wind-farm simulations in CNBLs require larger numerical domains than simulations that do not consider thermal stratification above the atmospheric boundary layer (ABL), since the farm has a larger footprint in such conditions \citep{Allaerts2017,Maas2023}. Second, the presence of gravity waves requires the use of appropriate methods for inflow and outflow conditions together with non-reflective upper boundary conditions. Only very recently, a first approach was proposed, based on a wave-free fringe-region technique \citep{Lanzilao2022b}, that does not excite spurious waves at in- and outlet of the simulation domain, while still allowing for a turbulent inflow in large-eddy simulations. In the current work, we use this approach to set-up a large simulation study that focuses on the influence of thermal stratification above the ABL on wind-farm performance and wind-farm blockage. Moreover, we carefully investigate the effect of domain size on possible artificial domain blockage in case of too small computational domains.

A large part of wind-farm--LESs performed in the past decade made use of pressure-driven boundary layers (PDBLs), which refer to ABLs without Coriolis forces, wind veer and free-atmosphere stratification. This simplified description of the atmosphere is reasonable when turbines are located in the surface layer, where the Coriolis force and boundary-layer height effects are negligible.  Early PDBL wind-farm simulations were, e.g., performed by \cite{Meyers2010} and \cite{Calaf2010}, \cite{Wu2011}, \cite{Lu2011}, \cite{Yang2014a} and \cite{Yang2014b}. These early studies were characterized by the assumption of `infinite' wind farms, using periodic boundary conditions in all directions, allowing for small simulation domains. With the increase in computational resources, semi-finite and finite wind-farm–LESs were performed, with the goal of investigating the flow behaviour also in regions surrounding the farm. Examples of this type of studies are given by \cite{PorteAgel2013}, \cite{Wu2013}, \cite{Wu2015}, \cite{Stevens2014}, \cite{Stevens2014b}, \cite{Stevens2015}, \cite{Wu2019}, and \cite{Stieren2022}. We note that many more PDBL simulations have been presented in the past, including those looking at stable or unstable surface-layer stratification. We refer to \cite{PorteAgel2020} for an extensive overview. 

With wind turbines growing in size, the assumption that wind farms operate in the inner part of the ABL is more and more questionable. Therefore, Coriolis forces need to be added to the governing equations, giving rise to the Ekman spiral in the ABL. This type of flow, if no (stable) stratification is present in the free atmosphere (nor the surface layer), is defined as a truly neutral boundary layer (TNBL) in the literature. For instance, the simulations in TNBLs performed by \cite{Goit2013} and \cite{Vanderlaan2015} clearly show the importance of considering the Coriolis force. However, the equilibrium height of the TNBL can be several kilometers high, scaling with the Rossby--Montgomery height. In practice, this situation rarely occurs, as the free atmosphere is usually stratified starting from 0.5 km to 1 km above the ground, damping turbulence and impeding further boundary layer development. The importance of the inversion layer and stable free atmosphere on the flow within the ABL was, e.g., noted by \cite{Csanady1974} and \cite{Zilitinkevich2002}. \cite{Hess2004} characterized the entrainment of momentum on the boundary layer as a function of the height of the capping inversion using LES and direct numerical simulation (DNS) while \cite{Zilitinkevich2003,Zilitinkevich2005} and \cite{Zilitinkevich2007} improved the equilibrium height formulation for the CNBL. More recently, \cite{Taylor2007,Taylor2008} and \cite{Pedersen2014} used LESs to investigate how the capping inversion and free-atmosphere stratification modify the temporal evolution of the CNBL profiles. 

Shortly after, CNBLs started to be used also in LESs of wind-farm. \cite{Churchfield2012} and \cite{Archer2013} were among the first to perform wind-farm  LES in CNBLs using SOWFA, an OpenFOAM based LES solver. However, both studies mostly focused on wind-farm wakes and turbine--turbine interactions, without reporting on the effects induced by the presence of a capping inversion and a stably-stratified free atmosphere. \cite{Abkar2013} and \cite{Allaerts2015} investigated the farm performance and the vertical entrainment of kinetic energy in the ABL under various free-atmosphere stratifications adopting an infinite farm (with periodic boundary conditions). Later, \cite{Allaerts2017} explored wind-farm operation in CNBLs using a farm with finite length in the streamwise direction. Here, the vertical domain dimension was extended up to 25 km, to allow for a proper Rayleigh damping layer at the top of the domain, since in a semi-finite wind-farm set-up, internal gravity waves can be triggered. They found that the flow divergence induced by the farm pushes upward the inversion layer, generating a cold anomaly  which in turn leads pressure feedbacks and a slow-down of the flow in front of the farm. This result was earlier predicted as well by \cite{Smith2010} based on a linear-theory model. Various more recent studies have further investigated this behaviour both using LES \citep{Allaerts2018,Wu2017,Maas2022,Maas2022b,Maas2023,Lanzilao2022,Lanzilao2022b} and much faster linearized wind-farm flow models \citep{Smith2022,Smith2023,Allaerts2019,Devesse2022}.

With the field measurement campaign of \cite{Bleeg2018}, and later \cite{Schneemann2021}, demonstrating upstream slow-down of the wind speed in the order of $4~\pm~2~\%$ in operational wind farms, a lot of research has started focusing on investigating wind-farm blockage. A large part of the literature has concentrated on hydrodynamic blockage (i.e. the joined induction of all turbines in the farm) as a root mechanism to explain blockage, using simple analytical models \citep{Branlard2020,Branlard2020b,Centurelli2021,Segalini2021}, numerical simulations \citep{Bleeg2022,Strickland2020,Strickland2022} and wind-tunnel experiments \citep{Medici2011,Segalini2019}. These studies report reductions in wind speed at turbine-hub height in the order of 1\% to 2\%. In CNBL simulations, where next to hydrodynamic effects, also hydrostatic effects arising from gravity waves are present, much larger wind-speed reductions in the order of 10$\%$ and more have been reported \citep{Allaerts2017,Maas2022b}. Although these studies were performed with a semi-infinite farm, \cite{Lanzilao2022,Lanzilao2022b} and \cite{Maas2023} noted similar behaviour in fully finite farms. Comparing these results with field measurements is rather difficult. In fact, gravity-wave effects extend over distances of several tens of kilometers, which makes them difficult to detect using traditional lidar systems. However, recently, analysis of Supervisory Control and Data Acquisition (SCADA) data from the Nordsee Ost and Amrumbank West wind farms located in the German Bight area have shown that velocity deficits and flow-blockage effects are strongly influenced by the capping-inversion height \citep{Canadillas2023}. In the current manuscript, we aim to further investigate relations between wind-farm blockage and capping-inversion height, strength, and free lapse rate, using the LES suite that we present. 

This manuscript is further structured as follows. The simulation setup is elaborated in Section~\ref{sec:methodology}. Thereafter, Section~\ref{sec:bl_initialization} discusses the boundary-layer initialization. Next, the sensitivity of the farm performance to the width and length of the numerical domain is reported in Section~\ref{sec:domain_sensitivity} while the sensitivity to the atmospheric state is shown in Section~\ref{sec:atm_sensitivity}. Finally, conclusions are drawn in Section~\ref{sec:conclusions}.

\section{Methodology}\label{sec:methodology}
The governing equations and the LES solver are described in Sections~\ref{sec:governing_equation} and \ref{sec:flow_solver}, respectively. Next, the boundary conditions and the buffer layers adopted to minimize wave reflection are discussed in Section~\ref{sec:boundary_conditions}. Finally, the numerical setup, wind-farm layout and atmospheric states are summarized in Sections~\ref{sec:numerical_setup}, \ref{sec:wf_layout} and \ref{sec:atm_state}, respectively.

\subsection{Governing equations}\label{sec:governing_equation}
In the current study, we make use of the incompressible filtered Navier--Stokes equations coupled with a transport equation for the potential temperature to investigate the flow in and around a large-scale wind farm \citep{Allaerts2017}. Such equations read as
\begin{align}
	& \frac{\partial \tilde{u}_i}{\partial x_i} = 0, \label{eq--continuity} \\
	& \frac{\partial \tilde{u}_i}{\partial t} + \frac{\partial}{\partial x_j} \bigl(\tilde{u}_j \tilde{u}_i \bigl) =  f_c \epsilon_{ij3} \tilde{u}_j + \delta_{i3} g \frac{\tilde{\theta}-\theta_0}{\theta_0} - \frac{\partial \tau_{ij}^\mathrm{sgs}}{\partial x_j} - \frac{1}{\rho_0} \frac{\partial \tilde{p}^\ast}{\partial x_i} - \frac{1}{\rho_0} \frac{\partial p_\infty}{\partial x_i} + f_i^\mathrm{tot}, \label{eq--momentum} \\
	& \frac{\partial \tilde{\theta}}{\partial t} + \frac{\partial }{\partial x_j} \bigl( \tilde{u}_j \tilde{\theta} \bigl) = - \frac{\partial q_j^\mathrm{sgs}}{\partial x_j}, \label{eq--thermodynamic}
\end{align}
where the horizontal directions are denoted with $i=1,2$ while the vertical one is indicated by $i=3$. Moreover, $\delta_{ij}$ denotes the Kronecker delta while $\epsilon_{ijk}$ is the Levi-Civita symbol. The filtered velocity and potential temperature fields are noted with $\tilde{u}_i$ and $\tilde{\theta}$, respectively. 

The first term on the right-hand side represents the Coriolis force due to planetary rotation, where the frequency $f_c = 2 \Omega_E \sin{\phi}$, with $\Omega_E$ the Earth angular velocity and $\phi$ the Earth latitude. The second component of the angular velocity vector $\Omega_E \cos{\phi}$ is neglected here since it is negligible when compared to the other terms in the momentum equations \citep{Wyngaard2010}. Thermal buoyancy is taken into account by the second term, where $g=9.81$ m s$^{-2}$ denotes the gravitational constant and $\theta_0$ is a reference potential temperature. Moreover, we make use of the Boussinesq approximation so that the incompressible continuity equation holds. This assumption has two implications. First, fluctuations in density are related to thermal effects rather than pressure ones, so that acoustic waves are filtered out. Second, all density variations from the background state are neglected except for the buoyancy term. Consequently, the thermodynamic equation has a direct influence only on the vertical momentum equation. This is a valid assumption for our study since the scale of the vertical motions is much smaller than the density scale height, which is typically in the order of 7 km \citep{Spiegel1960,AllaertsPhD}. Moreover, \cite{Maas2022b} performed two wind-farm–LESs, one with Boussinesq approximation and one with the anelastic assumption. He found nearly identical numerical results at turbine-hub height, with only minor differences several kilometers above the ABL. 

The flow is driven across the domain by applying a steady background pressure gradient $\partial p_\infty / \partial x_i$, with $i=1,2$. The latter is related to the geostrophic wind $G$ through the geostrophic balance. The pressure oscillations around $p_\infty$ are denoted with $\tilde{p}^\ast$. Moreover, the term $f_i^\mathrm{tot} = f_i + f_i^\mathrm{ra} + f_i^\mathrm{fr}$ groups all external forces exerted on the flow. Here, $f_i^\mathrm{ra}$ and $f_i^\mathrm{fr}$ represent the body forces applied within the RDL and fringe region, respectively, while $f_i$ denotes the wind-turbine drag force. Finally, the effects of unresolved scales are modelled by the subgrid-scale stress tensor $\tau_{ij}^\mathrm{sgs}$ and the subgrid-scale heat flux $q_j^\mathrm{sgs}$. The notations $(x_1,x_2,x_3)$ and $(x,y,z)$, $(\tilde{u}_1,\tilde{u}_2,\tilde{u}_3)$ and $(\tilde{u},\tilde{v},\tilde{w})$ and $\tilde{p}^\ast$ and $\tilde{p}$ are used interchangeably. Moreover, for the sake of simplicity, the tilde will not be used in the rest of the dissertation.

\subsection{Flow solver}\label{sec:flow_solver}
The governing equations (\ref{eq--continuity}-\ref{eq--thermodynamic}) are solved using the SP-Wind solver, an in-house software developed over the past 15 years at KU Leuven \citep{Meyers2007,Calaf2010,Goit2015,Allaerts2017,Munters2018,Allaerts2017b,Lanzilao2022,Lanzilao2022b}. The solver structure adopted here is mainly based on the version developed and used in \cite{Allaerts2017} and \cite{Lanzilao2022,Lanzilao2022b}. The equations are advanced in time using a classic fourth-order Runge--Kutta scheme with a time step based on a Courant--Friedrichs--Lewy number of $0.4$. The streamwise~($x$) and spanwise ($y$) directions are discretized with a Fourier pseudo-spectral method. This implies that all linear terms are discretized in the spectral domain while non-linear operations are computed in the physical domain, reducing the cost of convolutions from quadratic to log-linear \citep{Fornberg1996}. Further, the 3/2 dealiasing technique proposed by \cite{Canuto1988} is adopted to avoid aliasing error. For the vertical dimension~($z$), an energy-preserving fourth-order finite difference scheme is adopted~\citep{Verstappen2003}. The effects of subgrid-scale motions on the resolved flow are taken into account with the stability-dependent Smagorinsky model proposed by \cite{Stevens2000} with Smagorinsky coefficient set to $C_s=0.14$, similarly to previous study performed with SP-Wind. The constant $C_s$ is damped near the wall by using the damping function proposed by \cite{Mason1992}. Furthermore, continuity is enforced by solving the Poisson equation during every stage of the Runge--Kutta scheme. In regard to the turbine trust force, we model it using a non-rotating actuator disk model (ADM) \citep{Goit2015,Allaerts2015}. We refer to \cite{DelportPhD} for more details on the discretization of the continuity and momentum equations while the implementation of the thermodynamic equation and SGS model are explained in detail in \cite{AllaertsPhD}.

\subsection{Boundary conditions}\label{sec:boundary_conditions}
The effect of the bottom wall on the flow is modelled with classic Monin--Obukhov similarity theory for neutral boundary layers \citep{Moeng1984}. This wall-stress boundary condition is only dependent on the surface roughness $z_0$, which we assume to be constant. We refer to \cite{AllaertsPhD} for further details on the implementation. Periodic boundary conditions are naturally imposed at the streamwise and spanwise sides of the computational domain. At the top of the domain, a rigid-lid condition is used, which implies zero shear stress and vertical velocity and a fixed potential temperature. However, in case of stratified free atmospheres, a rigid-lid condition reflects back gravity waves triggered by the wind-farm drag force. To minimize gravity-wave reflection, we adopt a Rayleigh damping layer  (RDL) in the upper part of the domain \citep{Klemp1977,Durran1983,Allaerts2017,Lanzilao2022b}. This body force dissipates the upward energy transported by gravity waves before it reaches the top of the domain and it reads as:
\begin{equation}
	f_i^\mathrm{ra} (\boldsymbol{x}) = - \nu(z) \biggl( u_i(\boldsymbol{x}) - U_{G,i} \biggl)
\end{equation}
where $U_{G,1} = G \cos{\alpha}$, $U_{G,2} = G \sin{\alpha}$ and $U_{G,3} = 0$ with $\alpha$ the geostrophic wind angle. Moreover, $\nu(z)$ is a one-dimensional function which reads as:
\begin{equation}
	\nu(z) = \nu^\mathrm{ra} N \biggl[1 - \cos{\biggl( \frac{\pi}{s^\mathrm{ra}} \frac{z - \bigl(L_z-L_z^\mathrm{ra} \bigl)}{L_z^\mathrm{ra}}\biggl)} \biggl]
\end{equation} 
where $L_z$ and $L_z^\mathrm{ra}$ denote the height of the computational domain and of the RDL, respectively, while $N = \sqrt{g \Gamma/ \theta_0}$ represents the Brunt--V\"{a}is\"{a}l\"{a} frequency, with $\Gamma$ the lapse rate in the free atmosphere. Moreover, $\nu^\mathrm{ra}$ controls the force magnitude while $s^\mathrm{ra}$ regulates the function gradient along the vertical direction. \cite{Lanzilao2022b} have shown that the quality of the RDL strongly depends on these two tuning parameters, which are carefully tuned with the aim of minimizing gravity-wave reflection -- see Section~\ref{sec:numerical_setup}.

The periodic boundary condition along the streamwise direction recycles back the wind-farm wake. To break the periodicity and impose an inflow condition, we use a fringe technique \citep{Spalart1993,Lundbladh1999,Nordstrom1999,Stevens2014,Inoue2014,Munters2016,Lanzilao2022b}. This body force reads as:
\begin{equation*}
	f_i^\mathrm{fr}(\boldsymbol{x}) = -h(x) \biggl( u_i(\boldsymbol{x}) - u_{\mathrm{prec},i}(\boldsymbol{x})\biggl),
\end{equation*}
where $u_{\mathrm{prec},i}(\boldsymbol{x})$ denotes the statistically-steady inflow fields provided by a precursor simulation. Moreover, $h(x)$ is a one-dimensional non-negative function which is non-zero only within the fringe region, and is expressed as 
\begin{equation*}
	h(x) = h_\mathrm{max} \biggl[ F\biggl( \frac{x-x_s^h}{\delta_s^h}\biggl) -  F\biggl( \frac{x-x_e^h}{\delta_e^h} +1\biggl) \biggl]
\end{equation*}
with
\begin{equation*}
	F(x) = \begin{cases}
		0,& \text{if } x \leq 0\\
		\dfrac{1}{1+\text{exp}\biggl( \dfrac{1}{x-1} + \dfrac{1}{x}\biggl)},& \text{if } 0<x<1\\
		1,& \text{if } x \geq 1.
	\end{cases}
\end{equation*}
The parameters $x_s^h$ and $x_e^h$ denote the start and end of the fringe function support while its smoothness is regulated by $\delta_s^h$ and $\delta_e^h$. Moreover, $h_\mathrm{max}$ denotes the maximum value of the fringe function. 

\cite{Lanzilao2022b} noted that the standard fringe technique triggers spurious gravity waves which propagate through the domain of interest, significantly altering the numerical results. Therefore, in the current study we use the new wave-free fringe-region technique developed by \cite{Lanzilao2022b}. In addition to applying the body force $f_i^\mathrm{fr}$ described above, this technique also damps the convective term in the vertical momentum equation within the fringe region, multiplying it by the following damping function:
\begin{equation}
	d(x,z) = 1 -  \biggl[ F\biggl( \frac{x-x_s^d}{\delta_s^d}\biggl) -  F\biggl( \frac{x-x_e^d}{\delta_e^d} +1\biggl) \biggl] \mathcal{H}(z-H).
	\label{eq--damping_function}
\end{equation}
Here, $x_s^d$ and $x_e^d$ define the start and end of the damping function support while $\delta_s^d$ and $\delta_e^d$ control the function smoothness. Moreover, $H$ denotes the capping-inversion height while $\mathcal{H}$ represents a Heaviside function. For more details, we refer to \cite{Lanzilao2022b} and to Section \ref{sec:numerical_setup} below.

\subsection{Numerical set-up}\label{sec:numerical_setup}
The flow solver makes use of two numerical domains concurrently marched in time, i.e. the precursor and main domains. The precursor domain does not contain turbines and is only used for generating a turbulent fully developed statistically steady flow which drives the simulation in the main domain. Similarly to \cite{Allaerts2017,Allaerts2017b}, we fix the precursor domain length and width to $L_x^p=L_y^p=10$ km, with $L_z^p=3$ km. The wind farm is located in the main domain. The first-row turbine should be far enough from the inflow to properly capture the flow slow down in the farm induction region. Moreover, the last-row turbine should be far enough from the fringe region to minimize spurious effects and to let the farm wake to develop. Similarly, the domain width should be large enough to minimize sidewise blockage and to limit the channelling effects at the farm sides. In section \ref{sec:domain_sensitivity}, we present an extensive domain sensitivity study. Based on this, we select a domain size of $L_x \times L_y = 50 \times 30$ km$^2$, with distance between main domain inflow and first-row turbine of $L_\mathrm{ind}=18$ km. This domain size is further used for all simulations performed in Section \ref{sec:atm_sensitivity}. The vertical domain dimension is dictated by the presence of gravity waves. Following previous studies, we fix the main domain height to $L_z=25$ km \citep{Allaerts2017,Allaerts2017b,Lanzilao2022,Lanzilao2022b}. This allows us to insert a wide RDL at the top of the domain-- see below.

In SP-Wind, the precursor domain width and height should match the ones of the main domain when they are run concurrently. Therefore, we adopt the tiling technique to extend the precursor flow fields in the $y$-direction from 10 to 30 km \citep{Sanchez2023}. In regard to the $z$-direction, we extrapolate the flow fields from 3 to 25 km, using a constant geostrophic wind. At these altitudes the flow is laminar and no turbulence needs to be added.

In regard to the grid resolution, we fix $\Delta x = 31.25$ m and $\Delta y = 21.74$ m in the streamwise and spanwise direction, respectively. This leads to $N_x=1600$ and $N_y~=~1380$ grid points for the main domain and to $N_x^p=320$ and $N_y^p=460$ points for the precursor domain. A stretched grid is adopted in the vertical direction. The latter is composed of $300$ uniformly spaced grid points within the first $1.5$ km to capture the strong velocity gradients around the turbine-rotor disk, leading to a grid resolution of $\Delta z = 5$ m. This allows us to obtain a ratio between $\Delta z$ and the buoyancy length scale $l_b=1.69 \langle \overline{w'w'} \rangle^{0.5} N^{-0.5}$ in the capping inversion and free atmosphere above 2 for the majority of the simulation cases \citep{Otte2001,Pedersen2014}. Next, a first stretch is applied from $1.5$ to $15$ km, where $180$ points are used. A second one is applied in the last $10$ km of the domain, i.e. from $15$ to $25$ km. In summary, the domain is $25$ km high and the vertical grid contains a total of $490$ grid points. The combination of spanwise and vertical grid resolution allows us to have a total of $9$ and $40$ grid points along the turbine-rotor disk width and height, which is in accordance with simulations in the literature \citep{Calaf2010,Wu2011,Allaerts2017}. The combination of precursor and main numerical domains leads to a total of approximately $5.2 \times 10^9$ DOF. Finally, we also perform $4$ simulations on a domain which contains a single turbine. In these cases, the precursor and main domains have equal sizes. Those simulations are used for evaluating the power output of a turbine that operates in isolation, which will serve in Section~\ref{sec:atm_sensitivity} for scaling some of the results. More information about the single-turbine simulations is reported in Appendix~\ref{app:st_sim}.

\begin{figure}
	\centering
	\includegraphics[width=1.\textwidth]{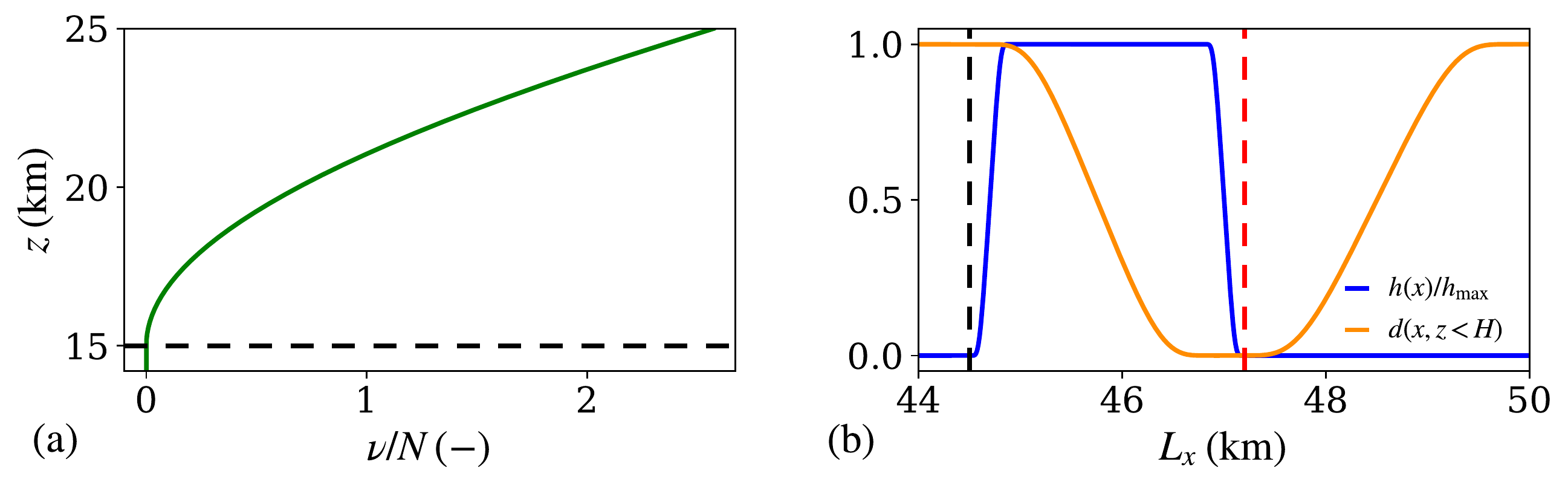}%
	\caption{(a) Rayleigh function obtained with $\nu^\mathrm{ra}=5.15$ and $s^\mathrm{ra}=3$ values. (b) Fringe and damping functions used with the wave-free fringe-region techniques. The black horizontal and vertical dashed lines denote the start of the RDL and the fringe region while the red vertical dashed line marks the end of the fringe forcing.}
	\label{fig:sponge_layers_setup}
\end{figure}

The vertical gravity-wave wavelength derived using gravity-wave linear theory under the hydrostatic assumption is given by $\lambda_z = 2 \pi G/N$. According to the free lapse rate values adopted in our study (see Section \ref{sec:atm_state}), the vertical wavelength varies between $3.8$ and $10.7$~km. Following \cite{Klemp1977}, who suggested that the depth of the RDL should be at least in the order of $\lambda_z$, we set $L_z^\mathrm{ra}=10$ km. Moreover, we fix $\nu^\mathrm{ra}=5.15$ and $s^\mathrm{ra}=3$. These parameters minimize gravity-wave reflection and are chosen following the procedure detailed in \cite{Lanzilao2022b}. The Rayleigh function is shown in Figure \ref{fig:sponge_layers_setup}(a).

The body force applied within the fringe region should be strong enough to impose the inflow condition without violating the stability constraint imposed by the $4$th order Runge–Kutta method \citep{Schlatter2005}. We carried out some tests (not shown) and we noted that $h_\mathrm{max}=0.3$~s$^{-1}$ satisfies both constraints. Moreover, we fix the fringe-region length to $L_x^\mathrm{fr}=5.5$ km. Given the wind-farm layout (see Section \ref{sec:wf_layout}), this means that there are gaps of $18$ km and $11.65$ km upwind and downwind of the farm. Further, we set $x_s^h=L_x-L_x^\mathrm{fr}$, $x_e^h=L_x-2.8$ km and $\delta_s^h=\delta_e^h= 0.4$ km while $x_s^d=x_s^h$, $x_e^d=L_x$, $\delta_s^d= 2.5$ km and $\delta_e^d= 3$ km. \cite{Lanzilao2022b} have shown that this set of parameters minimize the spurious gravity waves triggered by the fringe forcing. The fringe and damping functions are shown in Figure \ref{fig:sponge_layers_setup}(b).

\subsection{Wind-farm layout}\label{sec:wf_layout}
The wind-farm set-up is inspired by the work of \cite{Lanzilao2022,Lanzilao2022b}. Hence, we have $16$ rows and $10$ columns for a total of $N_t=160$ IEA offshore turbines \citep{Bortolotti2019} with a rated power of $P_\mathrm{rated}=10$~MW arranged in a staggered layout with respect to the main wind direction. The farm is relatively densely spaced with streamwise and spanwise spacings set to $s_xD=s_yD=5D$, where $D=198$ m denotes the turbine-rotor diameter. This corresponds to a density of roughly $P_\mathrm{rated}/s_x s_y D^2 \approx 10$ MW km$^{-2}$, which is a relatively dense scenario that is nonetheless considered nowadays in some development areas.

The turbine-hub height measures $z_h=119$~m while the thrust coefficient is selected from the turbine thrust curve using the undisturbed inflow wind speed at hub height measured in the precursor simulations, which results in $C_T=0.88$ \citep{Bortolotti2019}. The corresponding disk-based thrust coefficient is then $C_T'=1.94$ \citep{Calaf2010,Meyers2010}. Moreover, a simple yaw controller is implemented to keep all turbine-rotor disks perpendicular to the incident wind flow measured one rotor diameter upstream.

The farm has length and width of $L_x^f=14.85$~km and $L_y^f=9.4$~km, respectively. The ratios $L_\mathrm{ind}/L_x^f$, $L_x/L_x^f$ and $L_y/L_y^f$ measure 1.21, 3.37 and 3.19, respectively. We note that in the current work, we only focus on the effect of atmospheric conditions on the flow behaviour given a constant farm layout. Investigating the effects of farm density and shape is a topic for future research.

\begin{figure}
	\centerline{
		\includegraphics[width=0.78\textwidth]{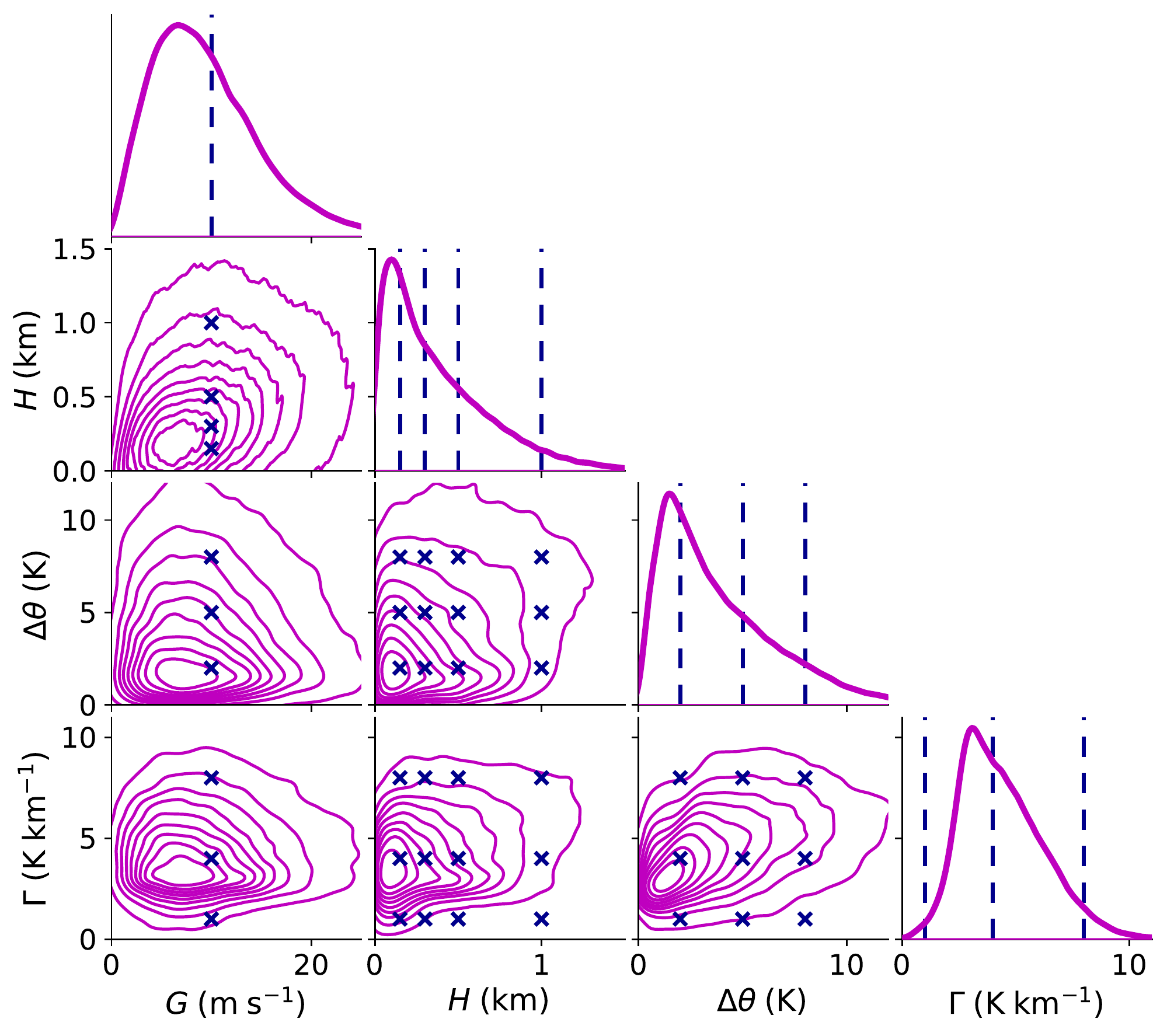}}% 
	\caption{Joint probability density function of the geostrophic wind $G$, capping-inversion height $H$, capping inversion strength $\Delta \theta$ and free-atmosphere lapse rate $\Gamma$. The parameters that characterize the CNBL profile are obtained by fitting the ERA5 vertical potential temperature profile between the surface level and $2.5$ km using the \cite{Rampanelli2004} model. The geostrophic wind is obtained by taking the mean velocity magnitude between the top of the capping inversion and $2.5$ km. The blue vertical dashed lines and crosses denote the parameters selected in the current study, i.e. $G=10$ m s$^{-1}$, $H=150,300,500,1000$ m, $\Delta \theta=2,5,8$ K and $\Gamma=1,4,8$ K km$^{-1}$. All combinations are considered, for a total of 36 cases. }
	\label{fig:case_setup}
\end{figure}

\subsection{Atmospheric state}\label{sec:atm_state}
To select a range of relevant atmospheric states, we analyzed $30$ years of ERA5 reanalysis data (from $1988$ to $2018$) measured at $51.6$N $3.0$E, which is the nearest grid point to the Belgian–Dutch offshore wind-farm cluster. We use the model proposed by \cite{Rampanelli2004} to fit the vertical potential temperature profile from the surface level up to $2.5$ km, using a least square fit to all levels in this range. The outputs of this model consist of an estimate of the capping inversion height $H$ and strength $\Delta\theta$ and lapse rate $\Gamma$ in the free atmosphere. To evaluate the geostrophic wind, we compute the mean velocity magnitude between the top of the capping inversion and $2.5$ km.

The subplots on the diagonal of Figure \ref{fig:case_setup} display the probability density functions of such parameters while the joint probability density functions are shown in the off-diagonal subplots. In this study, we fix the geostrophic wind to $10$ m s$^{-1}$, which is in line with previous studies \citep{Abkar2013,Wu2017,Allaerts2017,Allaerts2017b,Lanzilao2022}. Moreover, this choice leads to region~$2$ operation of our wind farm. In regard to the capping inversion height, we select the values of $150$, $300$, $500$ and $1000$ m, so that we can explore farm operations in shallow and deep boundary layers, including some cases where the inversion-layer base is located below the turbine-tip height. The capping inversion strength $\Delta \theta$ is set to $2$, $5$ or $8$ K while we fix $\Gamma$ to $1$, $4$ or $8$ K km$^{-1}$. This wide variety of capping-inversion strengths and free lapse rates allows us to study the influence of interfacial and internal waves on farm energy extraction and flow blockage. The ground temperature and the capping-inversion thickness are fixed to $\theta_0=288.15$~K and $\Delta H = 100$ m for all simulations. The blue crosses in Figure \ref{fig:case_setup} denote the $36$ atmospheric states that we selected. Finally, we fix the latitude to $\phi=51.6^\circ$, which leads to a Coriolis frequency of $f_c=1.14 \times 10^{-4}$ s$^{-1}$, and the surface roughness to $z_0 = 1 \times 10^{-4}$~m for all simulations. This value represents calm sea conditions and enters in the range of values observed over the North Sea, and more generally offshore \citep{Taylor2000,Allaerts2017,Lanzilao2022,Kirby2022}. More details on the atmospheric states selected and the suite of simulations performed are reported in Table~\ref{table:simulation_setup}.

In the remainder of the text, the state variables will be accompanied by a bar in case of time averages. For the horizontal averages along the full streamwise and spanwise directions, we use the angular brackets $\langle \cdot \rangle$. 

\begin{table}
	\begin{center}
		\def~{\hphantom{0}}
		\begin{adjustbox}{max width=\textwidth}
			\begin{tabular}{ccccccccccccc}
				\textbf{Cases}  & $\boldsymbol{H}$ \textbf{(m)} & $\boldsymbol{\Delta \theta}$ \textbf{(K)} & $\boldsymbol{\Gamma}$ \textbf{(K km$^{-1}$)} & $\boldsymbol{\Delta H}$ \textbf{(m)} & $\boldsymbol{M_\mathrm{hub}}$ \textbf{(m s$^{-1}$)} & $\boldsymbol{\mathrm{TI}_\mathrm{hub}}$ \textbf{(\%)} & $\boldsymbol{u_\star}$ \textbf{(m s$^{-1}$)} & $\boldsymbol{\alpha (^\circ)}$ & $\boldsymbol{\gamma_v}$ \textbf{(--)} & \textbf{\textit{Fr} (--)} & $\boldsymbol{P_N}$ \textbf{(--)} & $\boldsymbol{N_t}$ \textbf{(--)} \\[7pt]
				
				H150-$\Updelta \uptheta$0-$\Upgamma$0     & -  & 0.00 & 0 & -   & 9.47 & 3.30 & 0.277 & -18.28 & 1.06 & - & -    & 160\\
				H150-$\Updelta \uptheta$2-$\Upgamma$1     & 251  & 1.58 & 1 & 56  & 9.47 & 3.48 & 0.280 & -15.48 & 1.02 & 2.53 & 5.95 & 160\\
				H150-$\Updelta \uptheta$2-$\Upgamma$4     & 245  & 1.95 & 4 & 48  & 9.48 & 3.48 & 0.280 & -15.88 & 1.03 & 2.32 & 3.05 & 160\\
				H150-$\Updelta \uptheta$2-$\Upgamma$8     & 240  & 2.36 & 8 & 44  & 9.47 & 3.40 & 0.279 & -16.29 & 1.07 & 2.08 & 2.11 & 160\\
				H150-$\Updelta \uptheta$5-$\Upgamma$1     & 211  & 4.42 & 1 & 44  & 9.47 & 3.39 & 0.278 & -18.23 & 1.06 & 1.63 & 6.86 & 160\\
				H150-$\Updelta \uptheta$5-$\Upgamma$4     & 210  & 4.67 & 4 & 42  & 9.47 & 3.30 & 0.277 & -18.28 & 1.06 & 1.59 & 3.46 & 160\\
				H150-$\Updelta \uptheta$5-$\Upgamma$8     & 209  & 5.00 & 8 & 42  & 9.47 & 3.32 & 0.277 & -18.40 & 1.10 & 1.51 & 2.36 & 160\\
				H150-$\Updelta \uptheta$8-$\Upgamma$1     & 196  & 7.35 & 1 & 47  & 9.46 & 3.17 & 0.276 & -19.30 & 1.09 & 1.30 & 7.22 & 160\\
				H150-$\Updelta \uptheta$8-$\Upgamma$4     & 195  & 7.57 & 4 & 48  & 9.46 & 3.35 & 0.276 & -19.40 & 1.08 & 1.28 & 3.65 & 160\\
				H150-$\Updelta \uptheta$8-$\Upgamma$8     & 195  & 7.87 & 8 & 48  & 9.45 & 3.08 & 0.275 & -19.32 & 1.14 & 1.22 & 2.44 & 160\\
				H150-$\Updelta \uptheta$5-$\Upgamma$4-st  & 210  & 4.67 & 4 & 42  & 9.47 & 3.30 & 0.277 & -18.28 & 1.06 & 1.59 & 3.46 & 1\\[7pt]
				
				H300-$\Updelta \uptheta$0-$\Upgamma$0     & -  & 0.00 & 0 & -   & 9.42 & 3.60 & 0.281 & -12.60 & 0.97 & -    & -    & 160\\
				H300-$\Updelta \uptheta$2-$\Upgamma$1     & 348  & 1.95 & 1 & 57  & 9.38 & 3.66 & 0.280 & -11.83 & 0.98 & 1.97 & 4.41 & 160\\
				H300-$\Updelta \uptheta$2-$\Upgamma$4     & 348  & 2.19 & 4 & 52  & 9.38 & 3.68 & 0.280 & -11.87 & 0.98 & 1.85 & 2.20 & 160\\
				H300-$\Updelta \uptheta$2-$\Upgamma$8     & 347  & 2.49 & 8 & 49  & 9.39 & 3.63 & 0.281 & -11.95 & 0.97 & 1.75 & 1.57 & 160\\
				H300-$\Updelta \uptheta$5-$\Upgamma$1     & 325  & 4.93 & 1 & 67  & 9.41 & 3.61 & 0.281 & -12.59 & 0.97 & 1.29 & 4.76 & 160\\
				H300-$\Updelta \uptheta$5-$\Upgamma$4     & 325  & 5.15 & 4 & 66  & 9.42 & 3.60 & 0.281 & -12.60 & 0.97 & 1.26 & 2.38 & 160\\
				H300-$\Updelta \uptheta$5-$\Upgamma$8     & 326  & 5.42 & 8 & 63  & 9.41 & 3.59 & 0.281 & -12.55 & 0.98 & 1.22 & 1.66 & 160\\
				H300-$\Updelta \uptheta$8-$\Upgamma$1     & 316  & 7.93 & 1 & 78  & 9.43 & 3.58 & 0.281 & -12.91 & 0.98 & 1.03 & 4.87 & 160\\
				H300-$\Updelta \uptheta$8-$\Upgamma$4     & 316  & 8.15 & 4 & 77  & 9.43 & 3.59 & 0.281 & -12.92 & 0.98 & 1.01 & 2.43 & 160\\
				H300-$\Updelta \uptheta$8-$\Upgamma$8     & 316  & 8.46 & 8 & 76  & 9.43 & 3.58 & 0.282 & -12.88 & 0.98 & 0.99 & 1.71 & 160\\
				H300-$\Updelta \uptheta$5-$\Upgamma$4-st  & 325  & 5.15 & 4 & 66  & 9.42 & 3.60 & 0.281 & -12.60 & 0.97 & 1.26 & 2.38 & 1\\[7pt]
				
				H500-$\Updelta \uptheta$0-$\Upgamma$0     & -  & 0.00 & 0 & -   & 9.24 & 3.93 & 0.277 & -9.09  & 0.93 & -    & -    & 160\\
				H500-$\Updelta \uptheta$2-$\Upgamma$1     & 521  & 2.04 & 1 & 75  & 9.22 & 3.95 & 0.277 & -8.94  & 0.91 & 1.60 & 3.04 & 160\\
				H500-$\Updelta \uptheta$2-$\Upgamma$4     & 524  & 2.26 & 4 & 68  & 9.22 & 3.96 & 0.277 & -8.93  & 0.92 & 1.50 & 1.50 & 160\\
				H500-$\Updelta \uptheta$2-$\Upgamma$8     & 523  & 2.54 & 8 & 66  & 9.22 & 3.96 & 0.277 & -8.93  & 0.92 & 1.42 & 1.06 & 160\\
				H500-$\Updelta \uptheta$5-$\Upgamma$1     & 507  & 5.05 & 1 & 90  & 9.24 & 3.92 & 0.278 & -9.08  & 0.92 & 1.03 & 3.10 & 160\\
				H500-$\Updelta \uptheta$5-$\Upgamma$4     & 509  & 5.28 & 4 & 87  & 9.24 & 3.93 & 0.277 & -9.09  & 0.93 & 0.99 & 1.54 & 160\\
				H500-$\Updelta \uptheta$5-$\Upgamma$8     & 511  & 5.59 & 8 & 86  & 9.24 & 3.93 & 0.277 & -9.11  & 0.93 & 0.97 & 1.08 & 160\\
				H500-$\Updelta \uptheta$8-$\Upgamma$1     & 503  & 8.05 & 1 & 96  & 9.25 & 3.93 & 0.278 & -9.14  & 0.93 & 0.81 & 3.11 & 160\\
				H500-$\Updelta \uptheta$8-$\Upgamma$4     & 504  & 8.29 & 4 & 94  & 9.25 & 3.93 & 0.278 & -9.15  & 0.93 & 0.80 & 1.55 & 160\\
				H500-$\Updelta \uptheta$8-$\Upgamma$8     & 504  & 8.61 & 8 & 92  & 9.25 & 3.94 & 0.278 & -9.15  & 0.91 & 0.79 & 1.11 & 160\\
				H500-$\Updelta \uptheta$5-$\Upgamma$4-st  & 509  & 5.28 & 4 & 87  & 9.24 & 3.93 & 0.277 & -9.09  & 0.93 & 0.99 & 1.54 & 1\\[7pt]
				
				H1000-$\Updelta \uptheta$0-$\Upgamma$0    & - & 0.00 & 0 & -   & 9.13 & 4.18 & 0.275 & -7.90 & 0.86 & -    & -    & 160\\
				H1000-$\Updelta \uptheta$2-$\Upgamma$1    & 1003 & 2.08 & 1 & 96  & 9.12 & 4.16 & 0.275 & -7.93 & 0.86 & 1.80 & 1.63 & 160\\
				H1000-$\Updelta \uptheta$2-$\Upgamma$4    & 1003 & 2.32 & 4 & 95  & 9.12 & 4.15 & 0.275 & -7.97 & 0.86 & 1.80 & 0.82 & 160\\
				H1000-$\Updelta \uptheta$2-$\Upgamma$8    & 1003 & 2.65 & 8 & 95  & 9.14 & 4.15 & 0.275 & -7.88 & 0.86 & 1.80 & 0.58 & 160\\
				H1000-$\Updelta \uptheta$5-$\Upgamma$1    & 1001 & 5.08 & 1 & 99  & 9.13 & 4.16 & 0.275 & -7.87 & 0.86 & 0.74 & 1.65 & 160\\
				H1000-$\Updelta \uptheta$5-$\Upgamma$4    & 1001 & 5.33 & 4 & 99  & 9.13 & 4.18 & 0.275 & -7.90 & 0.86 & 0.73 & 0.82 & 160\\
				H1000-$\Updelta \uptheta$5-$\Upgamma$8    & 1001 & 5.33 & 8 & 99  & 9.13 & 4.17 & 0.275 & -7.91 & 0.86 & 0.70 & 0.58 & 160\\
				H1000-$\Updelta \uptheta$8-$\Upgamma$1    & 1000 & 8.08 & 1 & 100 & 9.14 & 4.16 & 0.275 & -7.92 & 0.86 & 0.59 & 1.64 & 160\\
				H1000-$\Updelta \uptheta$8-$\Upgamma$4    & 1001 & 8.33 & 4 & 100 & 9.13 & 4.16 & 0.275 & -7.98 & 0.86 & 0.58 & 0.82 & 160\\
				H1000-$\Updelta \uptheta$8-$\Upgamma$8    & 1001 & 8.67 & 8 & 100 & 9.13 & 4.15 & 0.275 & -7.99 & 0.86 & 0.57 & 0.58 & 160\\
				H1000-$\Updelta \uptheta$5-$\Upgamma$4-st & 1001 & 5.33 & 4 & 99  & 9.13 & 4.18 & 0.275 & -7.90 & 0.86 & 0.73 & 0.82 & 1\\[1pt]
			\end{tabular}
		\end{adjustbox}
		\caption{Overview of the spin-up cases used to drive 40 wind-farm simulations and 4 single-turbine simulations. The parameters are averaged over the last 4 h of the spin-up phase and include the capping-inversion height $H$, the capping-inversion strength $\Delta \theta$, the free atmosphere lapse rate $\Gamma$, the capping-inversion thickness $\Delta H$, the turbulence intensity measured at hub height $\mathrm{TI}_\mathrm{hub}$, the velocity magnitude measure at hub height $M_\mathrm{hub}$, the friction velocity $u_\star$, the geostrophic wind angle $\alpha$, the shape factor $\gamma_v$, the Froude number \textit{Fr}, the $P_N$ number and the number of turbines $N_t$. Note that the parameters $H$, $\Delta \theta$, $\Gamma$ and $\Delta H$ have been estimated by fitting the spin-up profiles averaged over the last 4 h of the precursor simulations with the \cite{Rampanelli2004} model.}
		\label{table:simulation_setup}
	\end{center}
\end{table}

\section{Boundary-layer initialization}\label{sec:bl_initialization}
The spin-up of the precursor simulations is discussed in Section \ref{sec:cnbls_spinup}. After the spin-up phase, we start the main domain simulation and we perform a wind-farm start-up phase driven by the precursor simulation, during which the flow adjusts to the presence of the farm. This phase is discussed in Section \ref{sec:windfarm_startup}. Finally, we discuss the methodology used to perform simulations in a neutral boundary layer (NBL) reference case in Section \ref{sec:cnbl2nbl}.

\subsection{Generation of a fully developed turbulent flow field}\label{sec:cnbls_spinup}
The various $H$, $\Delta \theta$ and $\Gamma$ values selected are combined together to form $36$ atmospheric states, which range from a shallow boundary layer with a strong inversion layer and free atmosphere stratification, to a deep boundary layer with low $\Delta \theta$ and $\Gamma$ values. The initial vertical potential-temperature profiles are generated giving the $H$, $\Delta \theta$ and $\Gamma$ values as input to the \cite{Rampanelli2004} model. For the initial velocity profile, we use a constant geostrophic wind above the capping inversion. Within the ABL, we use the \cite{Zilitinkevich1989} model with friction velocity $u_\ast = 0.26$~m s$^{-1}$. The velocity profiles below the capping inversion are then combined with the laminar profile in the free atmosphere following the method proposed by \cite{Allaerts2015}.
\begin{figure}
	\centerline{
		\includegraphics[width=1\textwidth]{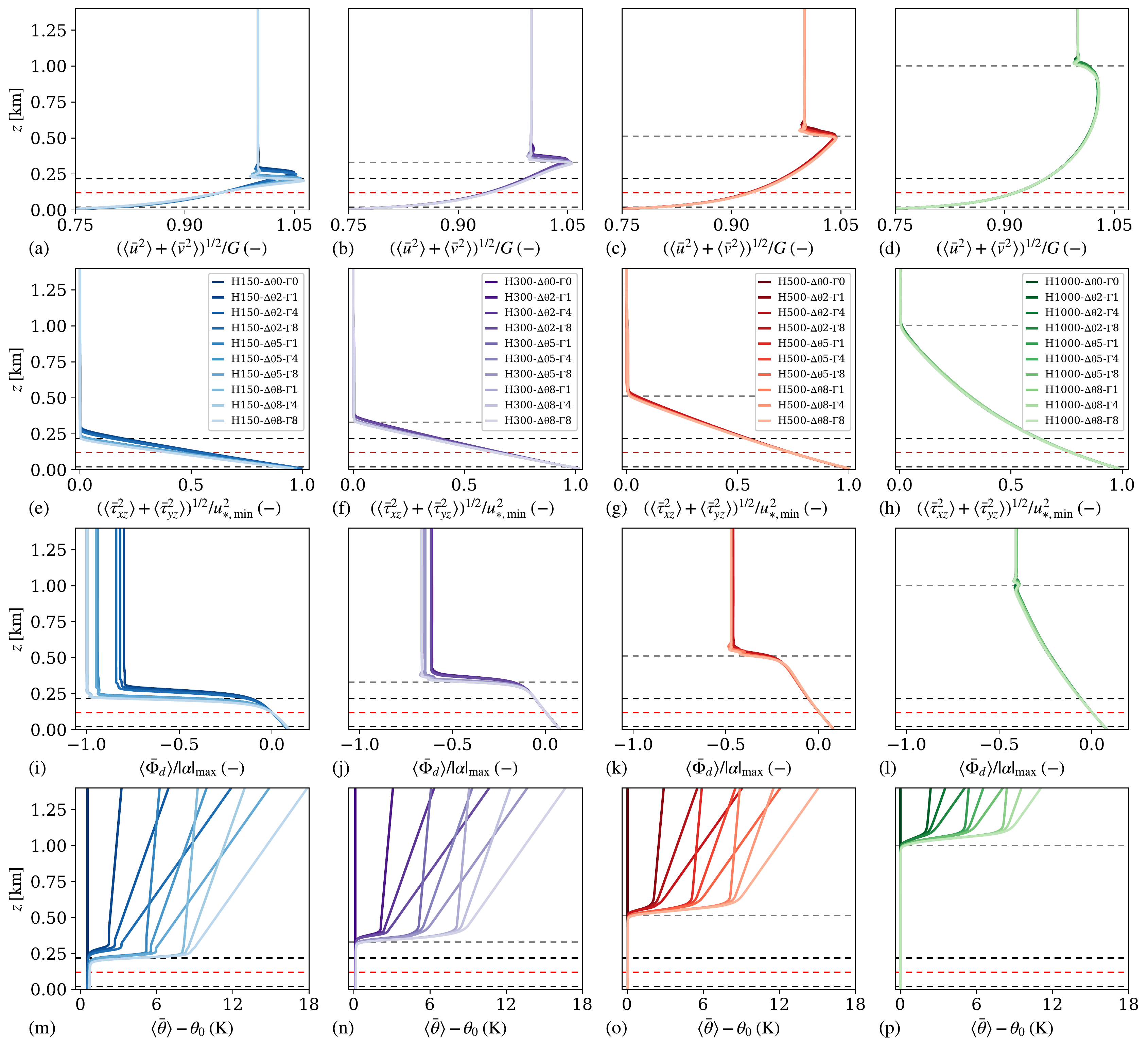}}% 
	\caption{Vertical profiles of (a-d) velocity magnitude, (e-h) shear stress magnitude, (i-l) wind direction and (m-p) potential temperature averaged along the full horizontal directions and over the last $4$ h of simulation. The results are shown for (a,e,i,m) H150, (b,f,j,n) H300, (c,g,k,o) H500 and (d,h,l,p) H1000 cases using various capping-inversion strengths and free-atmosphere lapse rates. Moreover, the profiles are further normalized with $G=10$ m s$^{-1}$, $u_{\star,\mathrm{min}}=0.275$ m s$^{-1}$, $|\alpha|_\mathrm{max}=19.4 ^\circ$ and $\theta_0=288.15$ K.  The red dashed line denotes the turbine-hub height while the black dashed lines are representative of the rotor dimension. Finally, the grey dashed line represents the averaged inversion-layer height.}
	\label{fig:precursor_results}
\end{figure}

Next, we add random divergence-free perturbations with an amplitude of $0.1G$ in the first $100$ m to the vertical velocity profiles. This initial state is given as input to the precursor simulation. Since no turbines are located in the domain, the only drag force acting on the flow is the wall stress. The flow is advanced in time for $20$ h, which is sufficient to obtain a turbulent fully developed statistically steady state \citep{Pedersen2014,Allaerts2017,Lanzilao2022b}. Figure \ref{fig:precursor_results} illustrates vertical profiles of several quantities of interest averaged over the last $4$ h of the simulations and over the full horizontal directions. Figure \ref{fig:precursor_results}(a-d) shows the velocity magnitude normalized with the geostrophic wind. The boundary layer extends up to the capping inversion, which limits its growth. All velocity profiles show a common feature, that is the presence of a super-geostrophic jet near the top of the ABL. The same behaviour was previously noted also by \cite{Hess2004} and \cite{Pedersen2014}. We note that this phenomenon is more accentuated for the H150 cases, where a stronger wind shear along with stronger velocity gradients at the top of the ABL are attained. Next, Figure \ref{fig:precursor_results}(e-h) displays the shear stress magnitude, which is non-zero only below the capping inversion, with a quasi-linear profile. We note that varying $\Delta \theta$ and $\Gamma$ results in very minor differences in terms of velocity and shear stresses. Next, Figure \ref{fig:precursor_results}(i-l) shows the flow angle. At turbine-hub height, the flow is parallel to the $x$-direction. This is achieved by using the wind-angle controller developed and tuned by \cite{Allaerts2015}, which is designed to ensure a desired orientation of the hub-height wind direction ($\Phi_d =0^\circ$ in this case). Below the turbine-tip height, the flow is almost unidirectional since most of the wind-direction change occurs within the inversion layer, expects for deep boundary layers. The geostrophic wind angle, which is the angle between the surface stress and the geostrophic wind velocity, is larger for shallow boundary layers, as noted by \cite{Allaerts2017}. Finally, the thermal stratification is illustrated in Figure \ref{fig:precursor_results}(m-p) by means of potential temperature profiles. 

The various spin-up cases together with some parameters of interest averaged over the last $4$ h of simulation are summarized in Table \ref{table:simulation_setup}. During the spin-up phase, the capping-inversion height moves slightly upward. The increase in inversion-layer height is more accentuated for the shallow boundary-layer cases. For instance, the H150 cases show a growth of $67$ m on average over the 20 h of spin-up. For the H1000 cases, the boundary-layer height remains unaltered. This result is consistent with the equilibrium theory of \cite{Csanady1974} and with previous LES findings \citep{Pedersen2014,Allaerts2015,Allaerts2017}. The capping-inversion strength slightly increases for the majority of the cases while the free-atmosphere stratification remains unchanged. The velocity magnitude at turbine-hub height varies between $9.1$ and $9.5$ m s$^{-1}$ among all cases, meaning that turbines operate in the region $2$ \citep{Bortolotti2019}. Moreover, the turbulence intensity at turbine-hub height varies from 3.3$\%$ to 4.1$\%$. These values are in line with the ones observed by \cite{Barthelmie2009} and \cite{Turk2010} over the North Sea.  The large variance in the Froude number, defined as $\textit{Fr}={U}_B/\sqrt{g'H}$ with ${U}_B$ the bulk velocity along the $x$-direction and $g'=g \Delta \theta/\theta_0$ the reduced gravity, will allow us to analyze the flow response to wind-farm forcing under critical ($\textit{Fr} \approx 1$), sub-critical ($\textit{Fr} \leq 1$) and super-critical ($\textit{Fr} \geq 1$) conditions with varying $P_N= {U}_B^2/NGH $ numbers. Both the $\textit{Fr}$ and $P_N$ number values are reported in Table \ref{table:simulation_setup}.

\subsection{Wind-farm start-up phase}\label{sec:windfarm_startup}
\begin{figure}
	\centerline{
		\includegraphics[width=1\textwidth]{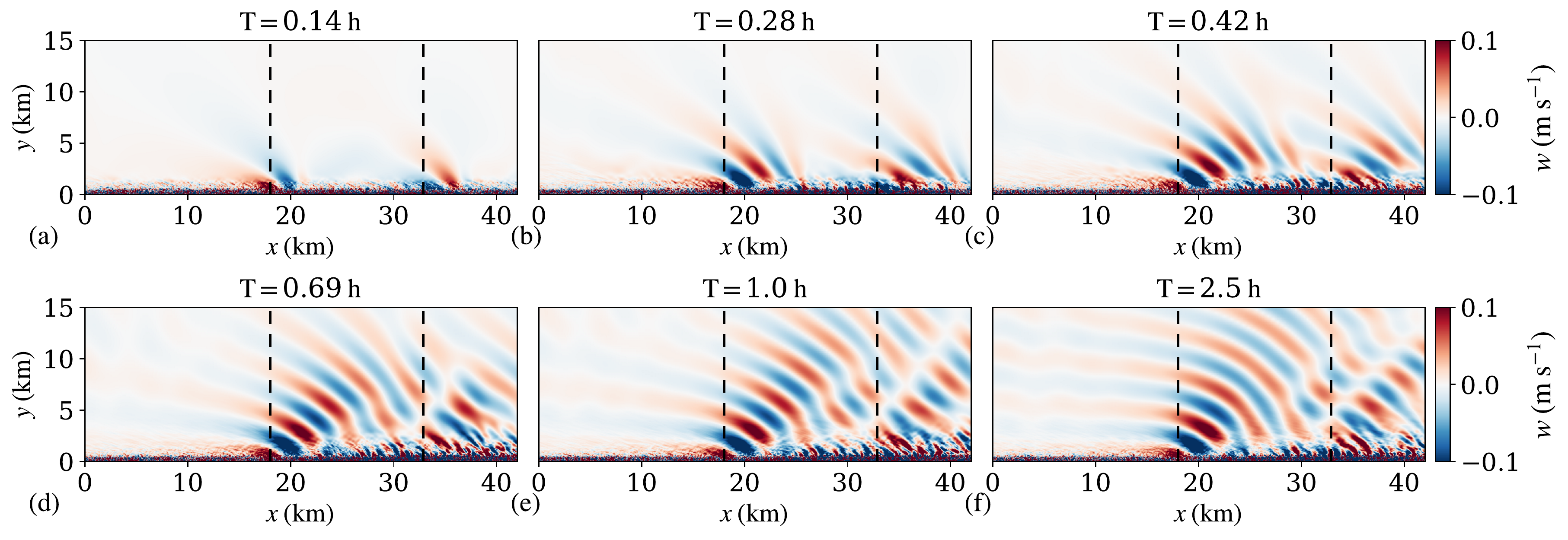}}% 
	\caption{Instantaneous contours of vertical velocity for case H500-$\Updelta \uptheta$2-$\Upgamma$8 obtained (a-d) during the wind-farm start-up phase, (e) at the end of the wind-farm start-up phase and (f) at the end of the simulation. The $x$--$z$ plane is taken along the centerline of the domain, i.e. at $y=15$ km. The black vertical dashed lines denote the location of the first- and last-row turbine. We note that the fringe region and RDL are not displayed in the plots.}
	\label{fig:windfarm_startup}
\end{figure}

The turbulent fully-developed inflow profiles previously discussed are now used to drive the simulation in the main domain, where the turbines impose a drag force on the flow. However, before collecting flow statistics over time, a second spin-up phase is required. In fact, the flow has to adjust to the presence of the farm in the main domain before reaching a new statistically-steady state. We name this phase wind-farm start-up. Figure \ref{fig:windfarm_startup} shows the time evolution of the vertical velocity field on an $x$--$z$ plane for case H500-$\Updelta \uptheta$2-$\Upgamma$8. The flow divergence induced by the farm drag force displaces upward the capping inversion, which in turn triggers a first train of internal gravity waves in proximity to the first-row turbine location. Vice versa, the flow convergence in the farm wake moves the capping inversion downward, generating a second train of internal waves at the last-row turbine location. This is clearly visible in Figure \ref{fig:windfarm_startup}(a). The two out-of-phase trains of waves are convected downstream, until they eventually merge at T$=$1~h, as shown in Figure \ref{fig:windfarm_startup}(e). At this point, the numerical solution is further advanced in time for $1.5$~h. The instantaneous vertical velocity flow field taken at T$=$2.5~h is displayed in Figure \ref{fig:windfarm_startup}(f). By comparing the numerical solution at T$=$1~h against the one obtained at T$=$2.5~h, we notice minimal differences. A similar behaviour is observed for the streamwise and spanwise velocity field (not shown). This means that 1 h of wind-farm spin-up time suffices for the flow to adjust to the farm drag force. Therefore, similarly to \cite{Allaerts2017} and \cite{Lanzilao2022}, we fix the duration of the wind-farm start-up phase to 1 h, which corresponds to roughly two and a half wind-farm flow-through times. Next, we switch off the wind-angle controller in the precursor simulation and we collect statistics during a time window of $1.5$ hour. 

\begin{figure}
	\centering
	\includegraphics[width=1\textwidth]{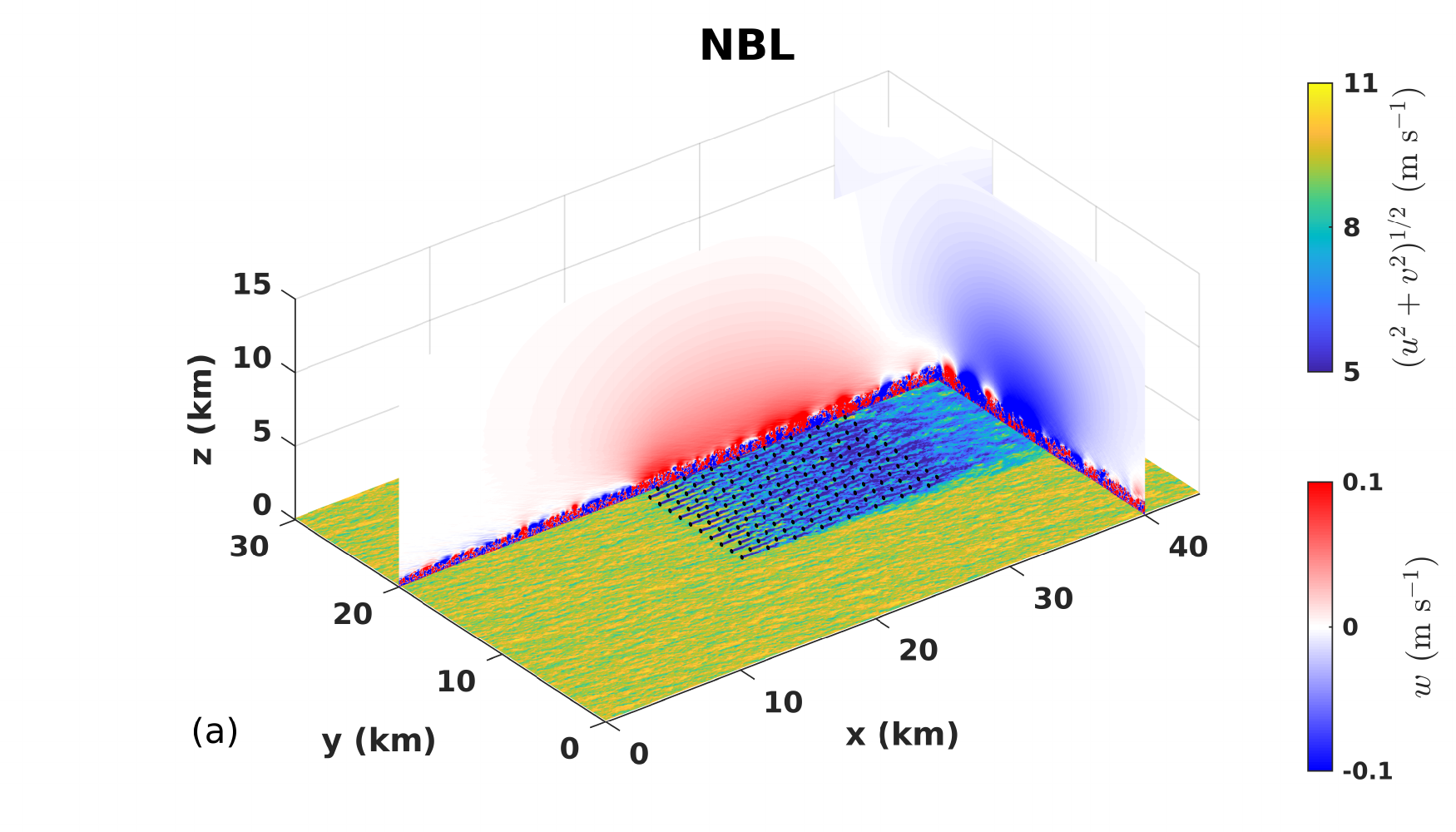}
	\includegraphics[width=1\textwidth]{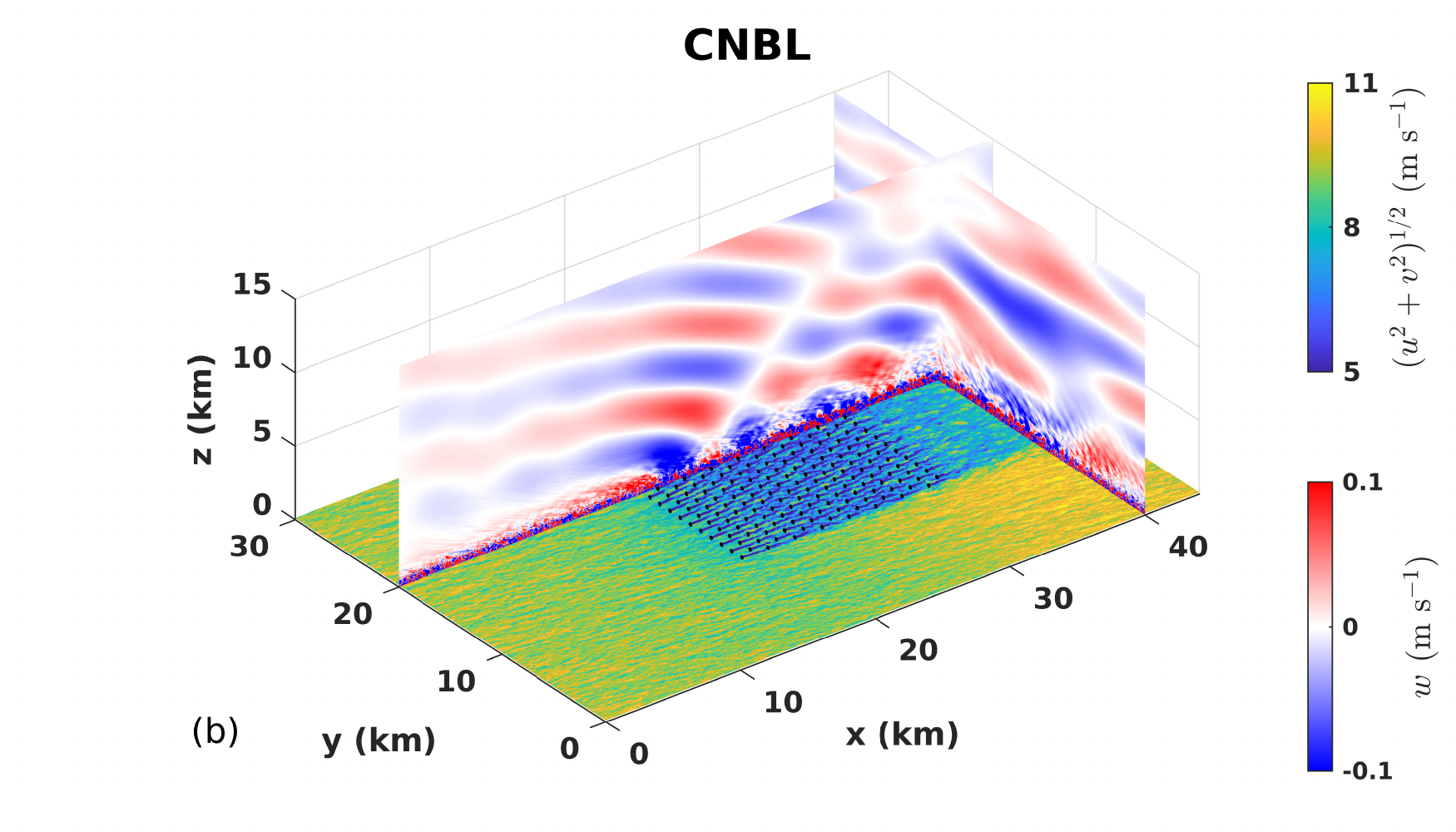}
	\caption{Snapshot of the instantaneous flow field obtained after 2.5 hour of simulation time in the (a) NBL reference case and in (b) CNBL conditions. The two simulations correspond to cases H500-$\Delta \theta$0-$\Gamma$0 and H500-$\Delta \theta$5-$\Gamma$4, respectively. The $x$--$y$ plane shows contours of velocity magnitude $(u^2+v^2)^{1/2}$ taken at turbine-hub height while the $x$--$z$ and $y$--$z$ planes display the vertical velocity field. The black disks denote the wind-turbine rotor locations. Finally, we note that the fringe region and RDL are not displayed. We remark that the only difference in the NBL reference case consists in the absence of thermal stratification above the ABL and therefore gravity waves.}
	\label{fig:3D_intro}
\end{figure}

\subsection{Construction of a neutral boundary layer reference case}\label{sec:cnbl2nbl}
For each inversion-layer height, we also include a case characterized by the absence of the capping inversion and a neutral free atmosphere, an idea originally proposed by \cite{Lanzilao2022}. To do so, the fringe forcing in the main domain only forces the velocity field to the one provided by the CNBL developed in the precursor domain, leaving the potential temperature constant with height. To give an example, case H500-$\Updelta \uptheta$0-$\Upgamma$0 is driven by the flow fields obtained in the precursor simulation of case H500-$\Updelta \uptheta$5-$\Upgamma$4, but sets the temperature profile in the main domain to a constant value. Consequently, in this case, the farm operates under the same turbulent inflow velocity profile but in the absence of atmospheric gravity waves. This is admittedly a numerical construction that does not really exist in reality but allows us to factor out the free-atmosphere stratification effects. It has however also some drawbacks. For instance, the capping-inversion height is much lower than the Ekman-layer equilibrium height that the boundary layer would attain in a TNBL. Therefore, in the main domain, the flow within the ABL inevitably starts mixing with the free atmosphere, varying its shear and veer already before reaching the first-row turbine location. In the H150-$\Updelta \uptheta$0-$\Upgamma$0 and H300-$\Updelta \uptheta$0-$\Upgamma$0 cases, this flow mixing is very high, varying considerably the flow profiles. For instance, the flow angle at the farm entrance measures approximately 10$^\circ$ in the H150-$\Updelta \uptheta$0-$\Upgamma$0 case. However, this effect becomes negligible in the H500-$\Updelta \uptheta$0-$\Upgamma$0 and H1000-$\Updelta \uptheta$0-$\Upgamma$0 cases, where the flow remains parallel to the streamwise direction and in general profiles show very similar results in front of and across the farm, with a farm efficiency of 47.0$\%$ and 46.7$\%$, respectively. Therefore, we use case H500-$\Updelta \uptheta$0-$\Upgamma$0 as a reference simulation for a farm operating in the absence of thermal stratification above the ABL in the rest of this manuscript, and we refer to it as the NBL reference case.

A working example of this idea is given in Figure \ref{fig:3D_intro}(a,b), which displays the flow field obtained in the main domain for the NBL reference case (i.e. H500-$\Updelta \uptheta$0-$\Upgamma$0) and the H500-$\Updelta \uptheta$5-$\Upgamma$4 case, respectively. The two simulations are driven by the same turbulent inflow profiles. However, the potential-temperature profile is set to a constant in Figure \ref{fig:3D_intro}(a) while Figure \ref{fig:3D_intro}(b) uses the potential temperature provided by the precursor simulations. By comparing the results of these two simulations, it is easy to investigate the effects induced by the thermal stratification above the ABL on the wind-farm flow behaviour. The various differences will be analyzed in detail in Section \ref{sec:atm_sensitivity}.

\section{Sensitivity of wind-farm performance to the domain length and width}\label{sec:domain_sensitivity}
The numerical domain should be wide enough to limit artificial sidewise blockage. The latter occurs when the lateral boundaries are too close to the farm. Moreover, enough streamwise distance should be present upwind of the farm, to properly account for the flow deceleration due to the wind-farm blockage effect. Similarly, the fringe region should be located far enough from the last-turbine row to avoid spurious effects on the flow development in proximity to the farm. In the presence of neutral atmospheres, these numerical artefacts have a very limited impact on the farm performance. In fact, the absence of free-atmosphere stratification reduces the wind-farm footprint, allowing for a smaller numerical domain. This is clearly visible in Figure \ref{fig:3D_intro}(a). However, the effects of the domain boundaries can significantly alter the wind-farm performance when thermal stratification above the ABL is considered. To date, there are no guidelines on how to select the domain length and width when performing simulation in CNBLs. The goal of this section is to define such criteria.
\begin{figure}
	\centerline{
		\includegraphics[width=0.6\textwidth]{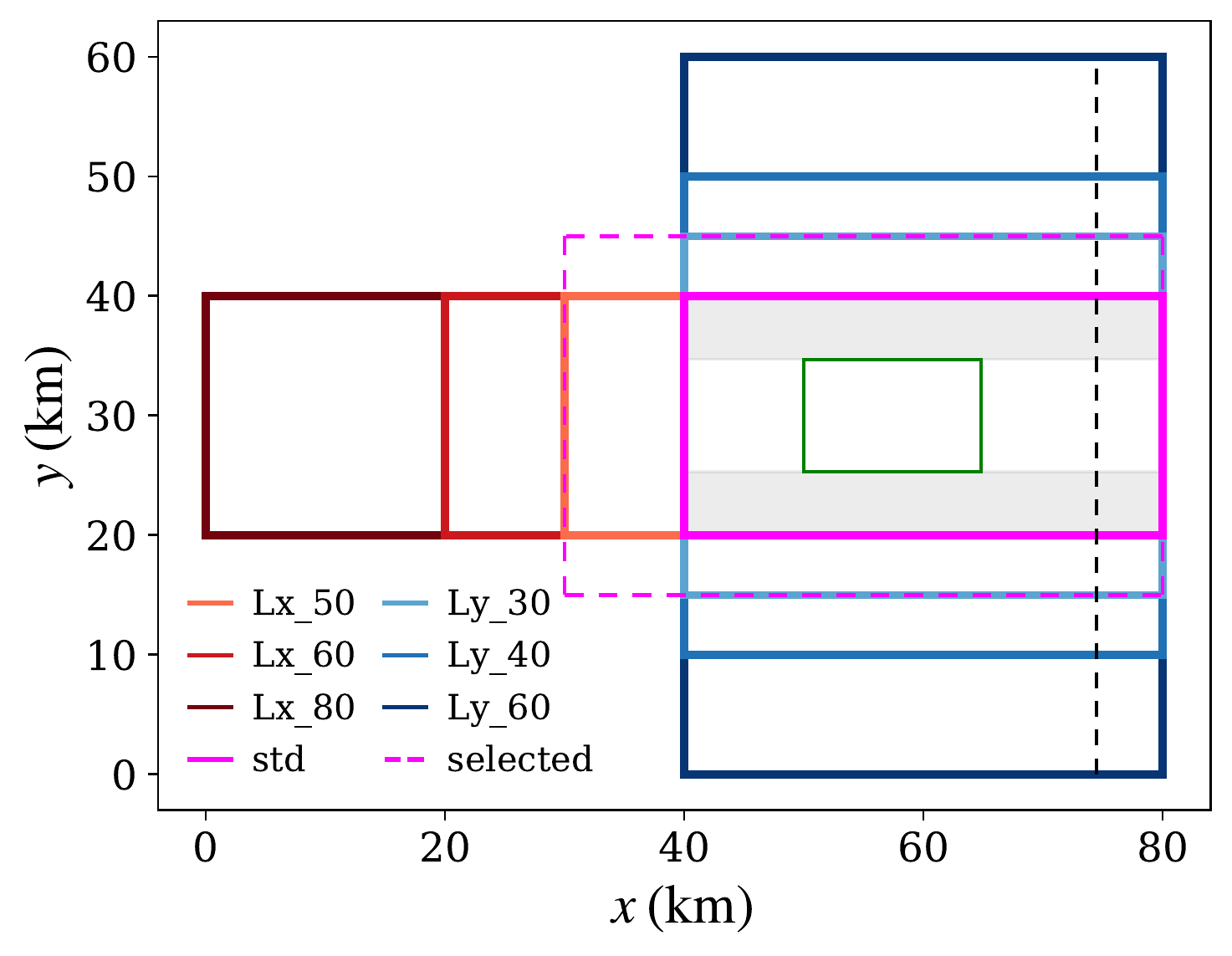}}% 
	\caption{Sketch of the numerical-domain length and width adopted in the sensitivity study. The dashed pink rectangle denotes the domain selected for performing the sensitivity study to the atmospheric state. Moreover, the green rectangle represents the wind-farm dimension while the vertical dashed black line denotes the starting point of the fringe region. Finally, the grey shaded areas and the region in between them represent the areas over which spanwise averages are taken by the operator $\langle \cdot \rangle_{\!s}$ and $\langle \cdot \rangle_{\!f}$, respectively.}
	\label{fig:domsens_layout}
\end{figure}

\begin{table}
	\def~{\hphantom{0}}
	\begin{center}
		\begin{adjustbox}{max width=\textwidth}
			\begin{tabular}{cccccccccc}
				\textbf{Cases}  & $\boldsymbol{L_x}$ \textbf{(km)} & $\boldsymbol{L_y}$ \textbf{(km)} & $\boldsymbol{L_z}$ \textbf{(km)} & $\boldsymbol{L_\mathrm{ind}/L_x^f}$ \textbf{(--)} & $\boldsymbol{L_y/L_y^f}$ \textbf{(--)} & $\boldsymbol{N_x}$ \textbf{(--)} & $\boldsymbol{N_y}$ \textbf{(--)} & $\boldsymbol{N_z}$ \textbf{(--)} & \textbf{DOF (--)}\\[7pt]
				std      & 40 & 20 & 25 & 0.67 & 2.13 & 1280 & 920  & 490 & $2.89$ $\times 10^9$\\
				Lx-50    & 50 & 20 & 25 & 1.35 & 2.13 & 1600 & 920  & 490 & $3.46$ $\times 10^9$\\ 
				Lx-60    & 60 & 20 & 25 & 2.02 & 2.13 & 1920 & 920  & 490 & $4.04$ $\times 10^9$\\
				Lx-80    & 80 & 20 & 25 & 3.37 & 2.13 & 2560 & 920  & 490 & $5.19$ $\times 10^9$\\
				Ly-30    & 40 & 30 & 25 & 0.67 & 3.19 & 1280 & 1380 & 490 & $4.33$ $\times 10^9$\\ 
				Ly-40    & 40 & 40 & 25 & 0.67 & 4.26 & 1280 & 1840 & 490 & $5.77$ $\times 10^9$\\
				Ly-60    & 40 & 60 & 25 & 0.67 & 6.38 & 1280 & 2760 & 490 & $8.66$ $\times 10^9$\\
				selected & 50 & 30 & 25 & 1.21 & 3.19 & 1600 & 1380 & 490 & $5.19$ $\times 10^9$\\[1pt]
			\end{tabular}
		\end{adjustbox}
		\label{table:domsens_setup}
	\end{center}
	\caption{Overview of the numerical domains used to perform the sensitivity study on the domain size. The parameters $L_x$, $L_y$ and $L_z$ denote the streamwise, spanwise and vertical domain dimensions while $N_x$, $N_y$ and $N_z$ denote the number of grid points used along the respective directions. Moreover, $L_\mathrm{ind}$ denotes the fetch of domain between the inflow and the first-row turbine location while $L_x^f$ and $L_y^f$ represent the wind-farm length and width, respectively. The last column reports the number of DOF comprehensive of both precursor and main domains. We note that each case is driven by three different CNBLs, that is H300-$\Updelta \theta$5-$\Upgamma$4, H500-$\Updelta \theta$5-$\Upgamma$4 and H1000-$\Updelta \theta$5-$\Upgamma$4, for a total of $21$ simulations. Finally, the last row reports the selected domain used to perform the sensitivity study to the atmospheric state.}
\end{table}

To this end, we first fix a relatively small reference domain with $L_x$ and $L_y$ equal to $40$ and $20$ km, respectively, which correspond to a typical domain length and width used in previous studies \citep{Allaerts2017,Allaerts2017b,Lanzilao2022,Lanzilao2022b}. Thereafter, we first keep $L_y$ constant and we vary $L_x$ between 40 and 80 km, so that the fetch of the domain from the inflow to the first-row turbine location (i.e. $L_\mathrm{ind}$) varies between 10 and 50 km, respectively. This corresponds to a ratio $L_\mathrm{ind}/L_x^f$ of 0.67 and 3.37. Next, we keep $L_x$ constant and we vary $L_y$ between 20 and 60 km, so that the ratio $L_y/L_y^f$ goes from 2.13 to 6.38, respectively. In total, we end up with seven domain configurations, which are illustrated in Figure \ref{fig:domsens_layout}. We note that all simulations adopt a vertical domain height of $25$ km and have the same grid resolution. Since we speculate that the domain size should scale with the height of the capping inversion, we drive these simulations with three different inflow profiles with equal $\Delta \theta$ and $\Gamma$ but different $H$, that is H300-$\Updelta \theta$5-$\Upgamma$4, H500-$\Updelta \theta$5-$\Upgamma$4 and H1000-$\Updelta \theta$5-$\Upgamma$4, for a total of $21$ simulations. The various cases analyzed in this section are summarized in Table~\ref{table:domsens_setup}. Finally, we note that the notations $\langle \cdot \rangle_{\!f}$ and $\langle \cdot \rangle_{\!s}$ are used to represent spanwise averages along the width of the farm (i.e. from first to last turbine column) and at its side, respectively. In case of a horizontal average over the full spanwise direction and along the turbine-rotor height or capping-inversion thickness, we adopt the notation $\langle \cdot \rangle_{\!r}$ and $\langle \cdot \rangle_{\!c}$, respectively. For instance, the notation $\langle \cdot \rangle_{\!f,r}$ represents a spanwise average over the farm width and a vertical average over the turbine-rotor height. We refer to Figure \ref{fig:domsens_layout} for more details.

\begin{figure}
	\centerline{
		\includegraphics[width=1\textwidth]{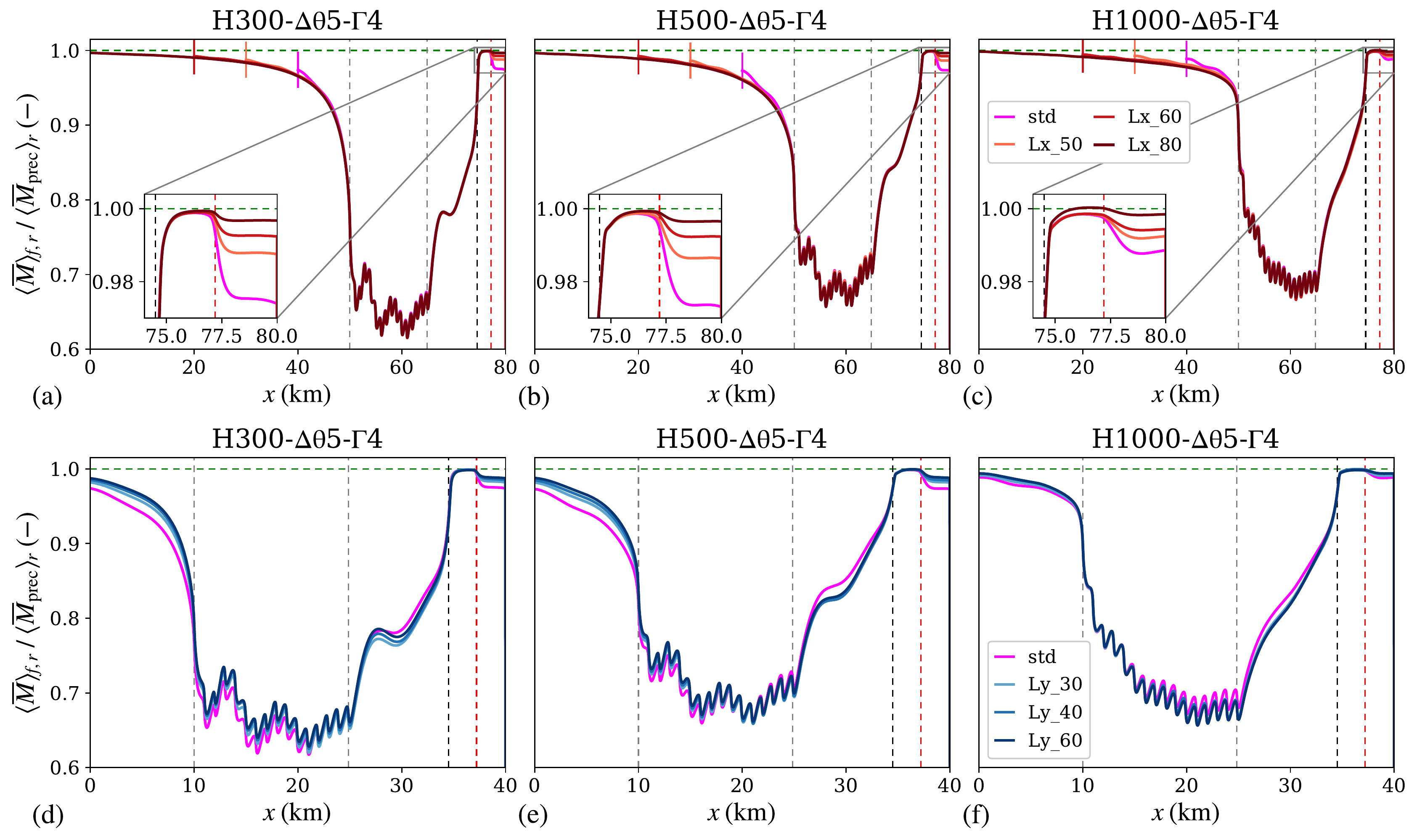}}% 
	\caption{Time-averaged velocity magnitude $M=(u^2+v^2)^{1/2}$ averaged in the horizontal direction along the farm width and in height along the turbine rotor. We normalize the results with the velocity magnitude obtained in the precursor simulation and we plot it as a function of the streamwise direction. Results are shown for different (a-c) domain lengths and (d-f) domain widths. The simulations are driven by the cases (a,d) H300-$\Updelta \theta$5-$\Upgamma$4, (b,e) H500-$\Updelta \theta$5-$\Upgamma$4 and (c,f) H1000-$\Updelta \theta$5-$\Upgamma$4. The grey vertical dashed lines are positioned in correspondence of the first- and last-row turbines. Moreover, the black and red vertical dashed lines denote the starting and ending point of the fringe forcing. The fringe region extends to the domain end to account for the damping of the convective term. In the bottom left corner of panels (a-c), the flow behaviour within the fringe region is magnified. Finally, the vertical segments in panels (a-c)  facilitate the identification of the starting points of the domains with smaller $L_x$.}
	\label{fig:domsens_vel_farm}
\end{figure}

Figure \ref{fig:domsens_vel_farm}(a) shows the time-averaged velocity magnitude further averaged in the horizontal directions along the farm width and in height along the turbine rotor. We scale the plot with the velocity magnitude obtained in the precursor simulation. Here, the results refer to domains of different lengths driven by the H300-$\Updelta \theta$5-$\Upgamma$4 case. Interestingly, the large $L_\mathrm{ind}$ value in the Lx\textunderscore80 case allows us to observe that the flow begins to slow down several tens of kilometers upstream of the farm. Within the farm, the effect of turbines and wake mixing is clearly visible. In the last 5.5 kilometers of the domain, the body force applied within the fringe region restores the inflow provided by the precursor simulation. Surprisingly, the solutions obtained on shorter domains follow the same trend and also have the same magnitude. In fact, the convective damping region within the fringe region seems to indirectly account for the additional flow slow down necessary to match the solution obtained on longer domains. Figure \ref{fig:domsens_vel_farm}(b,c) shows that the same behaviour is attained for deeper boundary layers.

The numerical solution is much more sensitive to the domain width. This is illustrated in Figure \ref{fig:domsens_vel_farm}(d), which shows the velocity magnitude obtained with domains of different widths driven by the H300-$\Updelta \theta$5-$\Upgamma$4 case. Here, the solution obtained on the standard domain has a velocity that is 1.5$\%$ lower than case Ly\textunderscore60 at $x=0$ km. This additional slow down is purely artificial and comes from the fact that $L_y/L_y^f$ measures only 2.12 in the standard domain. The same difference reduces to 0.5$\%$ for the deep boundary layer case -- see Figure \ref{fig:domsens_vel_farm}(f). This is expected since the influence of the inversion layer on the flow behaviour is inversely related to its height.

\begin{figure}
	\centerline{
		\includegraphics[width=1\textwidth]{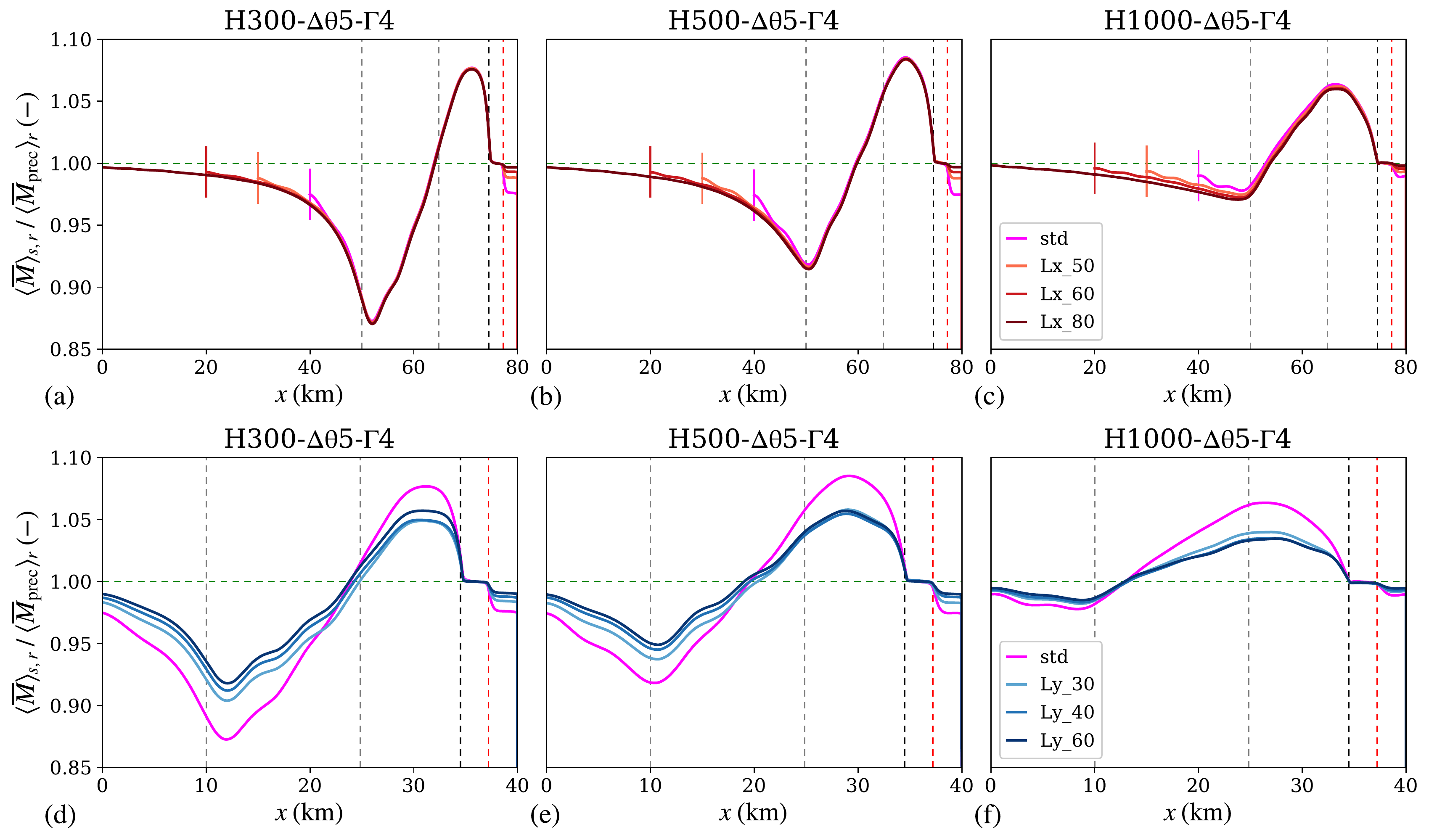}}% 
	\caption{Time-averaged velocity magnitude $M=(u^2+v^2)^{1/2}$ averaged in the horizontal directions along the farm sides and in height along the turbine rotor. We normalize the results with the velocity magnitude obtained in the precursor simulation and we plot it as a function of the streamwise direction. Results are shown for different (a-c) domain lengths and (d-f) domain widths. The simulations are driven by the cases (a,d) H300-$\Updelta \theta$5-$\Upgamma$4 , (b,e) H500-$\Updelta \theta$5-$\Upgamma$4  and (c,f) H1000-$\Updelta \theta$5-$\Upgamma$4. The grey vertical dashed lines are positioned in correspondence of the first- and last-row turbines. Moreover, the black and red vertical dashed lines denote the starting and ending point of the fringe forcing. The fringe region extends to the domain end to account for the damping of the convective term. Finally, the vertical segments in panels (a-c) facilitate the identification of the starting points of the domains with smaller $L_x$.}
	\label{fig:domsens_vel_side}
\end{figure}

The results in terms of velocity magnitude averaged in the horizontal directions along the farm sides (i.e. the shaded areas in Figure \ref{fig:domsens_layout}) and in height along the turbine rotor are shown in Figure \ref{fig:domsens_vel_side}. Again, when varying $L_x$ all solutions collapse onto the one obtained with the standard domain. However, considerable differences are observed in Figure \ref{fig:domsens_vel_farm}(d-e), where we vary the domain width. In fact, the presence of the inversion layer limits the boundary-layer thickening, causing the flow to accelerate at the farm sides. A narrow domain artificially enhances this channelling effect of about 2$\%$ in terms of velocity magnitude. As the inversion-layer height increases, solutions on wider domains tend to collapse onto each other.   

Finally, for a quantitative estimate of the influence of the domain length and width on wind-farm performance, we look at the farm efficiencies. Hence, similarly to \cite{Allaerts2017b}, we define the farm efficiency as $\eta_f = \eta_w \eta_{nl}$, with
\begin{equation}
	\eta_w = \frac{P_\mathrm{tot}}{N_t P_1}, \qquad \eta_{nl}=\frac{P_1}{P_\infty}
	\label{eq:farm_efficiencies}
\end{equation}
where $\eta_w$ and $\eta_{nl}$ denote the wake and non-local efficiency, respectively. Moreover, $N_t$ indicates the number of turbines, $P_\mathrm{tot}$ the total farm power, $P_1$ the averaged power output of first-row turbines while $P_\infty$ represents the power output of a turbine operating in isolation. The latter is evaluated using the single-turbine simulations discussed in Appendix \ref{app:st_sim}.

Figure \ref{fig:domsens_efficiency}(a) shows that the time-averaged non-local efficiency normalized with the value obtained in the standard domain is quasi-independent on the distance between the first-row turbine and the main domain inflow, i.e. on the ratio $L_\mathrm{ind}/L_x^f$. This is expected in light of the results shown in Figure \ref{fig:domsens_vel_farm}(a-c). However, considerable differences are observed when varying the domain width in shallow boundary layers. For instance, the time-averaged non-local efficiency obtained in case Ly\textunderscore60 driven by H300-$\Updelta \theta$5-$\Upgamma$4 is 1.12 times higher than the one measured in the standard domain. The wake efficiency, shown in Figure \ref{fig:domsens_efficiency}(b), is less dependent on the numerical domain size and it shows a negative correlation with respect to $\eta_{nl}$. 

In conclusion, the wind-farm performance is quasi-independent of the distance between the domain inflow and the first-row turbine location when using the wave-free fringe region technique. However, a narrow domain artificially enhances flow blockage. Finally, we observe that the domain size should depend upon the inversion-layer height. In fact, shallower boundary layers require wider domains than deeper ones. This also implies that wind-farm--LESs in CNBLs should adopt wider domains than simulations in neutral atmospheres. As a result of this study, we fix the main domain length and width to $50$ and $30$ km, respectively, with $L_\mathrm{ind}=18$ km. A sketch of this domain is reported in Figure~\ref{fig:domsens_layout}. We remark that this choice is mainly dictated by the computational resources at our disposal. In fact, an even wider domain would have been relevant. Based on the current analysis, we conclude that we may overestimate the effect of blockage on farm efficiency with a factor of roughly 1.05 for $H\leq500$ m, while the effect of blockage is properly represented for the H1000 cases. We note that the RDL and fringe region will be left out of the figures in the remainder of the text.

\begin{figure}
	\includegraphics[width=0.5\textwidth]{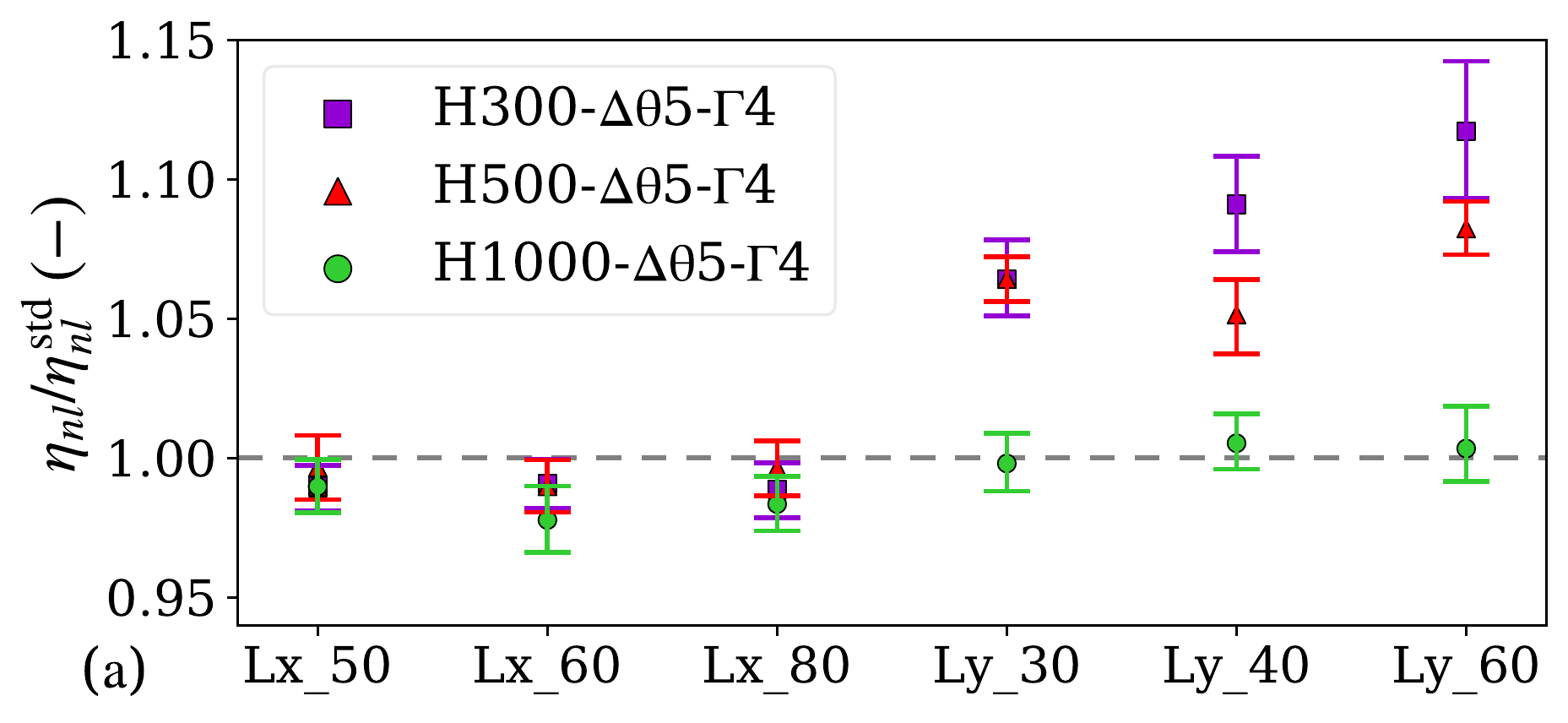}% 
	\includegraphics[width=0.5\textwidth]{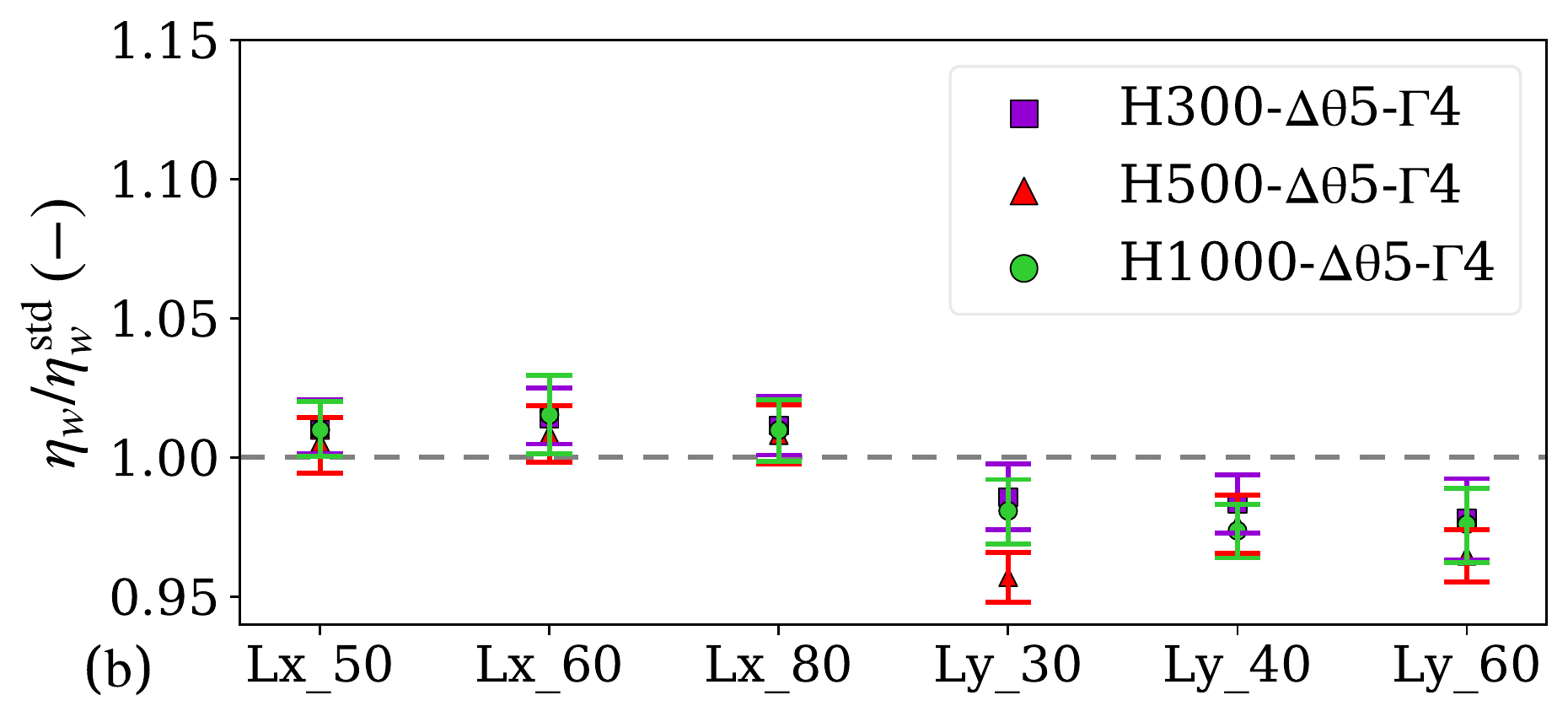}%
	\caption{(a) Non-local and (b) wake efficiency normalized by the values obtained in the standard domain. The symbol denotes the time-averaged value while the error bars represent the 95$\%$ confidence interval. The latter are computed using the moving block bootstrapping method developed by \cite{Beyaztas2017} and \cite{Boufidi2020}, using the procedure described by \cite{Bon2022}.}
	\label{fig:domsens_efficiency}
\end{figure}

\section{Sensitivity of wind-farm performance to the atmospheric state}\label{sec:atm_sensitivity}
The wind-farm flow development under various atmospheric conditions is examined in this section. First, we perform a qualitative analysis on the separate effects of varying the capping-inversion height, strength and free-atmosphere lapse rate on farm performance in Sections \ref{sec:ci_height} and \ref{sec:ci_strength}, respectively. Next, we compare the LES results against various one- and two-dimensional gravity-wave linear-theory models in Section \ref{sec:les_vs_lt}. Thereafter, we carry out a quantitative comparison in terms of state variables among all simulation cases in Section \ref{sec:bl_flow}. Finally, the column-averaged wind-farm power output is presented in Section \ref{sec:wf_power}, followed by a discussion on the non-local, wake and farm efficiencies in Section \ref{sec:wf_efficiencies}, where we also propose a new scaling parameter for the ratio of the non-local to wake efficiency.

\begin{figure}
	\centerline{
		\includegraphics[width=1\textwidth]{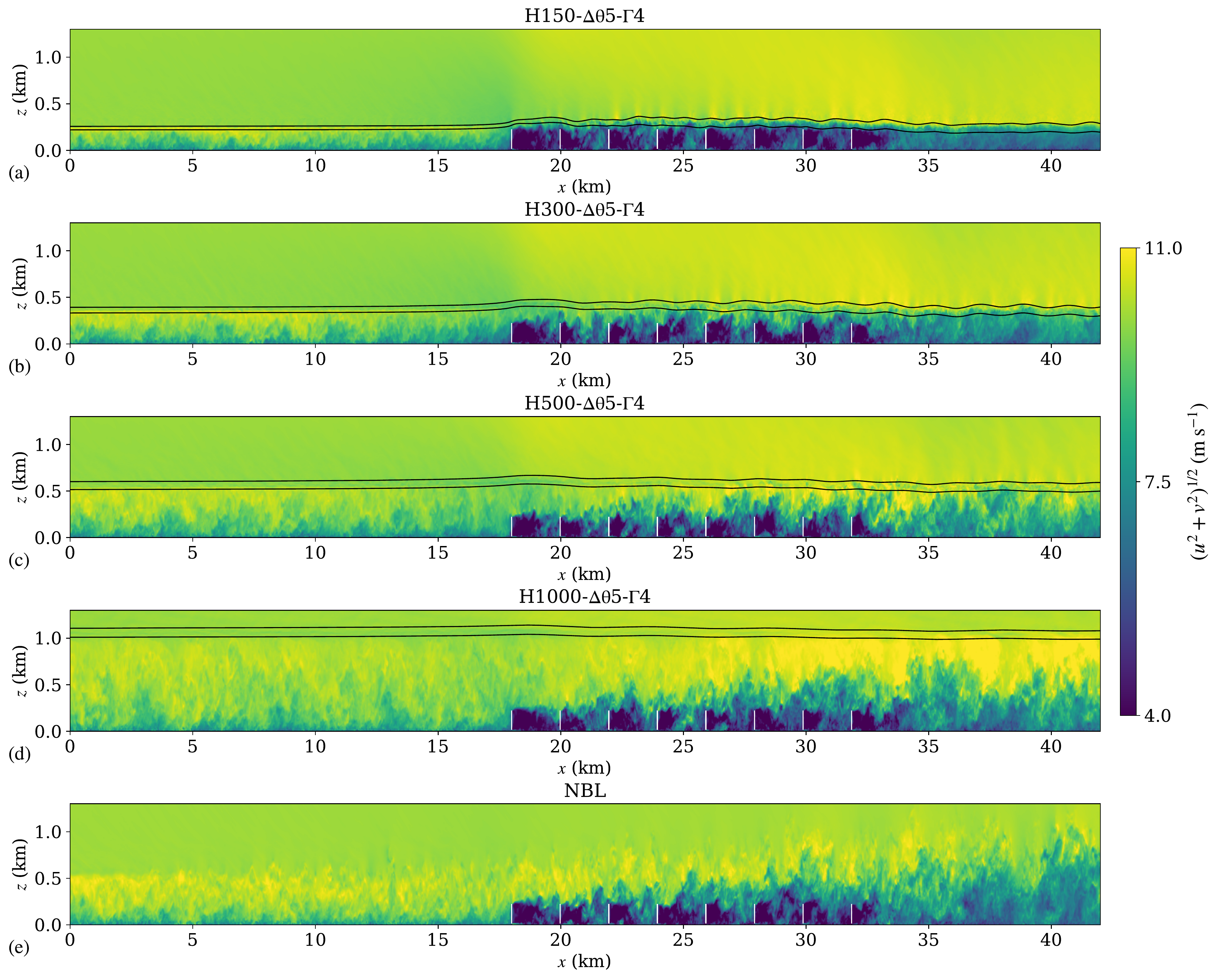}}% 
	\caption{Contours of the instantaneous horizontal velocity magnitude in an $x$--$z$ plane taken across the 6th column of turbines, i.e. at $y=15.25$ km, for cases (a) H150-$\Updelta \theta$5-$\Upgamma$4, (b) H300-$\Updelta \theta$5-$\Upgamma$4, (c) H500-$\Updelta \theta$5-$\Upgamma$4, (d) H1000-$\Updelta \theta$5-$\Upgamma$4 and (e) the NBL reference case. The black lines represent the bottom and top of the inversion layer computed with the \cite{Rampanelli2004} model. Finally, the location of the turbine-rotor disks is indicated with vertical white lines. The NBL reference case refers to simulation H500-$\Updelta \theta$0-$\Upgamma$0.}
	\label{fig:excases_m}
\end{figure}

\subsection{Effects of the inversion-layer height on the flow development}\label{sec:ci_height}
\begin{figure}
	\centerline{
		\includegraphics[width=1\textwidth]{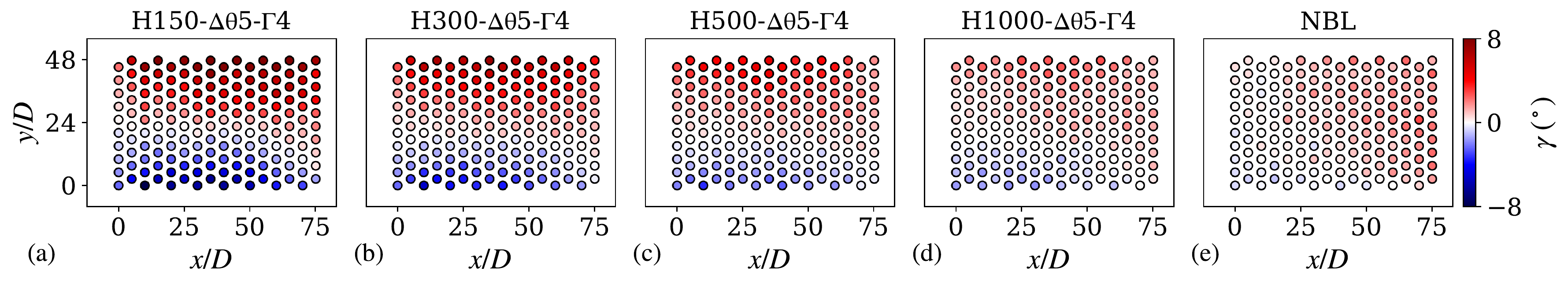}}% 
	\caption{Turbine-orientation angle with respect to the streamwise direction for cases (a) H150-$\Updelta \theta$5-$\Upgamma$4, (b) H300-$\Updelta \theta$5-$\Upgamma$4, (c) H500-$\Updelta \theta$5-$\Upgamma$4, (d) H1000-$\Updelta \theta$5-$\Upgamma$4 and (e) the NBL reference case. The latter refers to simulation H500-$\Updelta \theta$0-$\Upgamma$0.}
	\label{fig:excases_yaw}
\end{figure}

We start our analysis with Figure \ref{fig:excases_m}, which shows the $x$--$z$ plane across the 6th column of turbines of the instantaneous horizontal velocity magnitude together with the base and top of the inversion layer computed by fitting the LES data with the \cite{Rampanelli2004} model. Every turbine imparts a force on the flow, generating a patch of low wind speed downwind. Figure \ref{fig:excases_m}(a) shows that such velocity deficits are very strong in a shallow boundary layer. In fact, the turbine-tip height is in close proximity to the capping-inversion base, which limits energy entrainment and therefore flow mixing, slowing down the wake recovery process. Moreover, this also limits the growth of the internal boundary layer~(IBL) generated by the wind-turbine wakes expansion. Therefore, to conserve mass, the flow deceleration is mostly compensated by a flow redirection at the sides of the farm, generating high-speed flow channels. This is visible in Figure \ref{fig:excases_yaw}(a), which shows the time-averaged turbine-orientation angle with respect to the streamwise direction for all turbines in the farm. Here, the angles between the first- and last-turbine columns vary between $-8^\circ$ and $8^\circ$. The asymmetry with respect to the domain centerline observed in Figure \ref{fig:excases_yaw} is due to the presence of the Coriolis effects. The upward motion caused by the strong flow divergence results in a capping-inversion vertical displacement, which reaches a maximum in proximity to the second-row turbines, with a relative capping-inversion displacement of 42.9$\%$. A strong wind-speed reduction is also observed in the farm induction region. Further, the low level of energy entrainment also causes a strong wind-farm wake. The interaction between the IBL and capping inversion decreases as $H$ increases. We observe in Figure \ref{fig:excases_m}(c) that the vertical wake expansion reaches the inversion-layer height only in the farm wake. 

For a deep boundary layer, the capping-inversion effects on the wind-farm response become negligible. Here, the IBL does not interact with the inversion layer. This also allows a high-speed flow region to form between the IBL and capping inversion base, as shown in Figure \ref{fig:excases_m}(d), which enhances flow mixing and consequently the vertical transport of energy, therefore replenishing the turbine and farm wake at a higher rate. This also translates into a much smaller flow redirection at the farm sides since the vertical expansion of the IBL is not constrained by the inversion layer. Figure \ref{fig:excases_yaw}(d) shows variation in turbine-orientation angles only between $-2^\circ$ and $2^\circ$ for this case. In all atmospheres with thermal stratification above the ABL, Figure \ref{fig:excases_m} shows that the turbulent structures are damped by the inversion layer, leaving the free atmosphere non-turbulent. A different behaviour is observed in Figure \ref{fig:excases_m}(e), which illustrates the results obtained in a neutral atmosphere. The turbulence structures at the top of the boundary layer are not damped by buoyant forces in this case, allowing the IBL to reach higher altitudes. Here, the reduction in wind speed caused by the turbines is solely balanced by the boundary-layer thickening. In fact, Figure \ref{fig:excases_yaw}(e) shows that all turbines have positive orientation angles, which gradually increase towards the last row.

\begin{figure}
	\centerline{
		\includegraphics[width=1\textwidth]{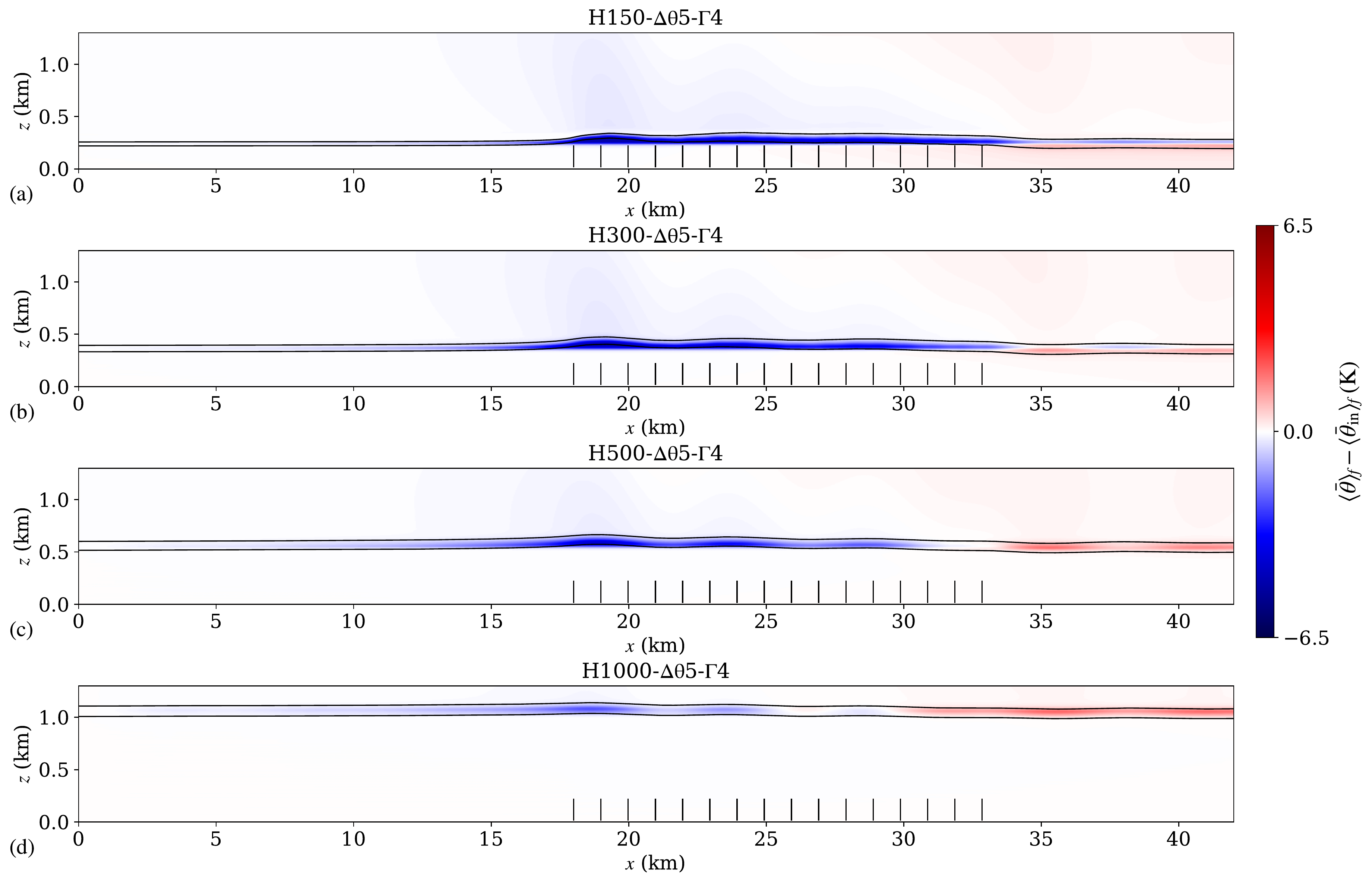}}% 
	\caption{Contours of the time-averaged potential-temperature perturbation with respect to the inflow in an $x$--$z$ plane further averaged in the $y$-direction along the width of the farm for cases (a) H150-$\Updelta \theta$5-$\Upgamma$4, (b) H300-$\Updelta \theta$5-$\Upgamma$4, (c) H500-$\Updelta \theta$5-$\Upgamma$4 and (d) H1000-$\Updelta \theta$5-$\Upgamma$4. The black lines represent the bottom and top of the inversion layer computed with the \cite{Rampanelli2004} model. Finally, the location of the turbine-rotor disks is indicated with vertical black lines.}
	\label{fig:excases_th}
\end{figure}

The upward displacement of the inversion layer, which has to be considered as an interfacial wave, brings air with a lower potential temperature to a higher altitude, generating a cold anomaly. This is illustrated in Figure \ref{fig:excases_th}, which shows the $x$--$z$ plane of the time-averaged potential-temperature perturbation field together with the base and top of the inversion layer, averaged in the horizontal directions along the farm width. The negative perturbation in the potential-temperature field is about 7 K and 2 K in cases H150-$\Updelta \theta$5-$\Upgamma$4 and H1000-$\Updelta \theta$5-$\Upgamma$4, respectively. This considerable difference is caused by the higher inversion-layer displacement attained in case H150-$\Updelta \theta$5-$\Upgamma$4 (and in shallow boundary-layer flows, in general), which measures 42.9$\%$ (i.e. 90 m) against the 3.8$\%$ (i.e. 40 m) obtained for case H1000-$\Updelta \theta$5-$\Upgamma$4. We note that the cold anomaly extends to the farm induction region in all cases. Moreover, the interfacial waves along the capping inversion are also clearly visible. In fact, Figure \ref{fig:excases_th} shows that the interfacial-wave crests correspond to high potential-temperature perturbations. Finally, the flow acceleration in the wind-farm wake pushes the inversion layer downward, generating a hot anomaly. Since the ABL itself is neutral, potential-temperature perturbations do not occur below the inversion layer. 
\begin{figure}
	\centerline{
		\includegraphics[width=1\textwidth]{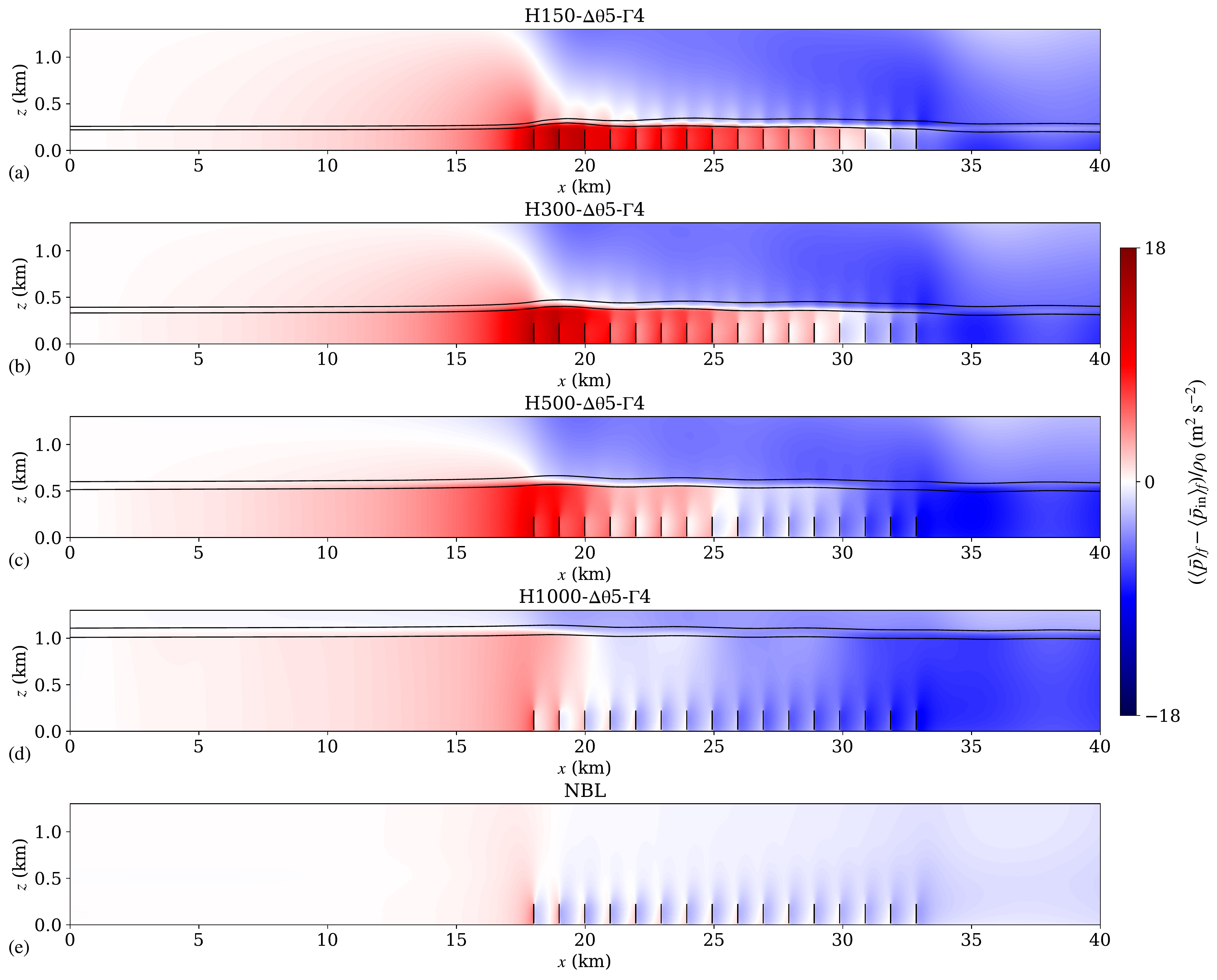}}% 
	\caption{Contours of the time-averaged pressure perturbation with respect to the inflow in an $x$--$z$ plane further averaged in the $y$-direction along the width of the farm for cases (a) H150-$\Updelta \theta$5-$\Upgamma$4, (b) H300-$\Updelta \theta$5-$\Upgamma$4, (c) H500-$\Updelta \theta$5-$\Upgamma$4, (d) H1000-$\Updelta \theta$5-$\Upgamma$4 and (e) the NBL reference case. The black lines represent the bottom and top of the inversion layer computed with the \cite{Rampanelli2004} model. Finally, the location of the turbine-rotor disks is indicated with vertical black lines. The NBL reference case refers to simulation H500-$\Updelta \theta$0-$\Upgamma$0.}
	\label{fig:excases_p}
\end{figure}

As noted by \cite{Smith2010} and \cite{Allaerts2017}, variations in the potential-temperature field are strongly correlated to pressure perturbations. In fact, a cold anomaly translates into a higher column of cold and heavy air, which locally increases pressure. Vice versa, a hot anomaly generates a region of low pressure. This behaviour is illustrated in Figure \ref{fig:excases_p}, which shows an $x$--$z$ plane of the time-averaged pressure-perturbation field together with the base and top of the inversion layer, averaged in the horizontal directions along the farm width. The strong cold anomaly generated in shallow boundary layers gives rise to strong increases in pressure, with a peak in the farm entrance region. This strong counteracting pressure gradient extends across the whole farm induction region. By comparing Figure \ref{fig:excases_p}(a,d), it becomes clear that the unfavourable pressure gradient is inversely proportional to the capping-inversion height. The low pressure region downwind of the farm generated by the hot anomaly gives rise to a favourable pressure gradient across the farm, which acts as an energy source and enhances the wake recovery mechanism. Finally, Figure \ref{fig:excases_p}(e) shows the pressure perturbation in an atmosphere that does not support gravity waves. Here, the unfavourable pressure gradient, which is solely due to the flow slow down caused by the cumulative turbine induction, is an order of magnitude lower than the one obtained in stratified atmospheres and only extends up to roughly six rotor diameters upstream of the first-row turbines. Moreover, also the favourable pressure gradient is negligible when compared to the one obtained in the presence of atmospheric gravity waves.

\subsection{Effects of the inversion-layer strength and free-atmosphere lapse rate on the flow development}\label{sec:ci_strength}
\begin{figure}
	\includegraphics[width=1\textwidth]{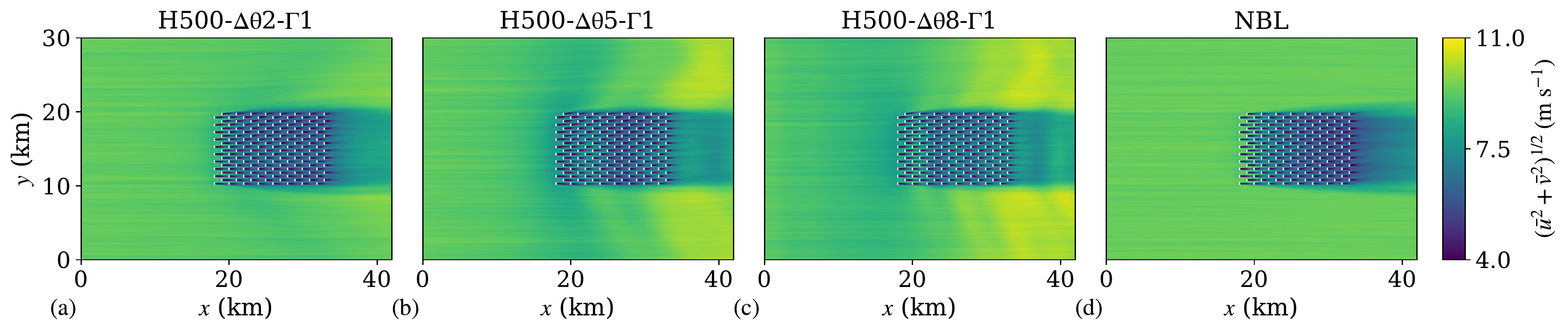} 
	\includegraphics[width=1\textwidth]{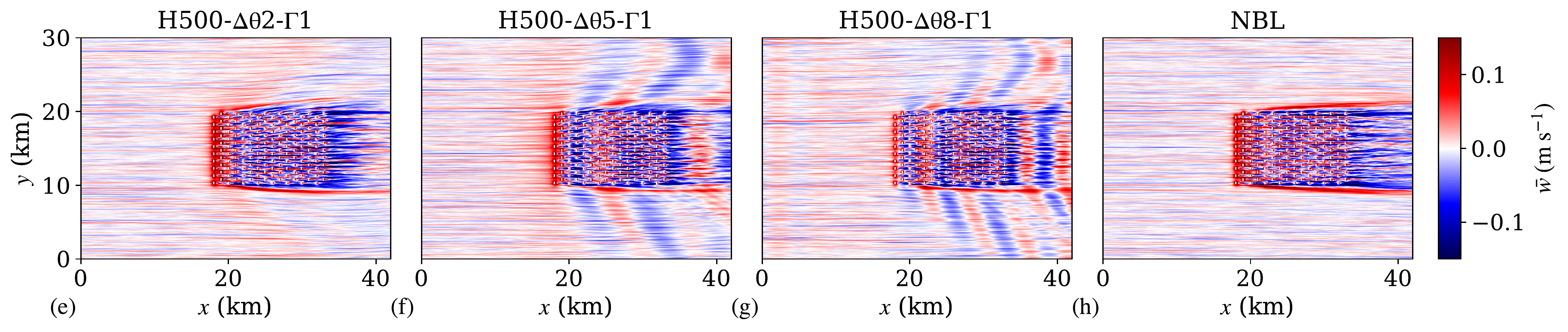} 
	\includegraphics[width=1\textwidth]{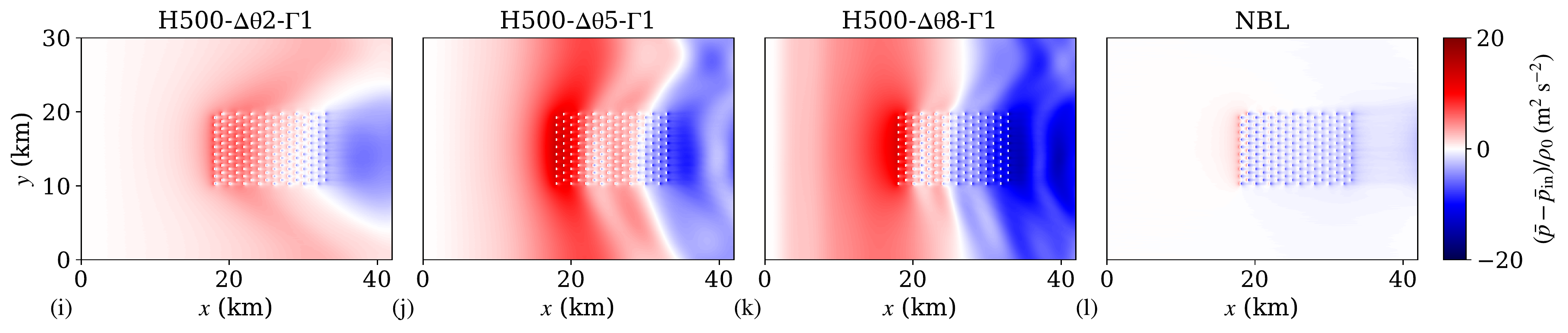} 
	\caption{Contours of the time-averaged (a-d) velocity magnitude, (e-h) vertical velocity and (i-l) pressure perturbation in an $x$--$y$ plane taken at turbine-hub height for cases (a,e,i) H500-$\Updelta \theta$2-$\Upgamma$1, (b,f,j) H500-$\Updelta \theta$5-$\Upgamma$1, (c,g,k) H500-$\Updelta \theta$8-$\Upgamma$1 and (d,h,l) the NBL reference case. The location of the turbine-rotor disks is indicated with vertical white lines. The NBL reference case refers to simulation H500-$\Updelta \theta$0-$\Upgamma$0.}
	\label{fig:excases_strength}
\end{figure}

Following the work of \cite{Klemp1975} and \cite{Vosper2004}, \cite{Sachsperger2015} derived a two-dimensional gravity-wave linear model in which they found out that the wavenumber $k_x$ of the interfacial waves depends upon the capping-inversion strength and Brunt--V\"{a}is\"{a}l\"{a} frequency as follows:
\begin{equation}
	k_x = \frac{g \Delta \theta}{2{U}_B^2\theta_0} + \frac{N^2 \theta_0}{2 g \Delta \theta}.
	\label{eq:lt_trapped_waves}
\end{equation}
Since the interfacial-wave horizontal wavelength is defined as $\lambda_x = 2 \pi / k_x$, they showed that $\lambda_x$ and $\Delta \theta$ are inversely related. Consequently, a stronger capping inversion supports interfacial waves with a lower wavelength. This is visible in Figure \ref{fig:excases_strength}, where we compare $x$--$y$ planes taken at turbine-hub height of velocity magnitude, vertical velocity and pressure perturbation of three simulations where the only varying parameter is $\Delta \theta$. Figure \ref{fig:excases_strength}(a-c) shows that a higher $\Delta \theta$ causes a higher flow speed up at the farm sides. This is due to the fact that, on average, a high inversion-layer strength reduces the capping-inversion upward displacement. Consequently, the flow rate at the farm sides has to increase to compensate for the limited thickening of the boundary layer. The vertical velocity fields shown in Figure \ref{fig:excases_strength}(e-g) clearly illustrate that the upward motion caused by these waves propagates down to turbine-hub height, generating patches of low and high wind speed. According to Equation~\ref{eq:lt_trapped_waves}, the interfacial-wave wavelength is 6.1 km and 4 km for cases H500-$\Updelta \theta$5-$\Upgamma$1 and H500-$\Updelta \theta$8-$\Upgamma$1, which is in line with the 7.3 km and 4.9 km observed in Figure \ref{fig:excases_strength}(f,g). However, $\lambda_x$ measures 10.1 km for case H500-$\Updelta \theta$2-$\Upgamma$1, which corresponds to 2/3 of the farm length. For this case, interfacial waves are not visible in Figure \ref{fig:excases_strength}(a,e,i). We speculate that a longer domain is necessary for these waves to become clearly visible. We also remark the presence of a V-shape pattern at the farm sides, which is very similar to the one noted by \cite{Allaerts2019}. This phenomenon will be further discussed in Section \ref{sec:les_vs_lt}.

Finally, Figure \ref{fig:excases_strength}(g-i) displays pressure perturbation with respect to the inflow value. The footprint of the counteracting pressure gradient in the farm induction region together with the favourable one across the farm is positively correlated with $\Delta \theta$. However, case H500-$\Updelta \theta$5-$\Upgamma$4 shows the highest pressure perturbation magnitude, the latter being roughly 1.5 times the one obtained in case H500-$\Updelta \theta$2-$\Upgamma$1. We speculate that this is due to the chocking effect \citep{Smith2010}, since $\textit{Fr}=0.99$ in this case. Moreover, the V-shape pattern causes a favourable pressure gradient also at the farm sides, which further enhances the channelling effects. We note that the recycling of pressure perturbations along the spanwise direction suggests that the domain width should be further increased in future studies. Finally, Figure \ref{fig:excases_strength}(d,h,l) shows the results obtained in the NBL reference case. As mentioned earlier, the pressure perturbations are at least one order of magnitude lower than in cases with thermal stratification above the ABL. This explains the absence of velocity reductions several kilometers upstream of the farm together with a monotonic decrease in velocity magnitude between the first- and last-row turbines.
\begin{figure}
	\includegraphics[width=1\textwidth]{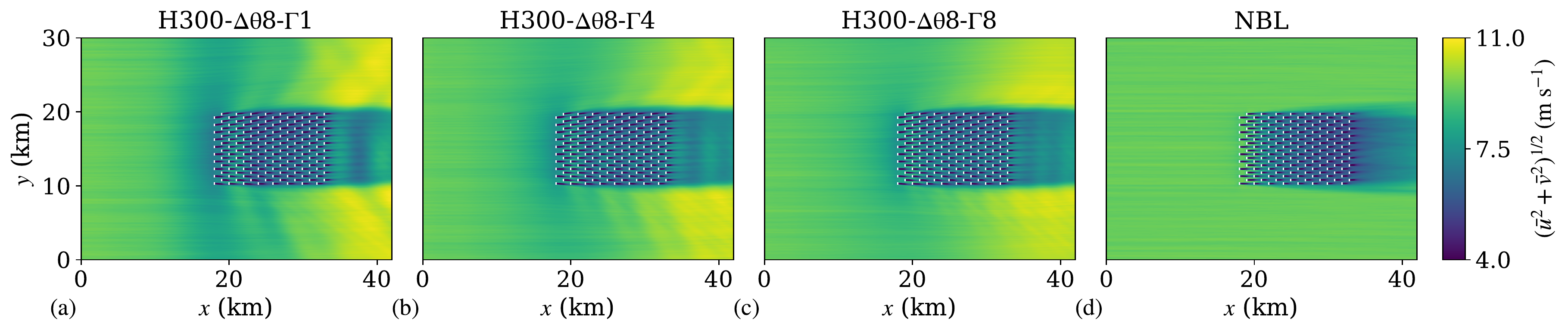} 
	\includegraphics[width=1\textwidth]{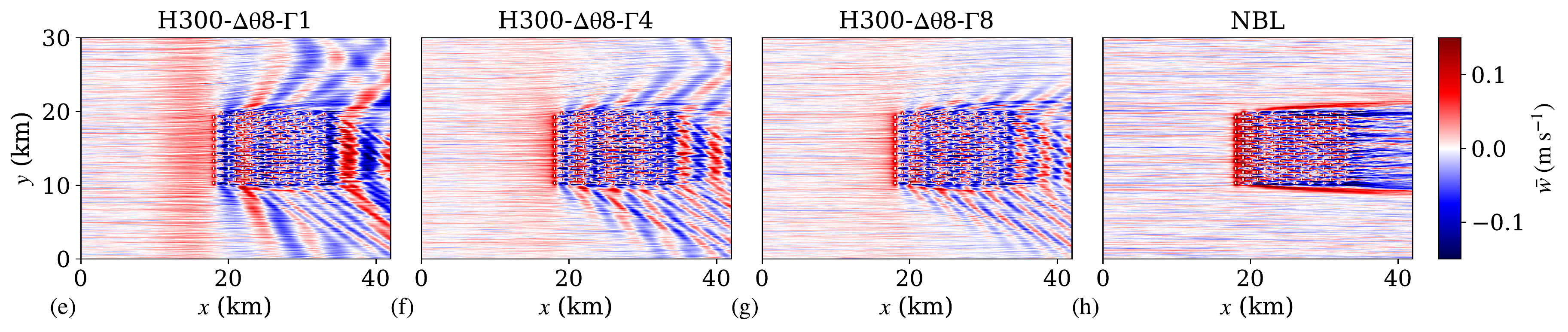} 
	\includegraphics[width=1\textwidth]{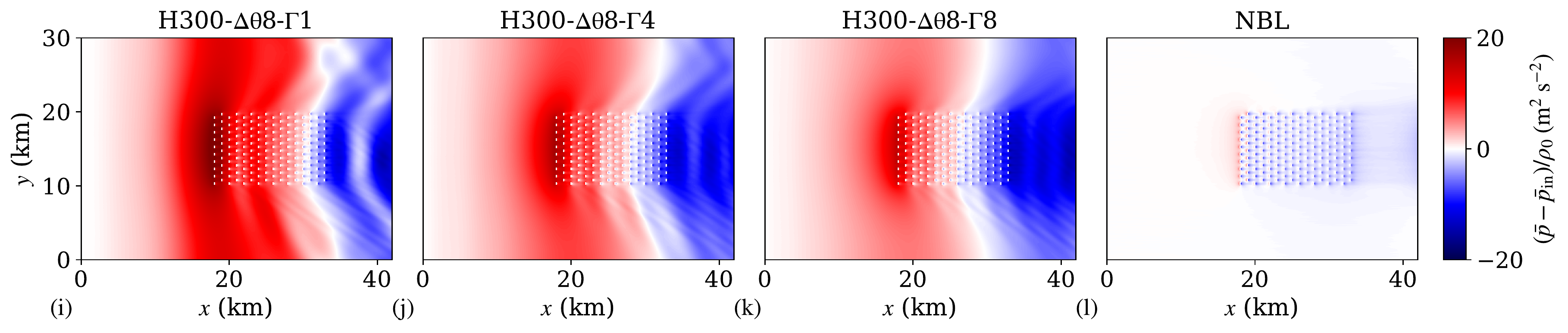} 
	\caption{Contours of the time-averaged (a-d) velocity magnitude, (e-h) vertical velocity and (i-l) pressure perturbation in an $x$--$y$ plane taken at turbine-hub height for cases (a,e,i) H300-$\Updelta \theta$8-$\Upgamma$1, (b,f,j) H300-$\Updelta \theta$8-$\Upgamma$4, (c,g,k) H300-$\Updelta \theta$8-$\Upgamma$8 and (d,h,l) the NBL reference case. The location of the turbine-rotor disks is indicated with vertical white lines. The NBL reference case refers to simulation H500-$\Updelta \theta$0-$\Upgamma$0.}
	\label{fig:excases_lapserate}
\end{figure}

We now turn our attention to the effects of changes in the free-atmosphere lapse rate. Equation \ref{eq:lt_trapped_waves} shows that the interfacial-wave horizontal wavelength is negatively correlated with the Brunt--V\"{a}is\"{a}l\"{a} frequency. This means that an increases in free-atmosphere stability results in a lower $\lambda_x$ value. Figure \ref{fig:excases_lapserate} clearly illustrates this behaviour. As $\Gamma$ increases, air parcels find more resistance in displacements along the vertical direction due to the stronger buoyant forces. Therefore, a stronger free-lapse rate damps waves along the inversion layer, limiting its vertical displacement. Consequently, a lower cold anomaly is generated, which leads to lower pressure perturbations. This is visible when comparing Figure \ref{fig:excases_lapserate}(i,k) which shows $x$--$y$ planes of pressure perturbations at turbine-hub height obtained with $\Gamma=1$~K~km$^{-1}$ and $\Gamma=8$~K~km$^{-1}$, respectively, while keeping all other parameters constant. This also implies that the hydrostatic flow blockage effect reduces as the stability of the free-atmosphere increases in the presence of a strong capping-inversion strength, as shown in Figure \ref{fig:excases_lapserate}(a-c). This result is in contrast with findings of \cite{Abkar2013} and \cite{Wu2017}, who observed that changing the free-lapse rate from 1~K~km$^{-1}$ to 10~K~km$^{-1}$ caused a power drop of about 35$\%$. However, in their study, they were at the same time varying the capping-inversion height and strength, which explains this difference in power output. Finally, Figure \ref{fig:excases_lapserate}(e-g) shows the vertical velocity fields, where the V-shape pattern at the farm sides together with interfacial-wave effects are clearly visible. Interestingly, we also observe slanted lines at the left side of the farm with an angle of 29$^\circ$ with respect to the streamwise direction. We note that this effect is related to the perturbation introduced by the single turbines and will be further discussed in Section \ref{sec:les_vs_lt}. For the sake of comparison, we report again the results obtained for the NBL reference case in Figure \ref{fig:excases_lapserate}(d,h,l).

\subsection{Comparison between LES results and gravity-wave linear-theory models}\label{sec:les_vs_lt}
The simplicity of gravity-wave linear-theory models contributed to their spread in analyzing mountain-wave phenomena. A wide variety of this type of models can be found in \cite{Gill1982}, \cite{Nappo2002}, \cite{Lin2007}, \cite{Sutherland2010} and \cite{Teixeira2014}. More recently, the same theory has been applied to wind-farm studies, since the latter can be considered as a ``permeable mountain''. Example of these models can be found in  \cite{Smith2010}, \cite{Allaerts2017b}, \cite{Allaerts2019}, \cite{Devesse2022} and \cite{Smith2022,Smith2023}. In the current section, we perform a comparison of our results against some of the simplest wind-farm gravity-wave linear-theory models.
\begin{figure}
	\centering
	\includegraphics[width=1\textwidth]{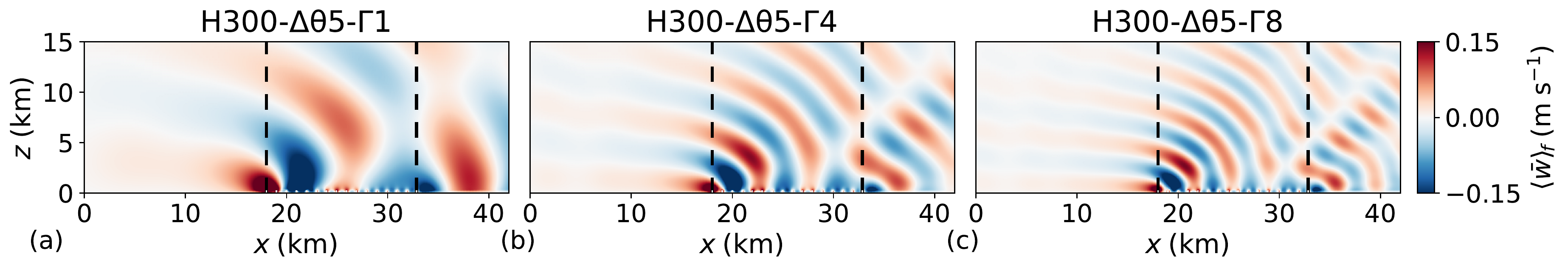} 
	\includegraphics[width=1\textwidth]{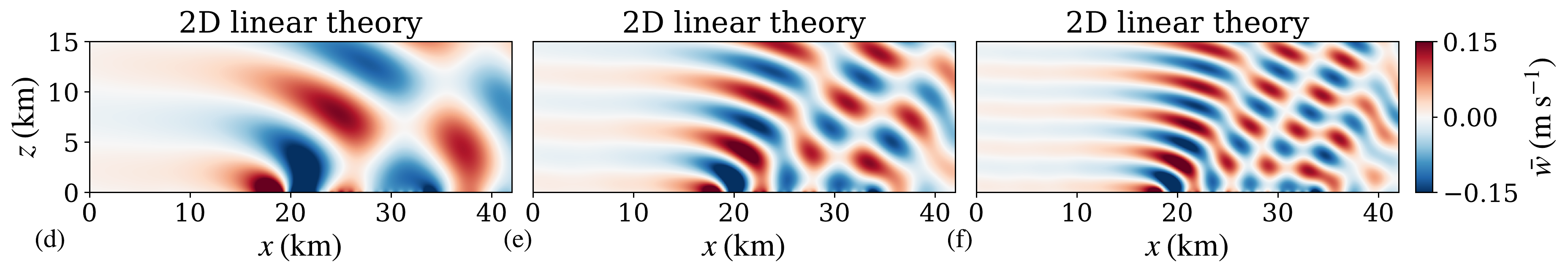}
	\caption{Contours of the time-averaged vertical velocity field in an $x$--$z$ plane further averaged along the farm width for cases (a) H300-$\Updelta \theta$5-$\Upgamma$1, (b) H300-$\Updelta \theta$5-$\Upgamma$4 and (c) H300-$\Updelta \theta$5-$\Upgamma$8. (d-f) Vertical velocity field obtained with the two-dimensional gravity-wave linear-theory model described in Appendix \ref{app:lt_model}. For instance, the internal waves in panel (d) are excited by an obstacle with shape given by the capping-inversion vertical displacement obtained in case H300-$\Updelta \theta$5-$\Upgamma$1. The vertical black dashed lines in panel (a-c) denote the location of the first- and last-row turbines while the location of the turbine-rotor disks in panels (a-c) are indicated with vertical white lines.}
	\label{fig:les_vs_lt_w}
\end{figure}

Figure \ref{fig:les_vs_lt_w} compares the internal gravity-wave pattern obtained in three LESs against the results obtained with a simple two-dimensional gravity-wave linear-theory model \citep{Nappo2002}. The latter is a quasi-analytical model which takes as an input the shape of the obstacle, assumed impermeable, which is given to the system of equations as a bottom boundary conditions.  In the current study, the inversion-layer displacement shown in Figure \ref{fig:blflow_all}(a-d) defines the obstacle shape. For the sake of clearness, we briefly explain this linear model in Appendix \ref{app:lt_model}. We remark the excellent agreement between our results and linear theory, particularly in terms of gravity-wave phase line and vertical wavelength. Moreover, the slanted region tilted downstream of zero vertical velocity that forms near the trailing edge of the farm is captured by both models. This supports the idea that this phenomenon is not a manifestation of internal gravity-wave reflection but is rather due to the interaction between two out-of-phase trains of internal waves triggered by the first- and last-row turbines -- see Figure \ref{fig:windfarm_startup}. We believe that this good comparison is mostly a result of the effort spent in properly developing and calibrating the buffer regions in the main domain. We note that a discussion on the internal gravity-wave reflectivity is reported in Appendix \ref{app:gw_reflectivity}.

Next, Figure \ref{fig:les_vs_lt_angle} displays a $x$--$y$ plane of the time-averaged vertical velocity field taken at turbine-tip height for case H300-$\Updelta \theta$5-$\Upgamma$1. The flow deceleration causes a strong upward motion in the proximity of the first rows of turbines, which results in interfacial waves along the capping inversion. The upward and downward motion that these waves generate is very visible in Figure \ref{fig:les_vs_lt_angle}, both in and around the farm.  
\begin{figure}
	\centerline{
		\includegraphics[width=0.6\textwidth]{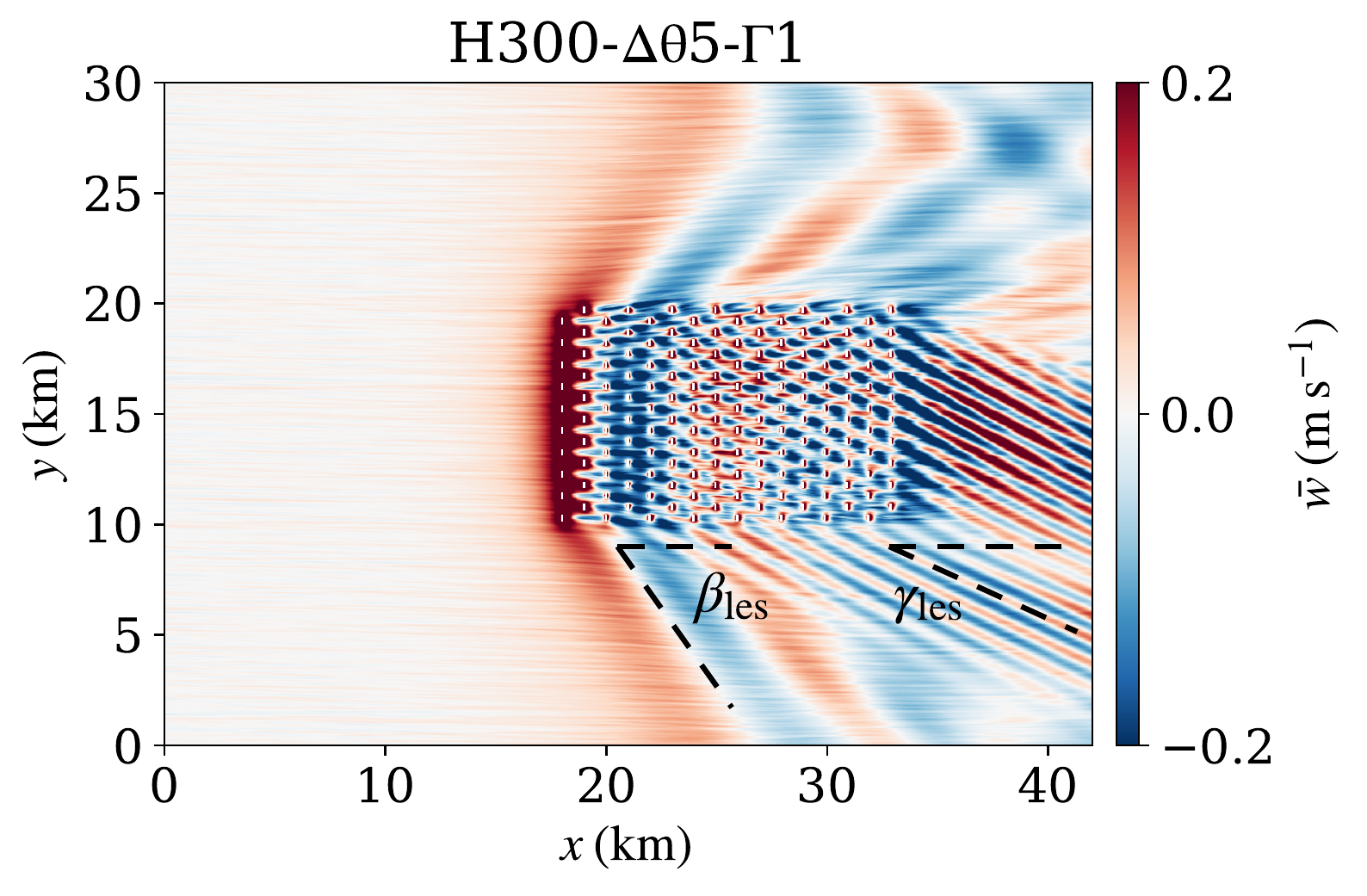}}% 
	\caption{Contours of the time-averaged vertical velocity field in an $x$--$y$ plane taken at turbine-tip height for case H300-$\Updelta \theta$5-$\Upgamma$1. The definition of the farm V-shape angle $\beta_\mathrm{les}$ together with the angle that the slanted lines make with the horizontal $\gamma_\mathrm{les}$ are reported in the figure. The location of the turbine-rotor disks is indicated with vertical white lines.}
	\label{fig:les_vs_lt_angle}
\end{figure}

Two additional distinct phenomena take place. First, we observe a V-shape pattern around the farm which has been already observed and investigated by \cite{Allaerts2019}. They reported that the angle of the pair of characteristic lines which gives rise to this pattern is only dependent on the Froude number and is given by 
\begin{equation}
	\beta_\mathrm{lin} = \arctan{\biggl(\bigl( \textit{Fr}^2-1\bigl)^{-1/2}}\biggl),
	\label{eq:v_shape_angle}
\end{equation}
where the Froude number is defined as a ratio between a boundary-layer velocity scale (i.e. ${U}_B$) and the interfacial-wave phase speed. In shallow-water wave theory, the phase speed is given by $\sqrt{g'H}$ \citep{Acheson1990,Sutherland2010}. This theory holds under the assumption that $L>4\pi H$, where $L$ denotes the length scale of the forcing. If we use the distance between first- and last-row turbines as a length scale (i.e. $L=L_x^f$), this relation holds and the Froude number is therefore defined as $\textit{Fr}={U}_B/\sqrt{g'H}$ \citep{Smith2010,Allaerts2019}. For instance, the Froude number is 1.29 for the case shown in Figure \ref{fig:les_vs_lt_angle}, which results in $\beta_\mathrm{lin}=50.8^\circ$. This value is in excellent agreement with the one observed in the LES results, which measures $\beta_\mathrm{les}=51^\circ$. Equation \ref{eq:v_shape_angle} shows that the angle is well defined only for supercritical flows. In subcritical cases, the characteristic lines become imaginary and no V-shape pattern occurs around the farm \citep{Allaerts2019}. The second pattern observed in Figure \ref{fig:les_vs_lt_angle} is a set of slanted lines, and is related to the perturbation introduced by the turbine spacing. In fact, if we use as a length scale the equivalent turbine spacing $S_e = D \sqrt{s_x s_y}$, then we should define the interfacial-wave phase speed using deep-water wave theory, which holds when $L<H/\pi$. Using this theory, the Froude number is defined as
\begin{equation*}
	\textit{Fr}_{dw} = \frac{{U}_B}{\sqrt{\dfrac{g'S_e}{2 \pi} \mathrm{tanh}\biggl( \dfrac{2 \pi H}{S_e}\biggl)}}
\end{equation*}
where the denominator represents the interfacial-wave phase speed. This Froude number measures 1.88 for case H300-$\Updelta \theta$5-$\Upgamma$1. Using Equation \ref{eq:v_shape_angle}, this results in an angle of 32.1$^\circ$, which we denote with $\gamma_\mathrm{lin}$. Also, this angle is in good agreement with the one observed in the LES results, which measures $\gamma_\mathrm{les}=28^\circ$. We note that the lines that make the angles $\beta_\mathrm{les}$ and $\gamma_\mathrm{les}$ with the horizontal are also reported in Figure \ref{fig:les_vs_lt_angle}. While the angle is explained by Equation \ref{eq:v_shape_angle}, we also see in Figure \ref{fig:les_vs_lt_angle} that these slanted lines only propagate along the left side of the farm. We speculate that this is due to the asymmetry generated by the Coriolis force.

\begin{figure}
	\centering
	\includegraphics[width=0.32\textwidth]{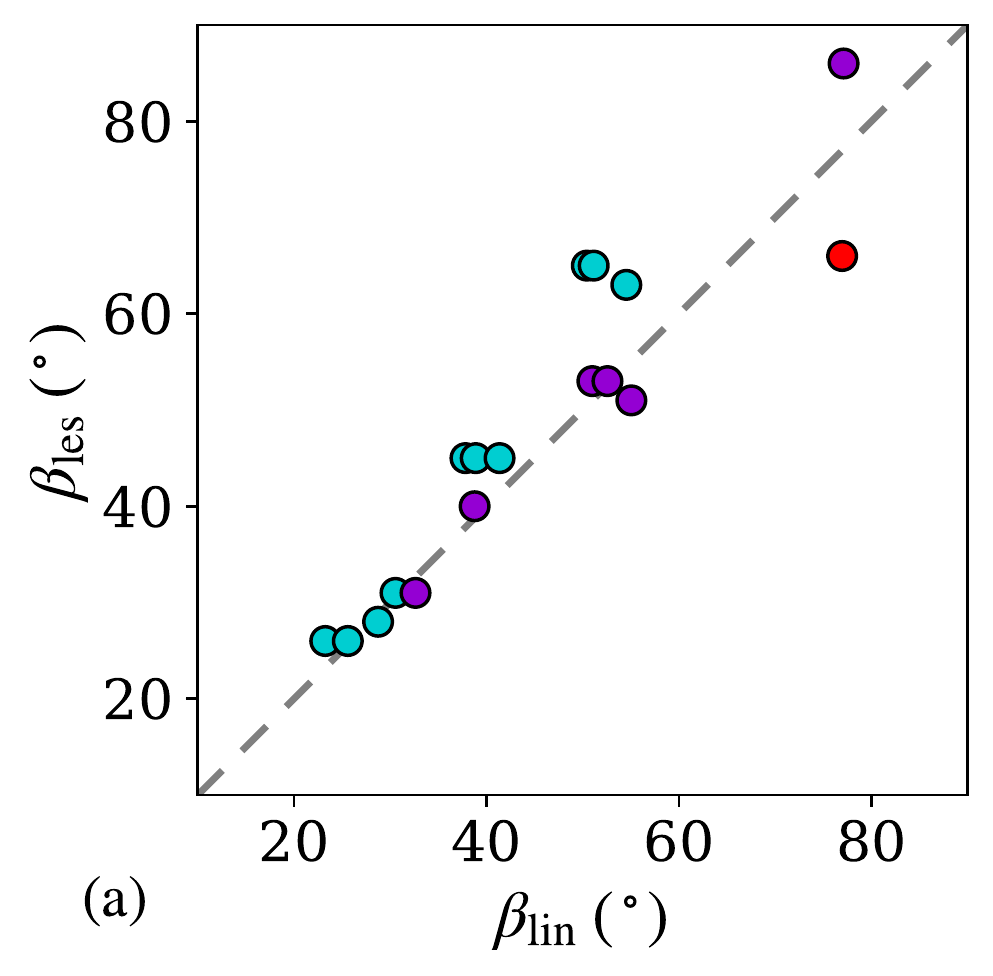}
	\hspace{0.012\textwidth}
	\includegraphics[width=0.33\textwidth]{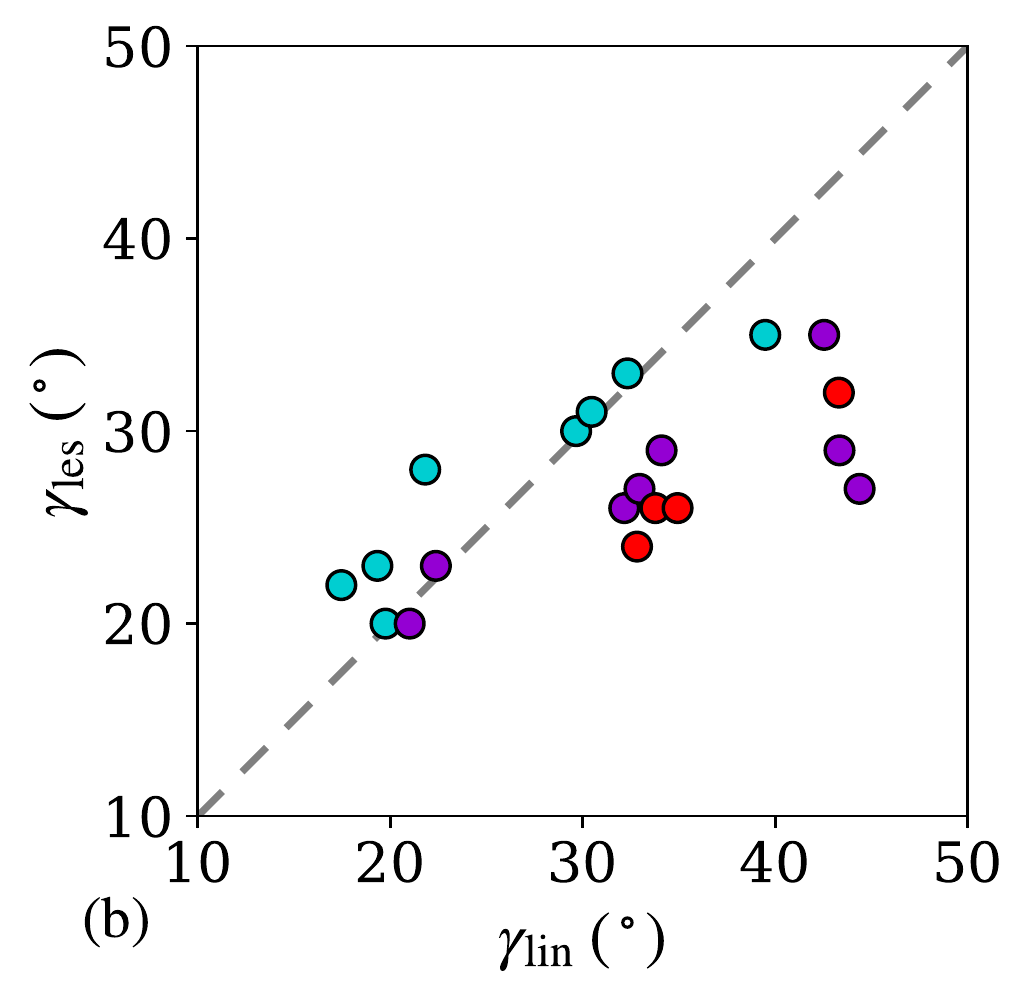}
	\qquad\qquad\qquad\qquad\qquad\qquad\qquad
	\vspace{0.008\textwidth}
	\includegraphics[width=0.319\textwidth]{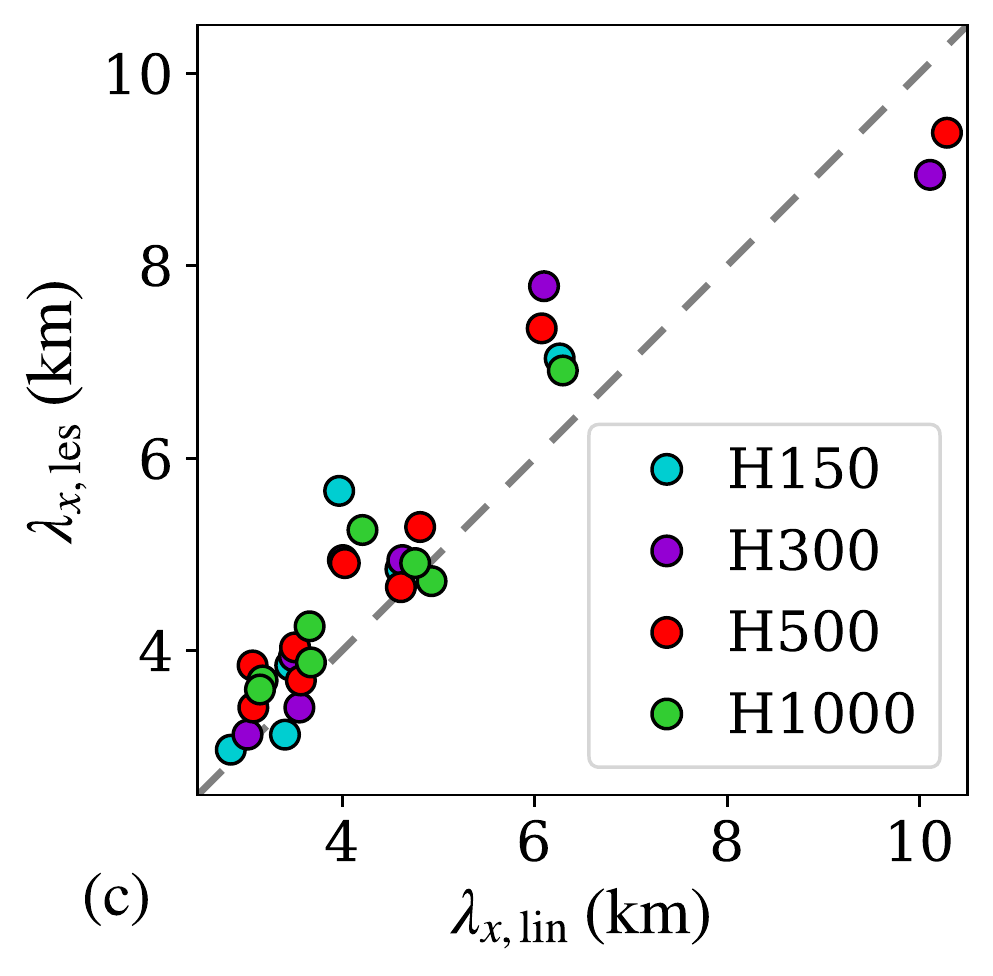}
	\hspace{0.012\textwidth}
	\includegraphics[width=0.345\textwidth]{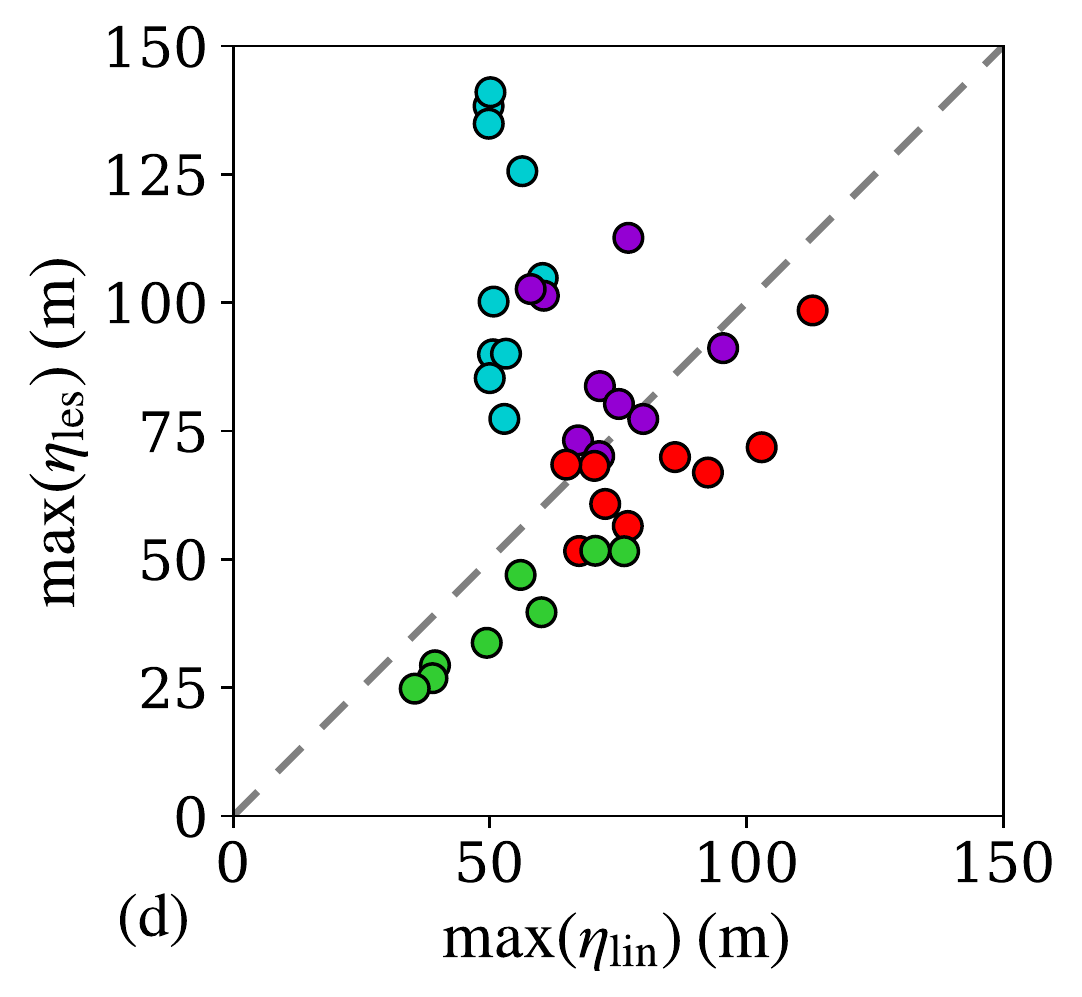}
	\caption{(a,b) Comparison of the $\beta$ and $\gamma$ angles measured in the LES results against the values predicted by linear theory \citep{Allaerts2019}. (c) Comparison of trapped-wave horizontal wavelength between LES results and the linear theory model of \cite{Vosper2004} and \cite{Sachsperger2015}. (d) Comparison of maximum inversion-layer vertical displacement between the LES results and the one-dimensional linear model developed by \cite{Allaerts2017b}.}
	\label{fig:les_vs_lt_axis}
\end{figure}

The angles $\beta_\mathrm{les}$ and $\gamma_\mathrm{les}$ are measured by carefully aligning a slanted line to the pattern observed in the vertical velocity field, as shown in Figure \ref{fig:les_vs_lt_angle}. A comparison between the $\beta_\mathrm{lin}$ and $\beta_\mathrm{les}$ angles for supercritical flows is shown in Figure \ref{fig:les_vs_lt_axis}(a). Overall, we observe an excellent agreement, with values collapsing along the diagonal. Next, Figure \ref{fig:les_vs_lt_axis}(b) compares the angle formed by the slanted lines triggered by the turbine spacing. We see again a good agreement, with $\gamma_\mathrm{les}$ slightly lower than $\gamma_\mathrm{lin}$ in deeper boundary layers. Finally, Figure \ref{fig:les_vs_lt_axis}(c) shows a comparison between the interfacial-wave horizontal wavelength obtained in LES against the one evaluated with linear theory, i.e. using Equation \ref{eq:lt_trapped_waves}. %Not only the values are in very good agreement, but also the trend (not shown).

\cite{Allaerts2017b} derived a linear one-dimensional model for the capping-inversion vertical displacement in response to a drag force imposed within the whole ABL. The governing equation reads as
\begin{equation}
	\biggl(1 - \frac{1}{\textit{Fr}^2} \biggl) \frac{\partial \eta}{\partial x} = c_t \mathrm{\Pi}(x) -2 \bigl( c_t \Pi(x) + c_d \bigl) \frac{\eta}{H} - \frac{1}{P_N} \mathcal{G}_N(\eta),
	\label{eq:dries_model}
\end{equation}
where $\Pi(x)$ denotes a step function with support going from the start to the end of the farm and $c_d=u_\star/U_B$ represents the friction with the ground. Moreover, the thrust force imposed on the flow scales linearly with $c_t$, which is defined as
\begin{equation}
	c_t=\frac{1}{2}\frac{\pi C_T}{4s_x s_y}\eta_w \gamma_v 
	\label{eq:dries_model_ct}
\end{equation}
with $\gamma_v$ a wind-profile shape factor determined using the precursor flow field and $\eta_w$ the wake efficiency. Finally, $\mathcal{G}_N(\eta)$ is a linear operator that accounts for internal-wave effects and is expressed as
\begin{equation}
	\mathcal{G}_N(\eta) = \frac{1}{2 \pi} \int |k| \biggl( \int \eta(x') e^{-ikx'} dx'\biggl) e^{ikx} dk.
\end{equation}
We refer to Appendix 2 in \cite{Allaerts2017b} for a full derivation of Equation \ref{eq:dries_model}. Given the atmospheric state, Equation \ref{eq:dries_model} predicts the capping-inversion displacement along the streamwise direction. Further, we name the maximum inversion-layer vertical displacement obtained with Equation \ref{eq:dries_model} and with the LES results as $\mathrm{max}(\eta_\mathrm{lin})$ and $\mathrm{max}(\eta_\mathrm{les})$, respectively. These two quantities are displayed in Figure \ref{fig:les_vs_lt_axis}(d), which shows a good match except for cases H150, for which $\mathrm{max}(\eta_\mathrm{lin}) < \mathrm{max}(\eta_\mathrm{les})$. We note that the values of $\eta_w$, $\beta_\mathrm{les}$, $\gamma_\mathrm{les}$, $\lambda_{x,\mathrm{les}}$ and $\mathrm{max}(\eta_\mathrm{les})$ for all cases are summarized in Table \ref{table:simulation_results}.

In this section, we have seen that our results match well when compared against various gravity-wave linear-theory models found in the literature. This further confirms that the LES results are not distorted by the domain boundaries and provide an interesting benchmark for the development, validation and calibration of existing and future low- and medium-fidelity wind-farm models.

\subsection{Comparison of flow profiles}\label{sec:bl_flow}
\begin{figure}
	\centering
	\includegraphics[width=1\textwidth]{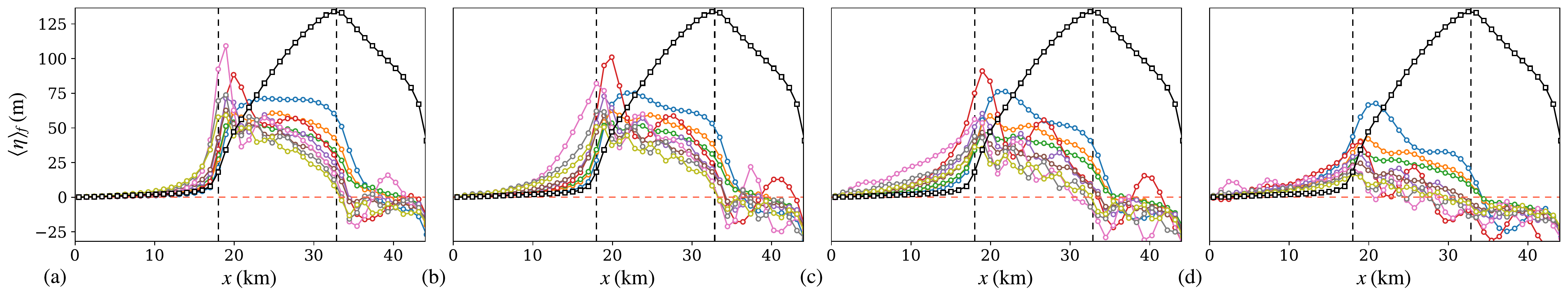} 
	\includegraphics[width=1\textwidth]{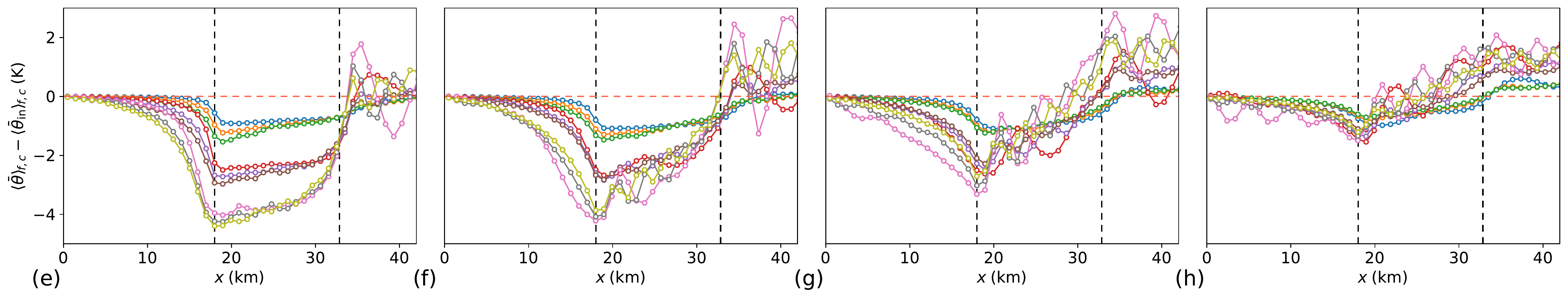} 
	\includegraphics[width=1\textwidth]{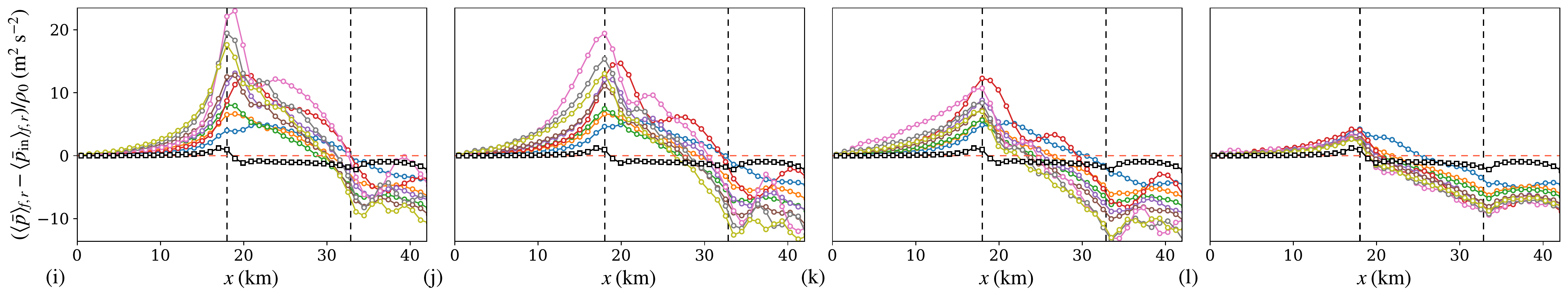} 
	\includegraphics[width=1\textwidth]{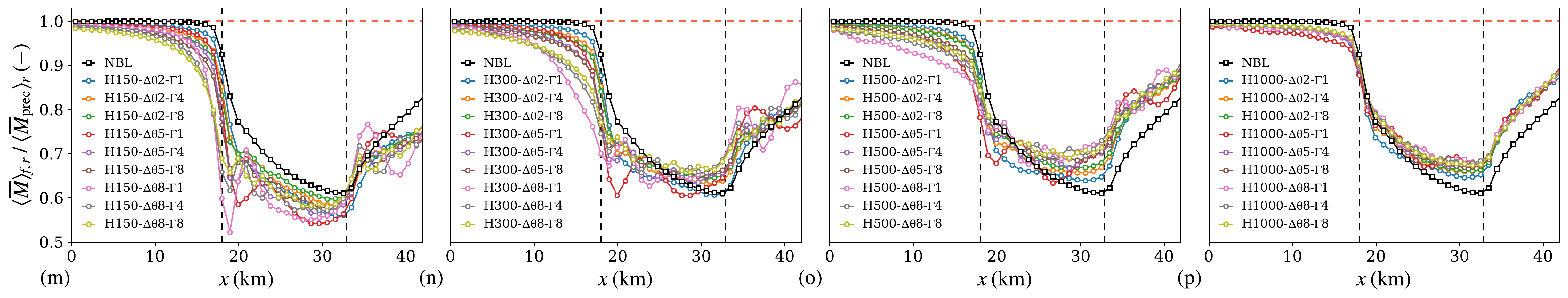} 
	\caption{Time-averaged (a-d) capping-inversion displacement, (e-h) potential-temperature perturbation, (i-l) pressure perturbation and (m-p) velocity magnitude averaged over the farm width as a function of the streamwise direction. The potential-temperature perturbation is further averaged over the capping-inversion thickness, while the other quantities are averaged over the turbine-rotor height. Moreover, the cases are organized per capping-inversion height. The vertical dashed black lines denote the location of the first- and last-row turbines.}
	\label{fig:blflow_all}
\end{figure}

We now investigate and compare flow profiles among all cases. Results are reported in Figure \ref{fig:blflow_all}, which displays time-averaged flow profiles as a function of the streamwise direction. We start the analysis with Figure \ref{fig:blflow_all}(a-d) which displays the streamwise variation of vertical inversion-layer displacement, here denoted with $\eta$, further averaged along the farm width. For the CNBL cases, $\eta$ is defined as the difference between the capping-inversion top evaluated in precursor and main domains using the \cite{Rampanelli2004} model. It is interesting to see that in the presence of free-atmosphere stratification, the flow reacts to the presence of the farm several kilometers upstream. Cases with a weak inversion layer show higher displacements, independently from the inversion-layer height. This is due to the weak resistance to vertical motion that air parcels have in such cases. As $\Delta \theta$ increases, interfacial waves with a lower wavelength get excited within the farm and in its wake. As noted previously, a stronger free atmosphere limits the displacement of the inversion layer, particularly within the first couple of turbine rows. Overall, shallow boundary layers tend to show higher inversion-layer vertical displacements than deeper ones. Moreover, $\eta$ sharply decreases downwind the last row of turbines in all cases, as a response to the flow acceleration in the wind-farm wake. In the NBL reference case, the absence of free-atmosphere stratification allows for higher growth of the ABL. In this case, $\eta$, which is computed as a streamline, shows a monotonic growth across the farm, with a maximum displacement obtained at the last-row turbine location. %We remark that $\eta$ has to be interpreted as an interfacial wave.

Perturbations in the potential-temperature field are shown in Figure \ref{fig:blflow_all}(e-f), averaged within the capping inversion height and along the farm width. As $H$ increases, the cold anomaly reduces in magnitude. For instance, the minimum negative temperature perturbation is $-1.4$ K for the H1000 cases, while it attains a value of $-4.6$ K in the H150 ones. Even when the inversion-layer displacement is limited by high $\Delta \theta$ values, the strong inversion that characterizes such cases generates higher temperature differences, which translate into stronger pressure perturbations. The latter are shown in Figure \ref{fig:blflow_all}(i-l), averaged along the farm width and within the turbine-rotor height. Here, we observe that the cases with strong unfavourable and favourable pressure gradients are the ones with a strong inversion layer. For instance, the counteracting pressure gradient in case H300-$\Updelta \theta$8-$\Upgamma$1 is 4.4 times the one obtained in case H300-$\Updelta \theta$2-$\Upgamma$1. Pressure feedbacks in the neutral reference case are negligible when compared to the ones obtained in the CNBL cases. For instance, case H300-$\Updelta \theta$5-$\Upgamma$4, which Figure \ref{fig:case_setup} defines as a highly probable atmospheric state over the North Sea, has a 14 times stronger unfavourable pressure gradient than the NBL reference case. However, we should note that all stratified cases are also characterized by a favourable pressure gradient through the farm, while the NBL reference case does not show this feature. 

The pressure-perturbation effects on the velocity magnitude are evident in Figure~\ref{fig:blflow_all}(m-p). The NBL reference case shows a reduction in wind speed only several rotor diameters upstream of the farm. Consequently, the velocity at the first-row turbine location is always higher than in the cases with thermal stratification above the ABL. Interestingly, the vertical motion generated by the interfacial waves in the CNBL cases has direct consequences in terms of velocity magnitude at the turbine-hub height. For instance, the velocity magnitude at the fourth-row turbine location is 5$\%$ higher than the one measured at the farm leading edge for case H300-$\Updelta \theta$8-$\Upgamma$1. As $H$ increases, the flow response becomes less sensitive to changes in $\Delta \theta$ and $\Gamma$. In fact, as the inversion-layer height approaches the equilibrium height of the TNBL, its effects on the farm–ABL interactions become smaller. Consequently, the velocity profiles and pressure perturbation of the CNBL cases get closer to the ones of the NBL reference case, as shown in Figure \ref{fig:blflow_all}(l,p). We remark that a stronger blockage effect and velocity deficits in shallow boundary layers have been observed also in SCADA data from the Nordsee Ost and Amrumbank West wind farms located in the German Bight area \cite{Canadillas2023} and in a previous study \citep{Allaerts2017}.
\begin{figure}
	\centerline{
		\includegraphics[width=1\textwidth]{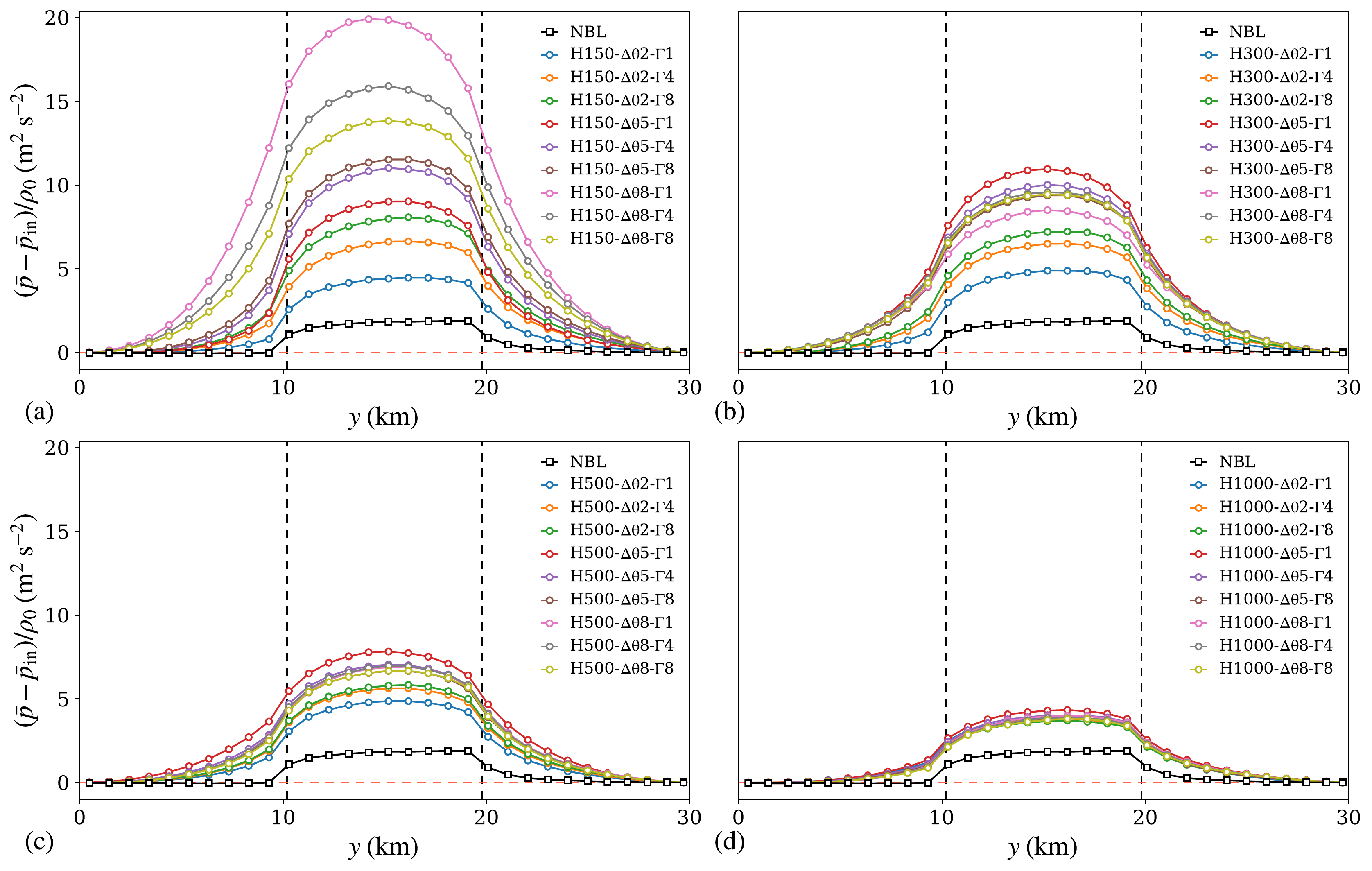}}% 
	\caption{Pressure perturbation as a function of the domain width taken at $x=18$ km (i.e. the location of the first-row turbines) and further averaged over the turbine-rotor height for cases (a) H150, (b) H300, (c) H500 and (d) H1000. The vertical dashed black lines denote the location of the first- and last-column turbines.}
	\label{fig:blflow_pressure}
\end{figure}

Cases H500-$\Updelta \theta$8-$\Upgamma$4, H1000-$\Updelta \theta$8-$\Upgamma$1 and H1000-$\Updelta \theta$8-$\Upgamma$4 in particular show oscillations in the capping-inversion displacement and temperature perturbation already in the farm induction region. Most likely, these are spurious effects introduced by the fringe region. In fact, \cite{Lanzilao2022b} have shown that their wave-free fringe-region technique can still introduce some perturbations in the domain of interest for a combination of low \textit{Fr} and $P_N$ numbers, which characterize these three cases -- see Table \ref{table:simulation_setup}.

Finally, Figure \ref{fig:blflow_pressure} focuses on the pressure build-up measured at the first-row turbine location along the spanwise direction. Here, the inverse relationship between capping-inversion height and counteracting pressure gradient is evident. Moreover, a deep boundary layer makes the simulation quasi-independent of changes in $\Delta \theta$ and $\Gamma$, as all profiles collapse in Figure \ref{fig:blflow_pressure}(d). However, the pressure build-up in all CNBL cases still remains higher than the one attained in the NBL reference case. Figure \ref{fig:blflow_pressure} also shows that turbines situated in the centre of the farm experience a higher counteracting pressure gradient than those located at the farm sides. The asymmetry of the profiles with respect to the domain centerline is caused by the presence of the Ekman layer. 

\begin{figure}
	\centering
	\includegraphics[width=1\textwidth]{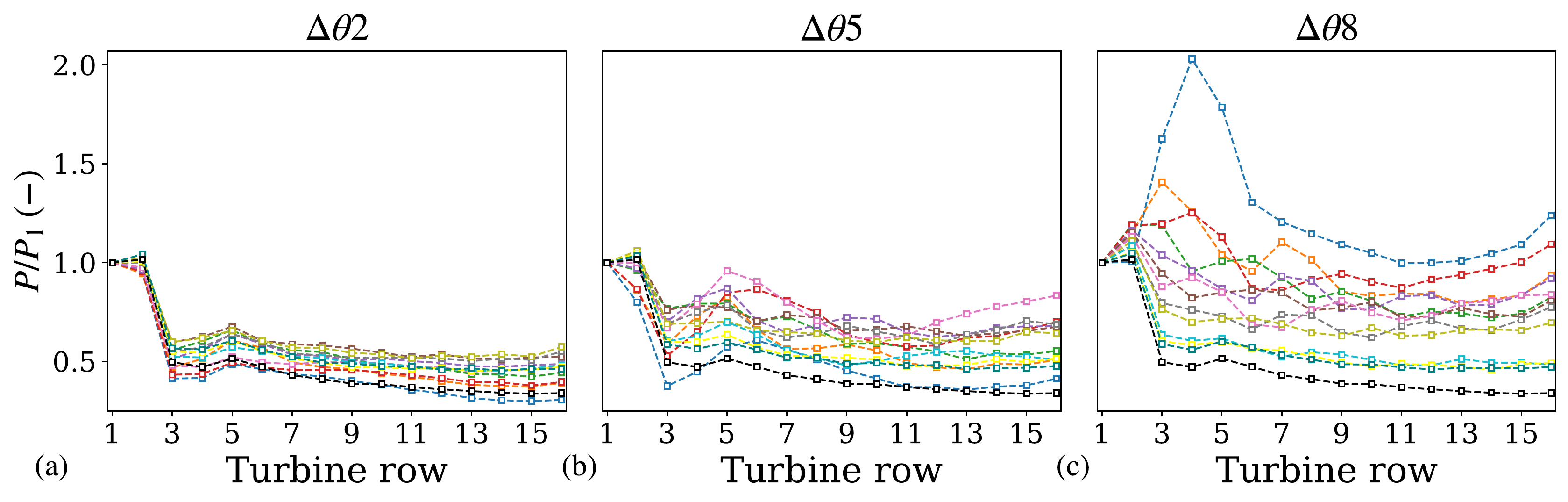} 
	\includegraphics[width=1\textwidth]{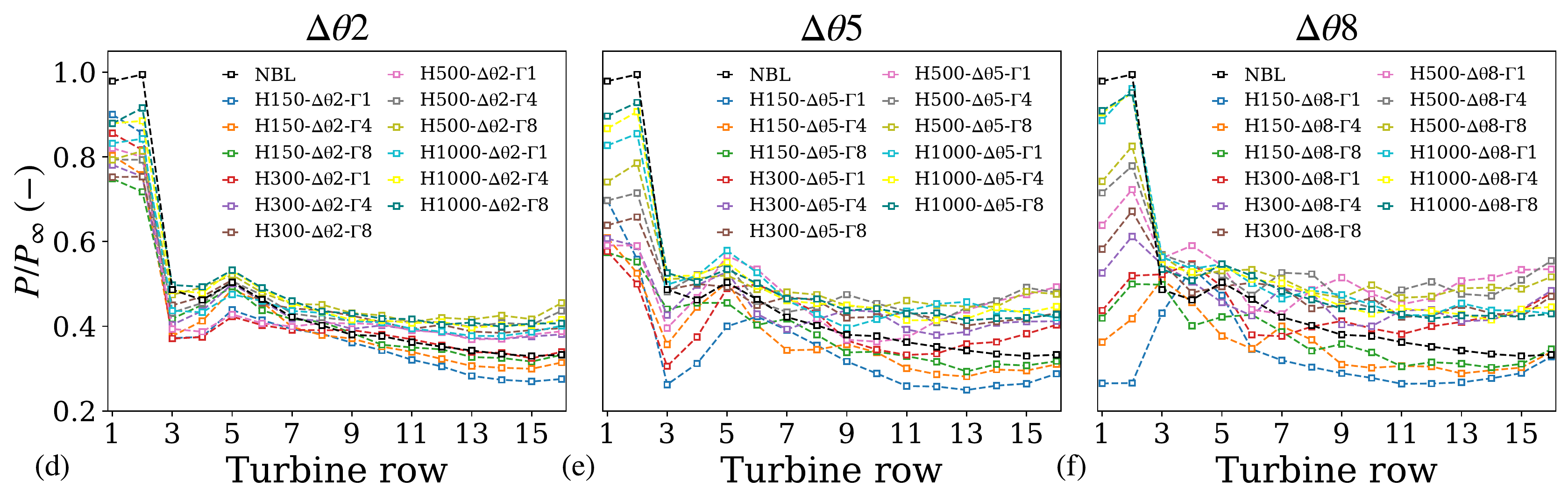} 
	\caption{Averaged power output per turbine row normalized by (a-c) $P_1$ of the respective simulation and (d-f) $P_\infty$ of the respective inversion-layer height obtained with a capping-inversion strength of (a,d) $\Delta \theta=2$ K, (b,e) $\Delta \theta=5$ K and (c,f) $\Delta \theta=8$ K under different inversion-layer heights and free-atmosphere lapse rates. Here, $P_1$ denotes the averaged power outputs of first-row turbines while $P_\infty$ represents the power output of a turbine operating in isolation obtained with the single-turbine simulations -- see Appendix \ref{app:st_sim}.}
	\label{fig:power_output}
\end{figure}

\subsection{Wind-farm power production}\label{sec:wf_power}
The time- and row-averaged turbine power output normalized with the first-row turbine power is displayed in Figure \ref{fig:power_output} organized per capping-inversion strength. A very similar trend in power output is observed for cases with weak inversion-layer strengths, as shown in Figure \ref{fig:power_output}(a). The farm has a staggered layout, therefore the high-speed channels that form between the turbines in the first row allow for a slightly higher power generation at the second row \citep{Mctavish2015,Meyerforsting2017}. After a drop of about 50$\%$ between the 2nd and 3rd row, the power remains approximately constant, with a minor increase towards the farm trailing edge for the cases with thermal stratification above the ABL. As the inversion-layer strength increases, power fluctuations are observed within the farm. These are mostly caused by the vertical motion generated by interfacial waves. Consequently, the power doesn't show a monotonic trend across the farm. Instead, the 5th row often extracts more power than the 3rd one when $\Delta \theta=5$~K, with differences up to 40$\%$. A more extreme behaviour is observed in Figure \ref{fig:power_output}(c), where in three cases a waked row extracts more power than the first row. Moreover, the strong favourable pressure gradient that develops across the farm in cases with a strong inversion layer also leads to an increase in power production towards the end of the farm. This phenomenon is accentuated in shallow boundary layers. For instance, the difference in power output between first- and last-row turbines is only about 8$\%$ in case H300-$\Updelta \theta$8-$\Upgamma$4, while the last-row turbines produce 23$\%$ more power than the first one in case H150-$\Updelta \theta$8-$\Upgamma$1. 

Figure \ref{fig:power_output}(d-e) shows the same results but normalized with $P_\infty$, that is the power output of a turbine operating in isolation. In the CNBL cases, the ratio $P_1/P_\infty$ (i.e. the non-local efficiency) is always much lower than 1, with a minimum of 0.26 attained in case H150-$\Updelta \theta$8-$\Upgamma$1. Moreover, the flow-blockage effect is much more sensitive to the inversion-layer height in the presence of strong capping inversion, as shown in Figure~\ref{fig:power_output}(f). In the neutral reference case, $P_1/P_\infty \approx 1$. Once more, this suggests that the flow deceleration in the farm induction region is mostly related to atmospheric gravity waves. Moreover, the power output keeps decreasing towards the farm trailing edge in the NBL reference case, while a power increase is observed for the majority of the CNBL cases. This is due to the favourable pressure gradient acting as an energy source across the farm.
\begin{figure}
	\centerline{
		\includegraphics[width=1\textwidth]{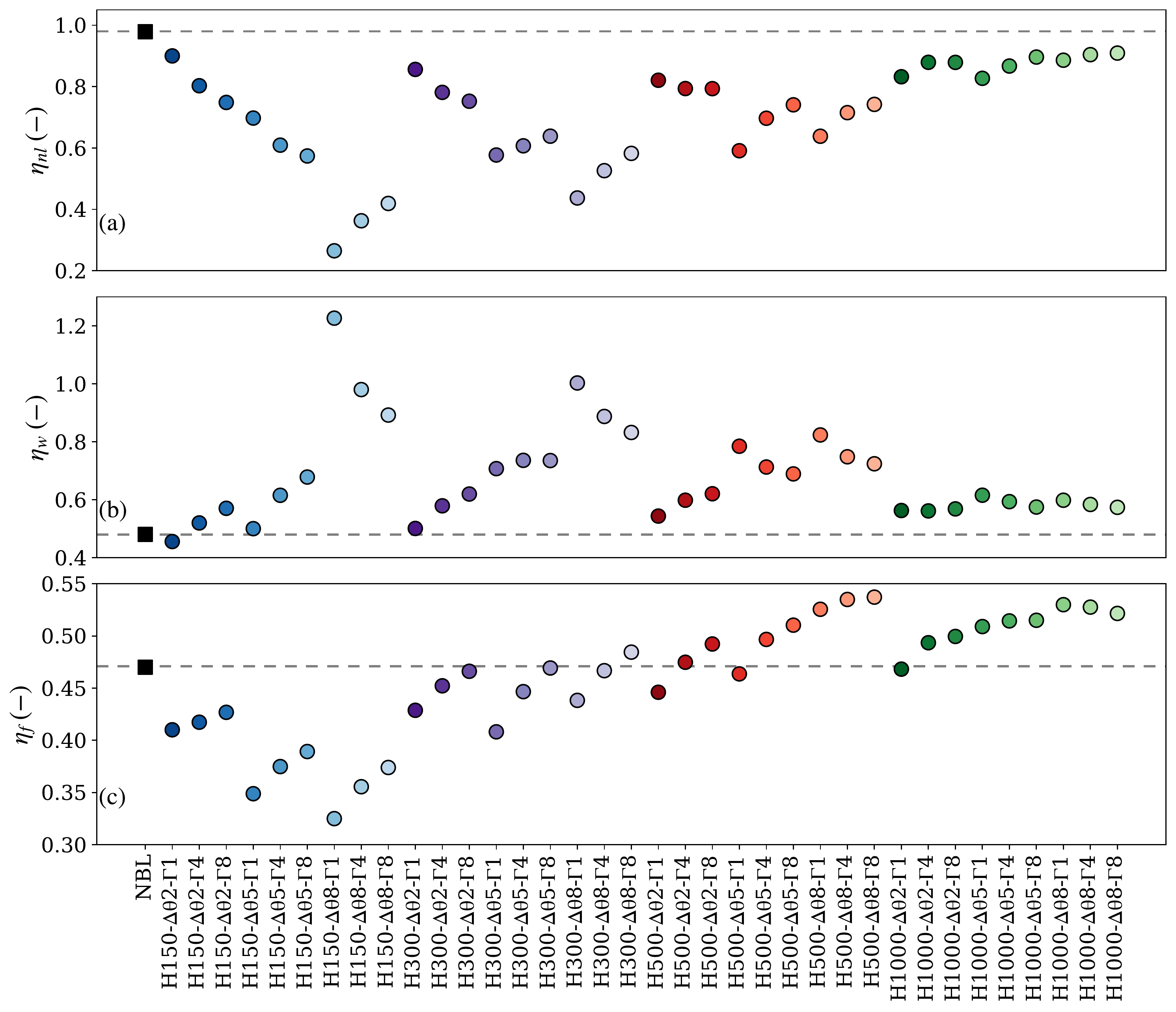}}% 
	\caption{(a) Non-local, (b) wake and (c) farm efficiency as a function of all simulation cases. The cases with thermal stratification above the ABL are represented by a circle while the NBL reference case is marked with a square. The horizontal grey dashed line indicates the relative efficiency value obtained in the NBL reference case.}
	\label{fig:farm_efficiency_dim}
\end{figure}

\subsection{Wind-farm efficiencies}\label{sec:wf_efficiencies}
We now turn our attention to the wind-farm efficiency, as well as the non-local and wake efficiencies, which are illustrated in Figure \ref{fig:farm_efficiency_dim} for all cases. We note that these efficiencies are defined in Equation \ref{eq:farm_efficiencies}. Figure \ref{fig:farm_efficiency_dim}(a) shows that the strong counteracting pressure gradient that characterizes shallow boundary layers causes $\eta_{nl}$ to be lower in the H150 cases than in the H1000 ones. The highest non-local efficiency is attained in the NBL reference case, showing that the cumulative turbine induction has a minor contribution to the flow-blockage effect for this case. The wake efficiency is shown in Figure \ref{fig:farm_efficiency_dim}(b). As hypothesized by \cite{Allaerts2017b}, a low non-local efficiency leads to a higher wake efficiency. In fact, the accumulation of potential energy caused by the flow slow-down in front of the farm is converted back into kinetic energy which accelerates the flow across the farm \citep{Allaerts2017b}. This negative correlation is clearly visible when comparing results in Figure \ref{fig:farm_efficiency_dim}(a,b). We note that case H150-$\Updelta \theta$8-$\Upgamma$1 has a wake efficiency greater than 1. For this atmospheric condition, the average power generated by the waked rows is higher than the one extracted by first-row turbines. The absence of favourable pressure gradients across the farm causes the wake efficiency of the NBL reference case to be among the lowest. Finally, Figure \ref{fig:farm_efficiency_dim}(c) displays the overall farm efficiency. The farm efficiency in the H150 and H300 cases is lower than the NBL reference case, while becomes higher for the H500 and H1000 cases, showing the strong influence that the capping-inversion height has on the wind-farm power output. Moreover, we observe that the farm efficiency is positively related with $\Gamma$. In fact, a free atmosphere with strong stratification leads to a higher wind-farm power output than a weakly stratified atmosphere, for a fixed $H$ and $\Delta \theta$ value. We note that the error bars representing the 95\% confidence interval computed with the moving block bootstrapping method for all efficiencies are in the order of $\pm 1\%$ and for this reason are not shown in Figure \ref{fig:farm_efficiency_dim}.

\begin{figure}
	\centering
	\includegraphics[width=1\textwidth]{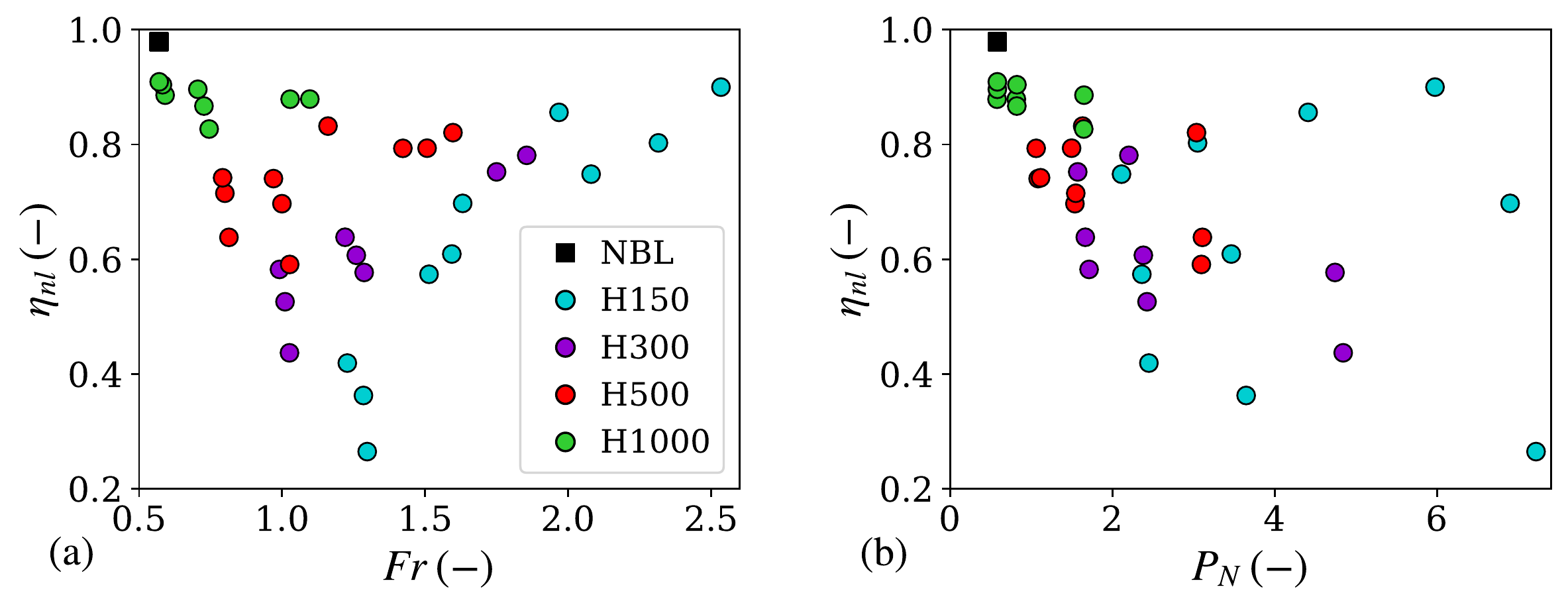}
	\includegraphics[width=1\textwidth]{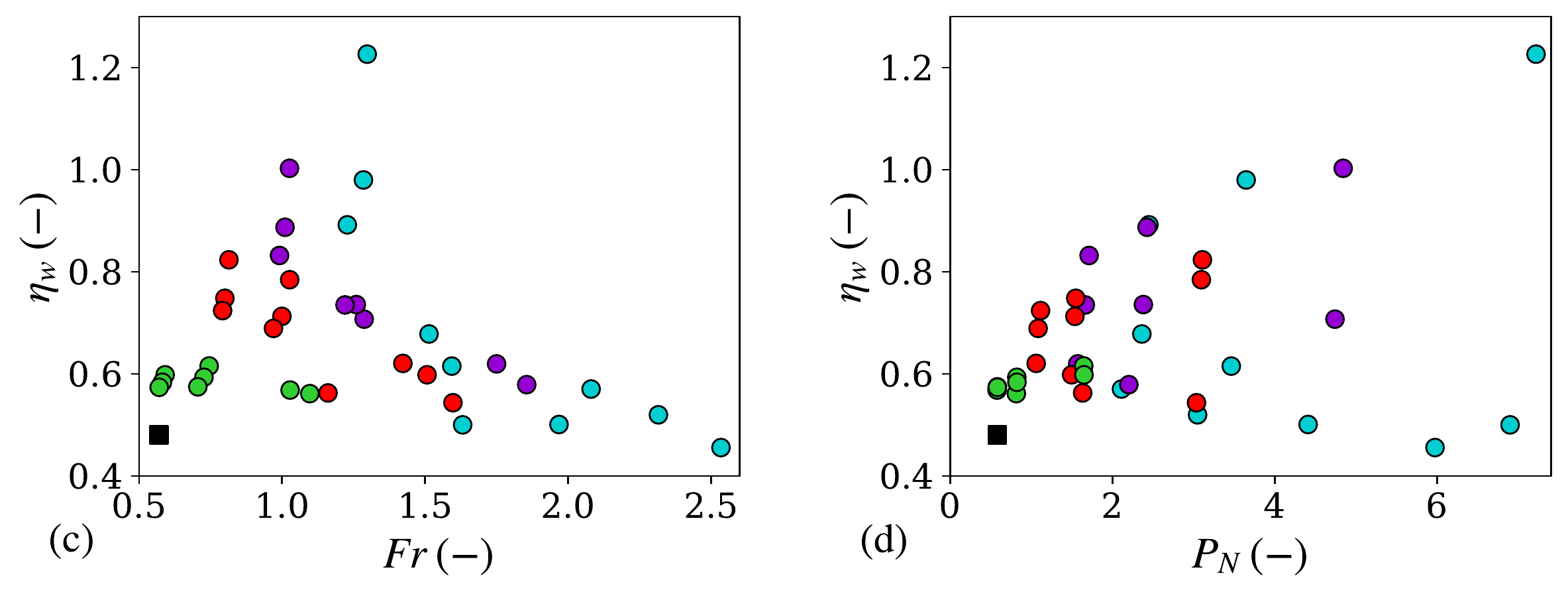}
	\caption{(a,b) Non-local and (c,d) wake efficiency as a function of the (a,c) \textit{Fr} and (b,d) $P_N$ numbers.}
	\label{fig:efficiency_vs_fr_pn}
\end{figure}

\begin{figure}
	\centering
	\includegraphics[width=1\textwidth]{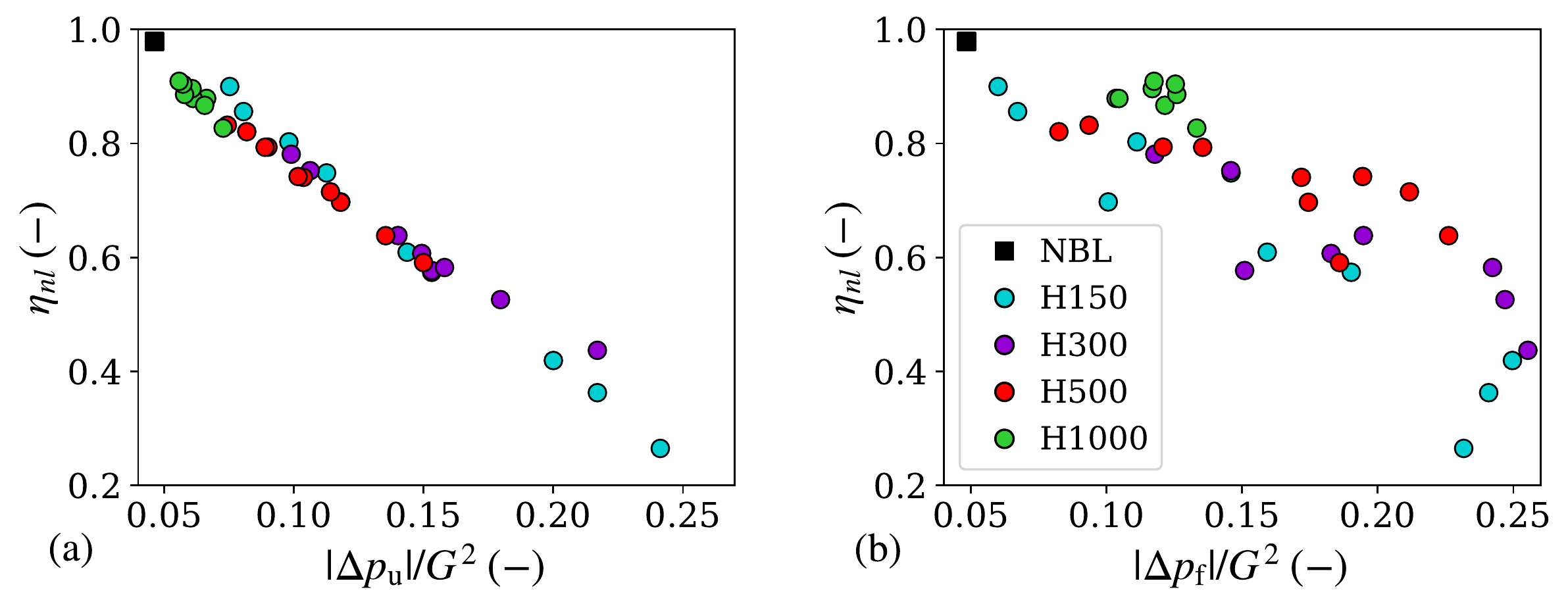}
	\includegraphics[width=1\textwidth]{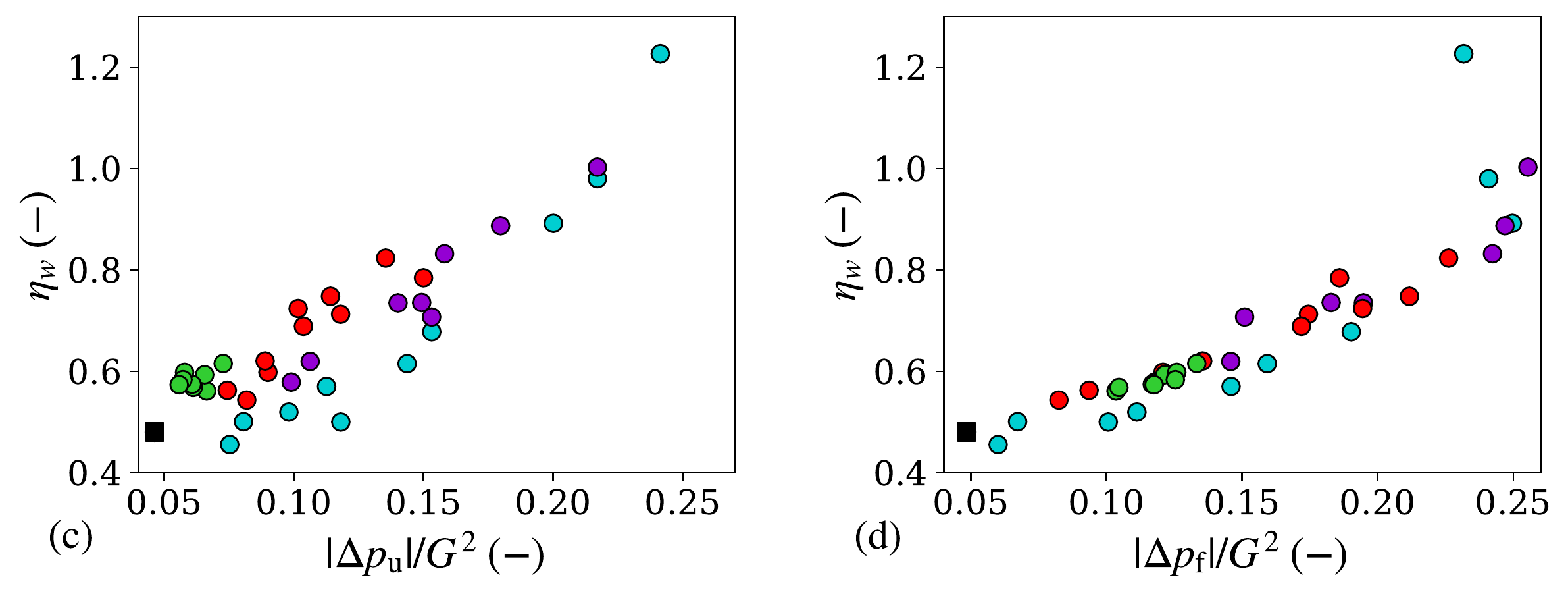}
	\caption{(a,b) Non-local and (c,d) wake efficiency as a function of the (a,c) unfavourable and (b,d) favourable pressure gradients magnitude normalized by the geostrophic wind.}
	\label{fig:efficiency_vs_pg}
\end{figure}

Next, we investigate how the non-local and wake efficiencies scale with the atmospheric state. In a first attempt, we plot the efficiencies against the $\textit{Fr}$ and $P_N$ numbers, which are the two non-dimensional groups that govern gravity-wave effects. Results are shown in Figure \ref{fig:efficiency_vs_fr_pn}. The lowest values of non-local efficiency are attained for $\textit{Fr} \approx 1.3$. Figure~\ref{fig:efficiency_vs_fr_pn}(b) shows that in general, the non-local efficiency decreases as $P_N$ increases. The wake efficiency shown in Figure \ref{fig:efficiency_vs_fr_pn}(c,d) shows an opposite trend, as expected. Overall, values remain scattered along the parameter space and strong trends are not observed. 

In the previous sections, we have seen that gravity-wave induced pressure gradients play a primary role in the flow dynamics. Therefore, in a second attempt, we try to scale the farm efficiencies against these pressure feedbacks. In particular, we define $\Delta p_u = \langle \bar{p} \rangle_{\!f,r}(x_\mathrm{ft}) - \langle \bar{p} \rangle_{\!f,r}(x_\mathrm{in})$ and $\Delta p_f = \langle \bar{p} \rangle_{\!f,r}(x_\mathrm{lt}) - \langle \bar{p} \rangle_{\!f,r}(x_\mathrm{ft})$ where $x_\mathrm{in}=0$ km represents the main domain inflow. Moreover, $x_\mathrm{ft}=18$~km and $x_\mathrm{lt}=32.85$~km denote the location of the first- and last-row turbines, respectively. Given this definition, $\Delta p_u$ and $\Delta p_f$ quantify the pressure feedback upstream and across the farm, respectively. Figure \ref{fig:efficiency_vs_pg}(a) shows the non-local efficiency as a function of $\Delta p_u$ normalized by the geostrophic wind. Here, we can observe a very strong negative correlation. This is expected since the non-local efficiency quantifies the flow slow down in front of the farm, which depends on the magnitude of the unfavourable pressure gradient. An opposite trend is visible in Figure \ref{fig:efficiency_vs_pg}(d), where the wake efficiency shows a strong positive correlation with $\Delta p_f$, confirming the role of the favourable pressure gradient in accelerating the wake recovery mechanism.

\begin{figure}
	\centerline{
		\includegraphics[width=0.525\textwidth]{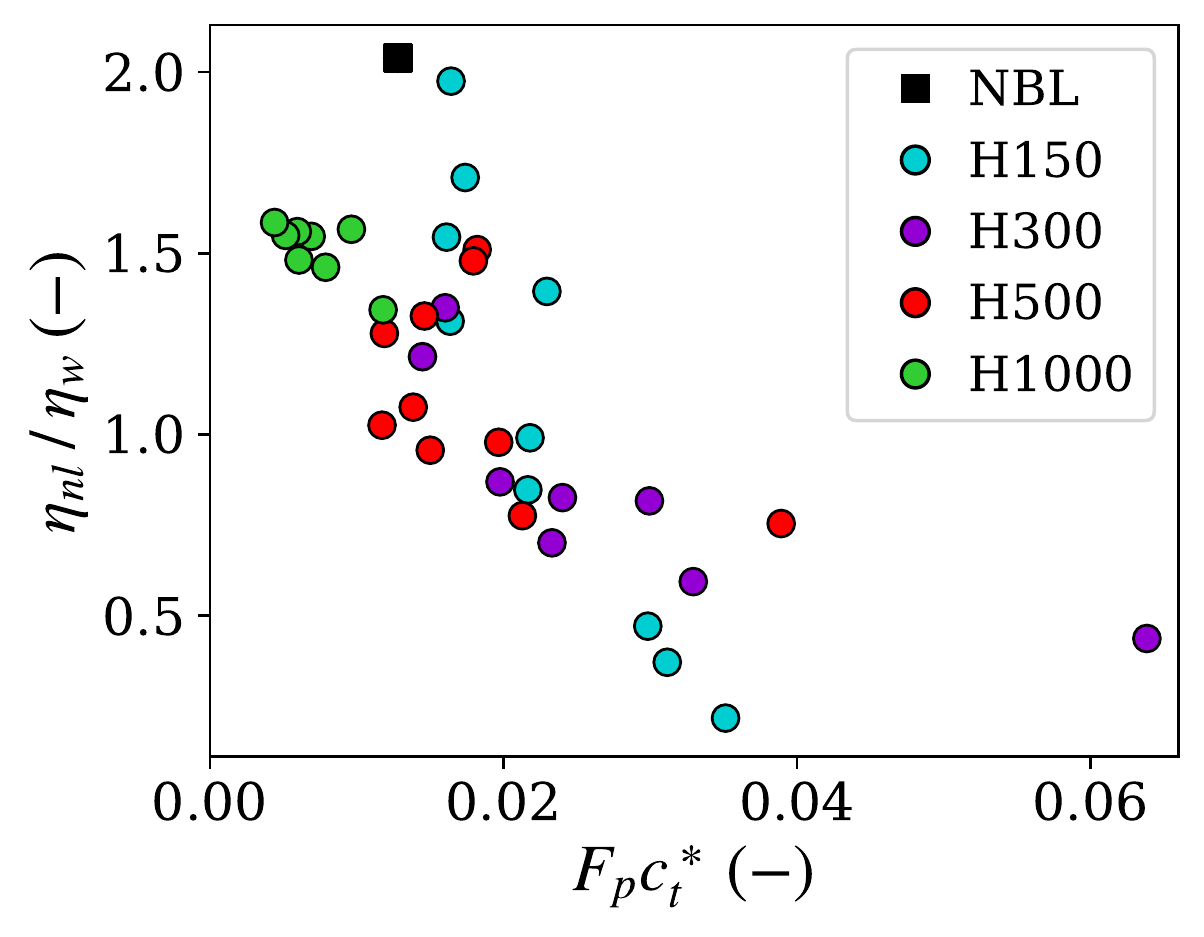}}%
	\caption{Ratio between the non-local and wake efficiency as a function of $F_p c_t^\ast$.}
	\label{fig:efficiency_vs_fp}
\end{figure}

Pressure perturbations induced by gravity waves scale with the capping-inversion vertical displacement $\eta$. In fact, \cite{Smith2010} has shown that in Fourier domain
\begin{equation*}
	\frac{\hat{p}}{\rho_0} = \biggl( g\frac{\Delta \theta}{\theta_0} + \frac{i \bigl(N^2 - \Omega^2 \bigl)}{m} \biggl) \hat{\eta}
\end{equation*}
with $m$ the vertical wave number and $\Omega$ the intrinsic wave frequency. Therefore, we believe that finding a good scaling parameter for $\eta$ would also result in a scaling parameter for the efficiencies. A simple model that provides an estimate for $\eta$ as a function of the atmospheric state has been proposed by \cite{Allaerts2017b} and is reported in Equation \ref{eq:dries_model}. In the Fourier domain, this model reads as
\begin{equation}
	\bigl(1 - \textit{Fr}^{-2} \bigl) i k \hat{\eta} = c_t \widehat{\mathrm{\Pi}} - \frac{2 c_t}{H} \widehat{\Pi} \ast \hat{\eta} -2 c_d \frac{\hat{\eta}}{H} - P_N^{-1} |k| \hat{\eta}
	\label{eq:dries_model_fourier}
\end{equation}
where the operator $\ast$ denotes a convolution. The second and third terms on the right-hand side of Equation \ref{eq:dries_model_fourier} represent a second-order term and the friction with the ground normalized with $H$, respectively, which are much smaller than the other terms. After neglecting these two terms, we can write
\begin{equation}
	\hat{\eta} = \frac{c_t}{k} \biggl(P_N^{-1} \frac{|k|}{k} + i \bigl(1-\textit{Fr}^{-2}\bigl)\biggl)^{-1} \widehat{\Pi}.
	\label{eq:eta_scaling}
\end{equation}
Using the dominant wavelength $k_d=2\pi/L_x^f$ in the streamwise direction, we define the following non-dimensional group
\begin{equation}
	F_p \triangleq \frac{k_d}{c_t \Vert \widehat{\Pi}(k_d) \Vert} \Vert \hat{\eta}(k_d) \Vert = \biggl(P_N^{-2} + \bigl(1-\textit{Fr}^{-2} \bigl)^2 \biggl)^{-0.5}
\end{equation}
where $\Vert \widehat{\Pi}(k_d) \Vert$ and $k_d$ are constant values since we do not vary the farm layout. By defining $c_t = c_t^\ast \eta_w$, we find that $\Vert \hat{\eta}(k_d) \Vert$ scales with $F_p c_t^\ast \eta_w$. We have previously shown that $\eta_{nl}$ is strongly related to $\Delta p_u$, which in turn depends upon the vertical displacement of the capping inversion. Hence
\begin{equation}
	\eta_{nl} \; \propto \; \Delta p_u \; \propto \; \Vert \hat{\eta}(k_d) \Vert \; \propto \; F_p c_t^\ast \eta_w
	\label{eq:eta_scaling_2}
\end{equation}
implying that $\eta_{nl}/\eta_w$ scales with $F_p c_t^\ast$. Figure \ref{fig:efficiency_vs_fp} confirms this result, showing a relatively strong negative correlation.  We note that the value of $\eta_{nl}$, $\eta_w$ and $\eta_f$ together with $P_1$, $P_\infty$, $\Delta p_u$ and $\Delta p_f$ for all cases are summarized in Table~\ref{table:simulation_results}.

\begin{table}
	\begin{center}
		\def~{\hphantom{0}}
		\begin{adjustbox}{max width=\textwidth}
			\begin{tabular}{ccccccccccccc}
				\textbf{Cases}  & $\boldsymbol{\beta_\mathrm{les} (^\circ)}$ & $\boldsymbol{\gamma_\mathrm{les} (^\circ)}$ & $\boldsymbol{\lambda_{x,\mathrm{les}}}$ \textbf{(km)} & $\boldsymbol{\mathrm{max}(\eta_\mathrm{les})}$ \textbf{(m)} & $\boldsymbol{r (\%)}$ & $\boldsymbol{\eta_{nl} (\%)}$ & $\boldsymbol{\eta_{w} (\%)}$ & $\boldsymbol{\eta_{f} (\%)}$ & $\boldsymbol{P_1}$ \textbf{(MW)} & $\boldsymbol{P_\infty}$ \textbf{(MW)} & $\boldsymbol{\Delta p_u}$ \textbf{(m\textsuperscript{2}s\textsuperscript{-2})} & $\boldsymbol{\Delta p_f}$ \textbf{(m\textsuperscript{2}s\textsuperscript{-2})} \\[7pt]		
				H150-$\Updelta \uptheta$2-$\Upgamma$1     & 26  & 22 & -    & 138.3  & 2.57 & 90 & 46 & 41 & 7.06 & 7.84 & 7.53  & 6.01 \\
				H150-$\Updelta \uptheta$2-$\Upgamma$4     & 26  & 23 & -    & 134.9  & 0.89 & 80 & 52 & 42 & 6.30 & 7.84 & 9.81  & 11.12\\
				H150-$\Updelta \uptheta$2-$\Upgamma$8     & 28  & 28 & -    & 141.0  & 0.58 & 75 & 57 & 43 & 5.87 & 7.84 & 11.26 & 14.59\\
				H150-$\Updelta \uptheta$5-$\Upgamma$1     & 45  & 30 & 7.03 & 100.2  & 4.61 & 70 & 50 & 35 & 5.47 & 7.84 & 11.81 & 10.07\\
				H150-$\Updelta \uptheta$5-$\Upgamma$4     & 45  & 31 & 4.85 & 90.0   & 1.78 & 61 & 62 & 37 & 4.78 & 7.84 & 14.37 & 15.93\\
				H150-$\Updelta \uptheta$5-$\Upgamma$8     & 45  & 33 & 3.13 & 85.4   & 1.12 & 57 & 68 & 39 & 4.50 & 7.84 & 15.32 & 19.02\\
				H150-$\Updelta \uptheta$8-$\Upgamma$1     & 65  & 35 & 5.66 & 125.6  & 8.06 & 26 & 123 & 32 & 2.08 & 7.84 & 24.13 & 23.17\\
				H150-$\Updelta \uptheta$8-$\Upgamma$4     & 65  & - & 3.85  & 90.1   & 2.59 & 36 & 98 & 36 & 2.85 & 7.84 & 21.70 & 24.09\\
				H150-$\Updelta \uptheta$8-$\Upgamma$8     & 63  & - & 2.97  & 77.4   & 1.58 & 42 & 89 & 37 & 3.29 & 7.84 & 20.01 & 24.97\\[7pt]
				
				H300-$\Updelta \uptheta$2-$\Upgamma$1     & 31  & 20 & -    & 104.8  & 3.33 & 86 & 50 & 43 & 6.85 & 8.00 & 8.06  & 6.72 \\
				H300-$\Updelta \uptheta$2-$\Upgamma$4     & 31  & 20 & -    & 101.4  & 0.82 & 78 & 58 & 45 & 6.25 & 8.00 & 9.90  & 11.78\\
				H300-$\Updelta \uptheta$2-$\Upgamma$8     & -   & 23 & -    & 102.7  & 0.49 & 75 & 62 & 47 & 6.02 & 8.00 & 10.63 & 14.58\\
				H300-$\Updelta \uptheta$5-$\Upgamma$1     & 53  & 26 & 7.79 & 112.6  & 7.56 & 58 & 71 & 41 & 4.62 & 8.00 & 15.31 & 15.09\\
				H300-$\Updelta \uptheta$5-$\Upgamma$4     & 53  & 27 & 4.94 & 83.8   & 1.92 & 61 & 74 & 45 & 4.86 & 8.00 & 14.93 & 18.28\\
				H300-$\Updelta \uptheta$5-$\Upgamma$8     & 51  & 29 & 3.41 & 73.2   & 1.13 & 64 & 74 & 47 & 5.11 & 8.00 & 14.02 & 19.47\\
				H300-$\Updelta \uptheta$8-$\Upgamma$1     & 86  & 35  & 4.94 & 91.2   & 7.17 & 44 & 100 & 44 & 3.50 & 8.00 & 21.70 & 25.54\\
				H300-$\Updelta \uptheta$8-$\Upgamma$4     & -   & 29 & 3.94 & 77.4   & 3.75 & 53 & 89 & 47 & 4.21 & 8.00 & 17.97 & 24.69\\
				H300-$\Updelta \uptheta$8-$\Upgamma$8     & -   & 27  & 3.13 & 70.1   & 2.84 & 58 & 83 & 48 & 4.66 & 8.00 & 15.81 & 24.23\\[7pt]
				
				H500-$\Updelta \uptheta$2-$\Upgamma$1     & 40  & -  & 8.94 & 80.3  & 3.39  & 82 & 54 & 45  & 6.49 & 7.90 & 8.19  & 8.25 \\ 
				H500-$\Updelta \uptheta$2-$\Upgamma$4     & -   & -  & 5.28 & 68.2  & 0.72  & 79 & 60 & 47  & 6.28 & 7.90 & 9.00  & 12.09\\
				H500-$\Updelta \uptheta$2-$\Upgamma$8     & -   & -  & 3.85 & 68.5  & 0.41  & 79 & 62 & 49  & 6.27 & 7.90 & 8.89  & 13.55\\
				H500-$\Updelta \uptheta$5-$\Upgamma$1     & 66  & 24 & 7.35 & 98.5  & 7.77  & 59 & 78 & 46  & 4.67 & 7.90 & 15.00 & 18.59\\
				H500-$\Updelta \uptheta$5-$\Upgamma$4     & 62  & 26 & 4.66 & 69.9  & 2.26  & 70 & 71 & 50  & 5.51 & 7.90 & 11.80 & 17.44\\
				H500-$\Updelta \uptheta$5-$\Upgamma$8     & -   & 26 & 3.69 & 60.8  & 0.97  & 74 & 69 & 51  & 5.86 & 7.90 & 10.37 & 17.19\\
				H500-$\Updelta \uptheta$8-$\Upgamma$1     & 70  & 32 & 4.91 & 66.9  & 8.75  & 64 & 82 & 53  & 5.05 & 7.90 & 13.54 & 22.62\\
				H500-$\Updelta \uptheta$8-$\Upgamma$4     & 70  & -  & 4.03 & 56.5  & 5.49  & 72 & 75 & 53  & 5.66 & 7.90 & 11.41 & 21.17\\ 
				H500-$\Updelta \uptheta$8-$\Upgamma$8     & -   & -  & 3.41 & 51.6  & 4.98  & 74 & 72 & 54  & 5.87 & 7.90 & 10.17 & 19.44\\[7pt]
				
				H1000-$\Updelta \uptheta$2-$\Upgamma$1    & -   & -  & 9.38 & 71.8  & 3.20  & 83 & 56 & 47 & 6.34 & 7.62 & 7.44  & 9.36 \\
				H1000-$\Updelta \uptheta$2-$\Upgamma$4    & -   & -  & 4.72 & 51.7  & 0.63  & 88 & 56 & 49 & 6.70 & 7.62 & 6.64  & 10.35\\
				H1000-$\Updelta \uptheta$2-$\Upgamma$8    & -   & -  & 3.69 & 47.0  & 0.55  & 88 & 57 & 50 & 6.70 & 7.62 & 6.12  & 10.46\\
				H1000-$\Updelta \uptheta$5-$\Upgamma$1    & -   & -  & 6.91 & 51.6  & 11.54 & 83 & 62 & 51 & 6.30 & 7.62 & 7.28  & 13.33\\
				H1000-$\Updelta \uptheta$5-$\Upgamma$4    & -   & -  & 4.91 & 39.7  & 3.53  & 87 & 59 & 51 & 6.61 & 7.62 & 6.57  & 12.15\\
				H1000-$\Updelta \uptheta$5-$\Upgamma$8    & -   & -  & 3.88 & 33.7  & 0.77  & 90 & 57 & 52 & 6.83 & 7.62 & 6.08  & 11.69\\
				H1000-$\Updelta \uptheta$8-$\Upgamma$1    & -   & -  & 5.25 & 29.4  & 15.82 & 89 & 60 & 53 & 6.75 & 7.62 & 5.79  & 12.59\\ 
				H1000-$\Updelta \uptheta$8-$\Upgamma$4    & -   & -  & 4.25 & 26.9  & 8.87  & 90 & 58 & 53 & 6.89 & 7.62 & 5.73  & 12.54\\
				H1000-$\Updelta \uptheta$8-$\Upgamma$8    & -   & -  & 3.60 & 24.8  & 5.96  & 91 & 57 & 52 & 6.93 & 7.62 & 5.58  & 11.76\\[7pt]
				NBL                                       & -   & -  & -    & -     & -     & 98 & 48 & 47 & 7.74 & 7.90 & 4.63  & 4.85\\[1pt]
			\end{tabular}
		\end{adjustbox}
		\label{table:simulation_results}
	\end{center}
	\caption[Overview of the wind-farm--LES results inherent to the sensitivity to the atmospheric state.]{Overview of the simulation results including the farm V-shape angle $\beta_\mathrm{les}$, the slanted lines angle $\gamma_\mathrm{les}$, the trapped-wave horizontal wavelength $\lambda_{x,\mathrm{les}}$, the maximum inversion-layer vertical displacement $\mathrm{max}(\eta_\mathrm{les})$, the reflectivity $r$, the non-local efficiency $\eta_{nl}$, the wake efficiency $\eta_w$, the farm efficiency $\eta_f$, the first-row turbine power output $P_1$, the power output of a turbine operating in isolation $P_\infty$ and the unfavourable and favourable pressure gradients magnitude $\Delta p_u$ and $\Delta p_f$, respectively. We remark that the NBL reference case refers to simulation H500-$\Updelta \theta$0-$\Upgamma$0.}
\end{table}

\section{Conclusions}\label{sec:conclusions}
This study set out to analyze the flow response to wind-farm forcing under atmospheres with different thermal stratification above the ABL. To this end, we performed 40 LESs of a $14.85 \times 9.4$~km$^2$ wind farm. Moreover, we run 21 additional LESs with varying domain sizes, with the aim of defining guidelines on how to select the domain length and width as a function of the inversion-layer height. The simulations were carried out with SP-Wind, an LES solver developed at KU Leuven. The main domain was driven by turbulent fully developed statistically steady flow fields obtained in precursor simulations. Moreover, the solver periodicity in the streamwise direction was broken using a wave-free fringe region technique while at the top of the domain an RDL was placed to damp gravity waves. Special effort in tuning the buffer regions was spent in order to minimize gravity-wave reflectivity and correctly impose the inflow conditions. 

The sensitivity study to the domain size has shown that the wind-farm performance is quasi-independent of the induction-region length when using the wave-free fringe region technique. In fact, by varying the fetch between the main domain inflow and the first-row turbine from 10 km to 50 km, the non-local and wake efficiency only varied by about 1 percentage point, independently from the capping-inversion height. However, we observed more pronounced differences in the flow behaviour as the domain width was increased. For instance, the non-local efficiency obtained in a domain with a ratio $L_y/L_y^f$ of 6.38 is 1.12 times the one measured in a domain with $L_y/L_y^f=2.13$ for a shallow boundary layer. Besides enhancing the flow-blockage effect, a domain with a low $L_y/L_y^f$ ratio also increases the channelling effect at the farm sides. Therefore, we concluded that it is necessary to have a domain with a high $L_y/L_y^f$ ratio when simulating wind-farm operations in shallow boundary-layer flows. Deeper boundary-layer flows are less sensitive to the domain width, but still require a wider space at the farm sides than simulations with a neutral atmosphere.

Next, we investigated the effect of changes in the capping-inversion height and strength, as well as free-atmosphere lapse rate, on the flow behaviour in and around the farm. In all cases, the flow convergence caused by the farm drag force displaces the capping inversion upward, creating a cold anomaly that induces pressure perturbations. We found very strong velocity deficits in shallow boundary layers, caused by the limited flow mixing and energy entrainment due to the close proximity of the inversion layer to the turbine-tip height. In turn, this resulted in higher inversion-layer vertical displacement and stronger cold anomalies. Therefore, shallow boundary-layer flows are characterized by a very strong counteracting pressure gradient in the farm induction region, further amplified by the presence of a strong inversion layer. Deeper boundary layers can accommodate the IBL growth, limiting flow redirection at the farm sides together with the vertical displacement of the capping inversion. Therefore, the counteracting pressure feedback has a lower magnitude in such cases. The pressure build-up in the farm entrance region is then released as a form of kinetic energy throughout the farm, which accelerates the wake recovery mechanism. This effect is more pronounced in shallow boundary layers. In regard to changes in $\Delta \theta$, we found that a stronger capping inversion reduces its vertical displacement, therefore increasing the flow rate at the farm sides. Moreover, inversion layers with higher strengths support interfacial waves with lower horizontal wavelengths. Finally, we noticed that a stronger free-atmosphere lapse rate limits air-parcel vertical motion, therefore reducing the magnitude of both the counteracting and favourable pressure gradients. We remark that the pressure perturbations in the NBL reference case were at least one order of magnitude lower than the ones in the CNBL cases.

The turbine power generation per row was also investigated. We observed that the flow divergence-convergence generated by interfacial waves causes considerable power fluctuations within the farm. Moreover, the turbine power tends to increase towards the farm trailing edge, particularly in cases with a shallow boundary layer and a high capping-inversion strength.

A non-local efficiency of 98$\%$ was observed for the NBL reference case. In contrast, the non-local efficiency attained values as low as 26$\%$ in the presence of thermal stratification above the ABL. This suggests that flow blockage is primarily related to atmospheric gravity waves. Moreover, we noticed that the non-local and wake efficiencies are inversely related. The farm efficiency in the H150 and H300 cases was lower than the NBL reference case, while it became higher for the H500 and H1000 cases. We also observed that the farm efficiency is positively related with $\Gamma$. In fact, a free atmosphere with strong stratification leads to a higher wind-farm power output than a weakly stratified atmosphere, for a fixed $H$ and $\Delta \theta$ value. These observations suggested that the shape of the vertical potential-temperature profile has a significant impact on the wind-farm performance. Finally, we found no simple scaling between the efficiencies and the $\textit{Fr}$ and $P_N$ numbers. Instead, a very good scaling was found with the pressure gradients. This observation allowed us to derive a new scaling parameter $F_p$ for the ratio of non-local to wake efficiencies.

Finally, we compared our results against various one- and two-dimensional gravity-wave linear-theory models. Here, we noticed an excellent agreement between the internal gravity-wave pattern and magnitude, underlining the importance of properly calibrating the fringe region and RDL. Moreover, the LES results were in line with linear theory also in terms of interfacial-wave horizontal wavelength and maximum inversion-layer vertical displacement.

The good agreement between gravity-wave linear-theory models and LES results and their consistency with previous findings entails that the numerical results are not distorted by the domain boundaries, and result in a valid benchmark for the development, validation and calibration of existing and future low- and medium-fidelity wind-farm models. In the future, we plan to extend the database to baroclinic atmospheres, where the geostrophic wind varies with height. Moreover, the presence of multiple capping inversions is also of interest. The farm geometry was kept constant in the current study, but we are aware that it can have significant effects on the flow behaviour. Therefore, pressure feedbacks under various farm layouts and densities will be investigated. Furthermore, we plan to study wind-farm operation in weak and strong SBLs, which are known to enhance the flow-blockage effect. 

\section*{Acknowledgements}
The authors acknowledge support from the Research Foundation Flanders (FWO, Grant No. G0B1518N), from the project FREEWIND, funded by the Energy Transition Fund of the Belgian Federal Public Service for Economy, SMEs, and Energy (FOD Economie, K.M.O., Middenstand en Energie) and from the European Union Horizon Europe Framework programme (HORIZON-CL5-2021-D3-03-04) under grant agreement no. 101084205. The computational resources and services in this work were provided by the VSC (Flemish Supercomputer Center), funded by the Research Foundation Flanders (FWO) and the Flemish Government department EWI.

\section*{Declaration of interests}
The authors report no conflict of interest.

\appendix
\section{Large-eddy simulation of a turbine operating in isolation}\label{app:st_sim}
Single-turbine simulations are necessary to evaluate the power output of a turbine operating in isolation, which we define as $P_\infty$. In the farm, first-row turbines are affected by flow blockage, so that their power output is naturally lower than $P_\infty$. Moreover, Figure \ref{fig:precursor_results} illustrates that the turbulent statistically steady velocity profiles obtained with the precursor simulations are independent of the capping-inversion strength and free-atmosphere lapse rate. They only differ as $H$ varies. Therefore, we perform only four single-turbine simulations, one per each value of $H$.

The presence of a single turbine allows us to use a much smaller domain, with an equal length and width of 10 km. This size corresponds to the one of the precursor domain. In the vertical direction, we keep $L_z=25$ km as in the wind-farm simulations. The horizontal and vertical grid resolution also correspond to the ones adopted in the wind-farm simulations, meaning that we have 320, 460 and 490 points in the $x$, $y$ and $z$-direction, respectively. The RDL and fringe region adopt the same setup described in Section \ref{sec:numerical_setup}. However, to account for the smaller domain, also the fringe region length has been reduced to 2.5 km compared to the 5.5 km adopted for the wind-farm simulations. Finally, the time horizon for the wind-farm start-up phase has been set to 1 h. Afterwards, statistics are collected over a simulation time of 1.5 h.

\begin{figure}
	\centerline{
		\includegraphics[width=1\textwidth]{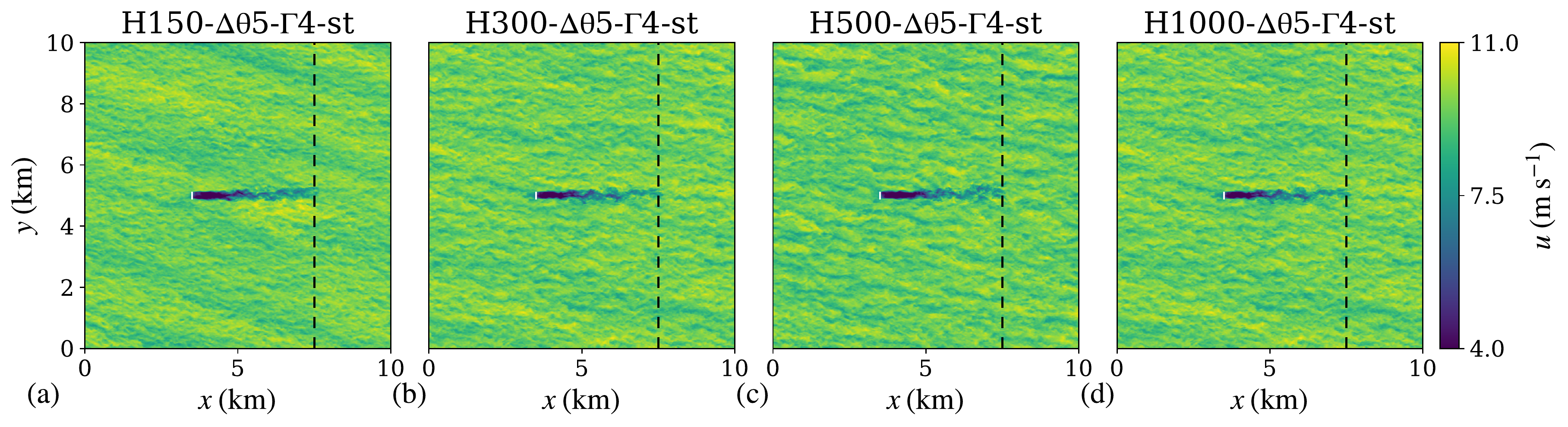}}% 
	\caption{Contours of the instantaneous streamwise velocity field in an $x$-$y$ plane taken at turbine-hub height for cases (a) H150-$\Updelta \theta$5-$\Upgamma$4-st, (b) H300-$\Updelta \theta$5-$\Upgamma$4-st, (c) H500-$\Updelta \theta$5-$\Upgamma$4-st and (d) H1000-$\Updelta \theta$5-$\Upgamma$4-st. The vertical black dashed line represents the start of the fringe region while the vertical white line denotes the turbine-rotor disk location.}
	\label{fig:single_turbine}
\end{figure}

Figure \ref{fig:single_turbine} illustrates instantaneous streamwise velocity fields in an $x$-$y$ plane taken at turbine-hub height for the four single-turbine simulations. In all cases, we can easily distinguish the turbine wake, which recovers downwind before being fully replenished by the fringe-region body force, which restores the inflow conditions. Overall, we observe that the perturbations that the turbine induces on the flow are too small to excite any noticeable gravity-wave effects, such as flow blockage upstream and flow speed-up at its side. Consequently, we believe that this set of simulations is suitable for providing an estimate for $P_\infty$. The free-stream power values are reported in Table \ref{table:simulation_results}.

\section{A simple two-dimensional gravity-wave linear-theory model}\label{app:lt_model}
We consider the Euler equations for irrotational, frictionless and adiabatic flow. Moreover, we neglect the Coriolis force and we assume that there is no variation in the $y$-direction, so that the problem can be considered to be two-dimensional. After manipulating the governing equations, the following solution in Fourier space for the streamwise and vertical velocity, pressure and potential temperature can be derived
\begin{align}
	&\hat{u}(k,z) = - i G m \hat{h}_o e^{i m z} \label{eq:single_turbine_1}\\ 
	&\hat{w}(k,z) = i G k\hat{h}_o e^{i m z} \\
	&\hat{p}(k,z) = i G^2 \rho_0 m \hat{h}_o e^{i m z} \\
	&\hat{\theta}(k,z) = -\frac{ N^2 \theta_0}{ g} \hat{h}_o e^{i m z}, \label{eq:single_turbine_2}
\end{align}
where the hat denotes a variable in the Fourier domain, $k$ represents the streamwise wavenumbers while $i$ is the imaginary unit. Moreover, the vertical wavenumber $m$ depends upon $k$ and the atmospheric state, being defined as
\begin{equation*}
	m = 
	\begin{cases}
		\sqrt{N^2/G^2-k^2} \qquad &N^2/G^2>k^2 \\[5pt]
		i\sqrt{k^2-N^2/G^2} \qquad &N^2/G^2<k^2
	\end{cases}
\end{equation*}
where $N$ is the Brunt--V\"{a}is\"{a}l\"{a} frequency and $G$ the geostrophic wind. We refer to \cite{Nappo2002}, \cite{Lin2007}, \cite{Sutherland2010} and \cite{Cushman2011} for more details on the derivation of Equations (\ref{eq:single_turbine_1}-\ref{eq:single_turbine_2}). The only input to the model consists of the obstacle shape in Fourier space, here denoted with $\hat{h}_o$. The latter is assumed impermeable. Moreover, it is assumed that the streamline slope equals the terrain slope locally. Finally, an inverse Fourier transform is necessary to express Equation (\ref{eq:single_turbine_1}-\ref{eq:single_turbine_2}) in real space.

\begin{figure}
	\centerline{
		\includegraphics[width=1\textwidth]{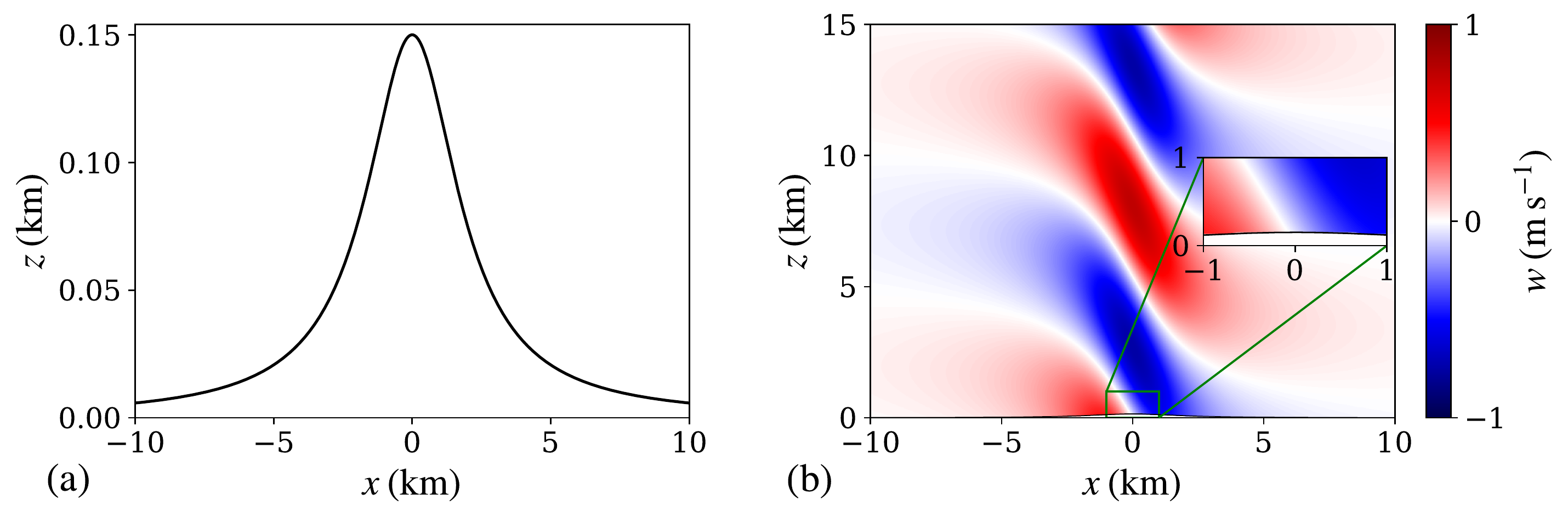}}% 
	\caption{(a) Bell-shaped mountain known in the literature as the \textit{Witch of Agnesi}. (b) Vertical velocity field predicted by the two-dimensional linear-theory model as a response to the obstacle shown in panel (a). In the center right of panel (b), the solution in the region close to the obstacle is magnified.}
	\label{fig:linear_theory_model}
\end{figure}

To conclude, we briefly show an example case. We consider a bell-shaped mountain, also known as the \textit{Witch of Agnesi}. The latter is defined as
\begin{equation*}
	h_o(x) = \frac{h_m a^2}{x^2+a^2}, \qquad\;\;  \hat{h}_o(k) = \frac{h_m a}{2} e^{-ka} \quad \text{for} \quad k>0,
\end{equation*}
where $h_m=0.15$ km is the mountain height while $a=2$ km denotes the mountain half-width. Moreover, we fix the geostrophic wind to 10 m s$^{-1}$ and the free-atmosphere lapse rate to 1 K km$^{-1}$. Figure \ref{fig:linear_theory_model}(a) shows the obstacle shape while the vertical velocity field is illustrated in Figure \ref{fig:linear_theory_model}(a), where internal waves triggered by the obstacle are clearly visible.

\section{Internal gravity-wave reflectivity}\label{app:gw_reflectivity}
In this appendix, we investigate the performance of the RDL by analyzing the internal gravity-wave reflectivity $r=E_\downarrow/E_\uparrow$, where $E_\downarrow$ and $E_\uparrow$ denote the vertical kinetic energy per unit mass (i.e. $0.5w^2$) associated with upward and downward internal gravity waves. To compute $r$, we use the algorithm proposed by \cite{Taylor2007} which makes use of the fact that the sign of the vertical phase velocity is opposite to the sign of the vertical group velocity for internal gravity waves. Moreover, similarly to \cite{Allaerts2017,Allaerts2017b} and \cite{Lanzilao2022b}, the upward and downward vertical kinetic energy are computed over a $x$-$z$ plane including the free atmosphere only (i.e., starting at a height of 1.5 km above the capping inversion) without the buffer regions. 

\begin{figure}
	\centerline{
		\includegraphics[width=0.55\textwidth]{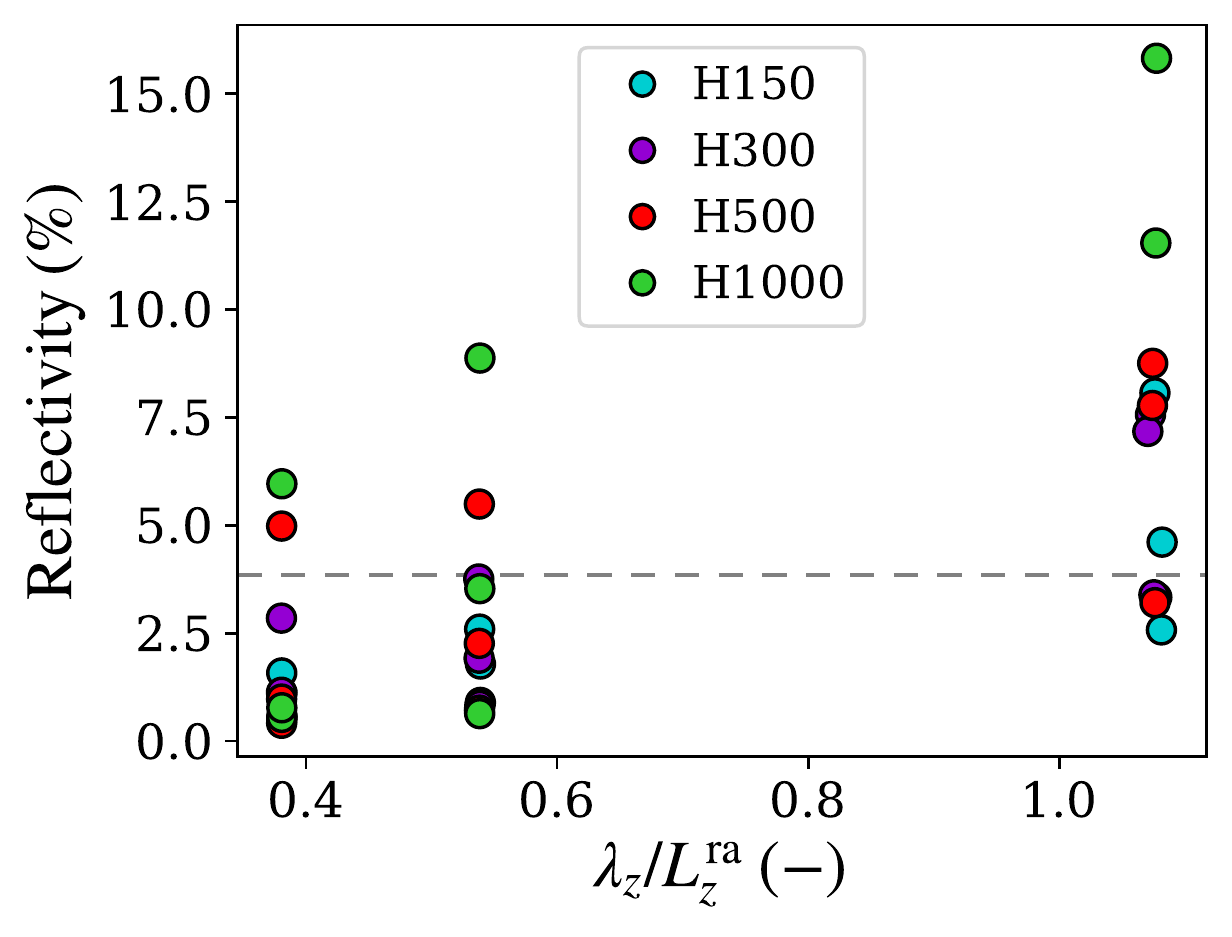}}% 
	\caption{Reflectivity as a function of the ratio between the gravity-wave vertical wavelength and the RDL vertical extension.}
	\label{fig:reflectivity_1}
\end{figure}

Results are shown in Figure \ref{fig:reflectivity_1}, which illustrates the reflectivity in percentage points as a function of the ratio $\lambda_z/L_z^\mathrm{ra}$. The internal gravity-wave vertical wavelength only depends upon the Brunt--V\"{a}is\"{a} frequency since the geostrophic wind is constant in our study. Therefore, the results are naturally divided into three groups defined by a free-atmosphere lapse rate of 1, 4 and 8 K/km. For the RDL to be effective, \cite{Klemp1977} mentioned that $L_z^\mathrm{ra}$ should be at least one time the internal gravity-wave vertical wavelength. Therefore, it is no surprise to see on average higher values of reflectivity when $\Gamma=1$ K/km, which corresponds to a ratio $\lambda_z/L_z^\mathrm{ra}$ of 1.08. As the stability of the free atmosphere increases, the ratio $\lambda_z/L_z^\mathrm{ra}$ reduces to 0.38, and so does the reflectivity. Overall, $r$ spans from a minimum of 0.41$\%$ to a maximum of 15.82$\%$, with an average value over all cases of 3.85$\%$. The NBL reference case does not support gravity waves, therefore it is not included in the results. We note that the reflectivity values for all cases are summarized in Table \ref{table:simulation_results}.

\begin{figure}
	\centerline{
		\includegraphics[width=1.\textwidth]{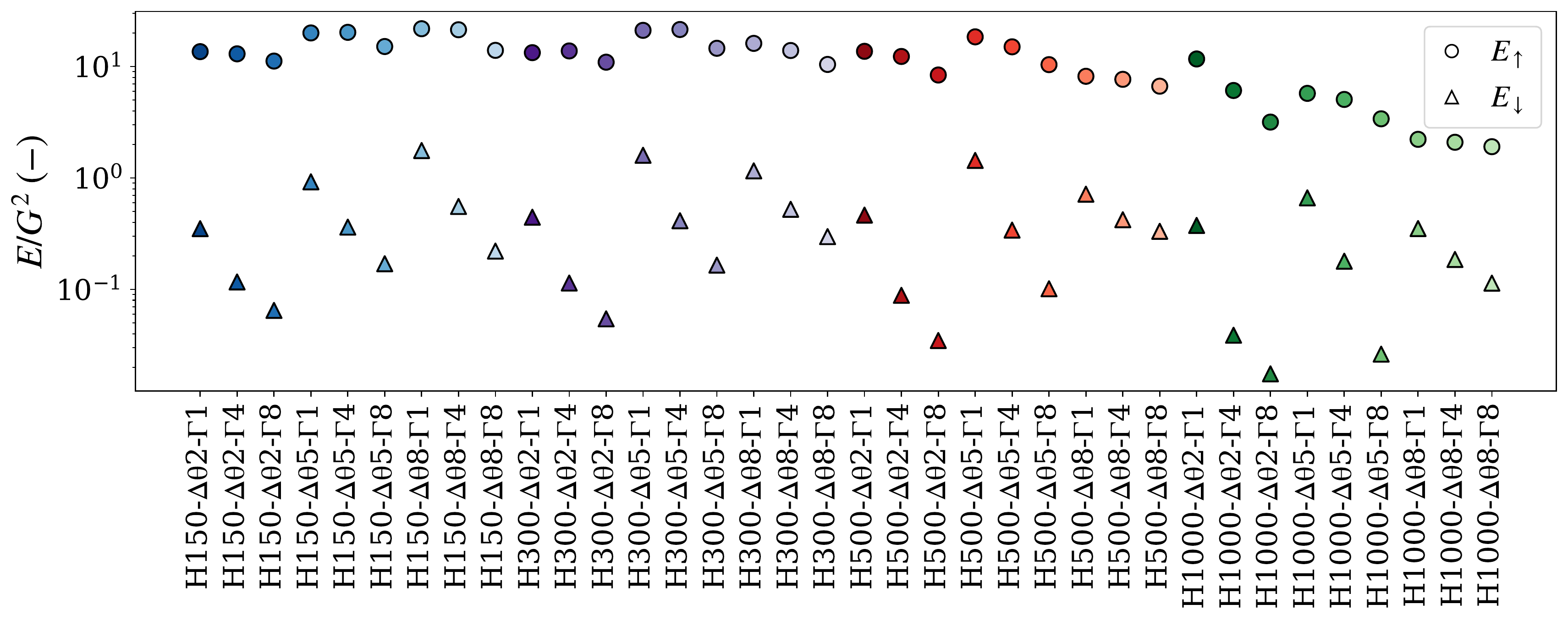}}% 
	\caption{Vertical kinetic energy associated with upward ($E_\uparrow$ -- circle) and downward ($E_\downarrow$ -- triangle) internal waves scaled with the geostrophic wind.}
	\label{fig:reflectivity_2}
\end{figure}

Next, the vertical kinetic energy associated with upward and downward waves is shown in Figure \ref{fig:reflectivity_2}. Overall, $E_\downarrow$ is roughly an order of magnitude lower than $E_\uparrow$. Moreover, the highest values of $E_\downarrow$ are always attained in weakly stratified atmospheres, i.e. when $\Gamma=1$ K km$^{-1}$. This is expected since the RDL does not perform optimally in such a case. Finally, we note that $E_\uparrow$ gradually reduces as $H$ increases. This is related to the fact that the capping-inversion displacement diminishes with $H$, therefore inducing vertical motions with lower magnitude in the free atmosphere.

\newpage

\bibliographystyle{jfm}

\end{document}